%% file: astroph.tex
\documentclass{aa}  

\usepackage{savesym}
\usepackage{graphicx}
\usepackage{txfonts}
\savesymbol{rotate}
\usepackage{deluxetable}
\restoresymbol{DTBL}{rotate}
\usepackage{times}
\usepackage{epsfig}
\usepackage{mathrsfs}
\usepackage{natbib}
\usepackage{amssymb}
\usepackage{amstext}
\bibpunct{(}{)}{;}{a}{}{,}
\usepackage{caption}
\usepackage[english]{babel}
\usepackage{longtable}
\usepackage{url}
\usepackage{color}

\def\ifundefined#1{\expandafter\ifx\csname#1\endcsname\relax}
\ifundefined{ensuremath}\def\ensuremath#1{\relax\ifmmode{#1}}
\else${#1}$\fi\else\relax\fi
\ifundefined{nuc}\def\nuc#1#2{\relax\ifmmode{}^{#1}{\protect\mathrm{#2}}
\else${}^{#1}$#2\fi}\else\relax\fi

\usepackage{morefloats}
\usepackage{inputenc}

\newcommand{\Msun}{\mathrm{M}_{\odot}}

\newcommand{\dmfiften}{\Delta {m}_{15}}
\newcommand{\sbv}{s_{{BV}}}
\newcommand{\Ni}{$\element[][56]{Ni}$}
\begin{document}
%
%
%
\title{Two transitional type~Ia supernovae located in the Fornax
cluster member NGC~1404: SN~2007on and SN~2011iv\thanks{This
 work is based in part on observations made at the Las Campanas 
Observatory, including the 6.5~m Magellan Telescope. It is also based 
in part on spectra collected at the European 
Organization for Astronomical Research in the Southern 
Hemisphere, Chile (ESO Programmes 184.D-1151, 184.D-1152, 
088.D-0222, 184.D-1140, 080.A-0516, 080.C-0833); 
the 8.1~m Gemini-S Telescope (Program GS-2011B-Q-1); 
the Nordic Optical Telescope (Program 44-024); and the 
NASA/ESA {\em Hubble Space Telescope} (GO-12592), 
obtained at the Space Telescope Science Institute (STScI), which is 
operated by the Association of Universities for Research in 
Astronomy, Inc., under National Aeronautics and Space Administration 
(NASA) contract NAS 5-26555 
(Programs GO-12298, GO-12582, GO-12592, 
GO-13286, and GO-13646). {\it Swift} spectroscopic 
observations were performed under program GI-5080130.
Some of the data presented herein were obtained at the W. M. Keck
Observatory, which is operated as a scientific partnership among the
California Institute of Technology, the University of California, and
NASA; the observatory was made possible by the generous financial
support of the W. M. Keck Foundation.
}}

 \author{
 C. Gall\inst{1,2}
 \and 
 M. D. Stritzinger\inst{1}
 \and
 C. Ashall\inst{3}
\and 
 E. Baron\inst{4}
\and
C. R. Burns\inst{5}
\and
P. Hoeflich\inst{6}
 \and
 E.  Y. Hsiao\inst{6,1}
\and
P. A. Mazzali\inst{3}
 \and 
 M. M. Phillips\inst{7}
 \and
  A. V. Filippenko\inst{8}
 \and 
 J. P. Anderson\inst{9}
 \and
 S. Benetti\inst{10}
 \and
P. J. Brown\inst{11}
\and 
A. Campillay\inst{7}
\and
P. Challis\inst{9}
 \and 
C. Contreras\inst{7,1}
\and 
N. Elias de la Rosa\inst{10}
\and
G. Folatelli\inst{11}
\and
 R. J. Foley\inst{12} 
 \and
 M. Fraser\inst{13}
 \and 
  S. Holmbo\inst{1}
  \and
  G. H. Marion\inst{14}
  \and
 N. Morrell\inst{7}
 \and
 Y.-C. Pan\inst{12}
 \and  
 G. Pignata\inst{15,16}
 \and
 N. B. Suntzeff\inst{11}
\and
F. Taddia\inst{17}
 \and 
 S. Torres Robledo\inst{7}
 \and 
S. Valenti\inst{18}
}
%
\institute{Department of Physics and Astronomy, Aarhus University, 
Ny Munkegade 120, DK-8000 Aarhus C, Denmark 
\and 
Dark Cosmology Centre, Niels Bohr Institute, 
University of Copenhagen, Juliane Maries Vej 30, 
2100 Copenhagen \O, Denmark
\and
Astrophysics Research Institute, Liverpool John Moores University, IC2, Liverpool Science Park, 146 Brownlow Hill, Liverpool L3 5RF, UK
\and
Homer L. Dodge Department of Physics and Astronomy, University of Oklahoma, 440 W. Brooks, Rm 100, Norman, OK 73019-2061, USA
\and 
Observatories of the Carnegie Institution for Science, 813 Santa Barbara St., Pasadena, CA 91101, USA 
\and
Department of Physics, Florida State University, Tallahassee, FL 32306, USA
\and 
Carnegie Observatories, Las Campanas Observatory, 601 Casilla, La Serena, Chile 
\and
Department of Astronomy, University of California, Berkeley, CA 94720-3411, USA
\and
European Southern Observatory, Alonso de C\'ordova 3107, Casilla 19, Santiago, Chile
\and
INAF-Osservatorio Astronomico di Padova, vicolo dell Osservatorio 5, 35122 Padova, Italy
\and
George P. and Cynthia Woods Mitchell Institute for Fundamental Physics \& Astronomy, Texas A\&M University, Department of Physics, 4242 TAMU, College Station, TX 77843
\and
 Department of Astronomy and Astrophysics, University of California, Santa Cruz, CA 95064, USA
 \and
School of Physics, O'Brien Centre for Science North, University College Dublin, Belfield, Dublin 4, Ireland
\and
Department of Astronomy, University of Texas, Austin, TX 78712, USA
\and
Departamento de Ciencias Fisicas, Universidad Andres Bello, Avda. Republica 252, Santiago, Chile
\and
Millennium Institute of Astrophysics, Chile
\and
The Oskar Klein Centre, Department of Astronomy, Stockholm University, AlbaNova, 10691 Stockholm, Sweden
\and
Department of Physics, University of California, Davis, CA 95616, USA }
   \date{Received; accepted}
%
\abstract
{We present an analysis of ultraviolet (UV) to near-infrared 
observations of the fast-declining  Type Ia supernovae (SNe~Ia) 
2007on and 2011iv, hosted by the Fornax cluster member 
NGC~1404. The $B$-band light curves of SN~2007on and 
SN~2011iv  are characterised by $\Delta m_{15}(B)$ decline-rate 
values of 1.96 mag and 1.77 mag, respectively. Although they 
have similar decline rates, their peak $B$- and $H$-band 
magnitudes differ by $\sim 0.60$ mag and $\sim 0.35$ mag, 
respectively. After correcting for the luminosity vs. decline rate 
and the luminosity vs. colour relations, the peak $B$-band and 
$H$-band light curves provide distances that differ by 
$\sim14$\% and $\sim9$\%, respectively. These findings 
serve as a cautionary tale for the use of transitional 
SNe~Ia located in early-type hosts in the quest to measure 
cosmological parameters. Interestingly, even though SN~2011iv 
is brighter and bluer at early times, by three weeks past maximum 
and extending over several months, its $B-V$ colour is 0.12 mag 
redder than that of SN~2007on. To reconcile this unusual behaviour, we 
turn to guidance from a suite of spherical one-dimensional Chandrasekhar-mass delayed-detonation
explosion models. In this context, $^{56}$Ni production depends 
on both the so-called transition density and the  central density of 
the progenitor white dwarf. To first order, the transition density drives 
the luminosity--width relation, while the central density is an important 
second-order parameter. 
Within this context, the differences in the $B-V$ color evolution along the Lira regime suggests the progenitor of SN~2011iv had a higher central density than SN~2007on.}
 %
 \titlerunning{CSP observations of SN~2007on and SN~2011iv}    
 \authorrunning{Gall et al.}     
   \maketitle
%
%
\section{Introduction}

Type~Ia supernovae 
 are fundamental extragalactic 
distance indicators used to map out the expansion history of the 
Universe. In doing so, they provide an accurate estimate of the 
Hubble constant \citep{2016arXiv160401424R} and a means to 
study the nature of dark energy \citep[e.g.,][]{2012ApJ...746...85S, 
2014A&A...568A..22B, 2016ApJS..224....3N}. Although details on 
their origin remain a mystery, SNe~Ia have long been considered 
to arise from the thermonuclear disruption of carbon-oxygen 
white dwarfs in binary star systems \citep{1960ApJ...132..565H}.

Today, SN~Ia cosmology is no longer limited by sample size but 
rather by a subtle matrix of systematic errors. In order for future 
SN~Ia experiments to differentiate between various static and/or 
time-dependent types of dark energy, models will require an improvement 
in the precision of SN~Ia peak luminosity measurements to 
$\sim 1$\% out to $z \approx 1.0$ \citep{2006astro.ph..9591A}. 
Contemporary SN~Ia experiments achieve a precision
 of $\sim 5$\% \citep{2012MNRAS.425.1007B, 
2015Sci...347.1459K}. To reach percent-level distances will 
require a rest-frame near-infrared (NIR) SN~Ia sample located in the 
Hubble flow where relative peculiar motions are small \citep{2010AJ....139..120F,
2011ApJ...731..120M}, along with improved photometric calibrations and 
(ultimately) a deeper theoretical understanding of the progenitors 
and explosion mechanism than currently exists. 

Current efforts to improve upon the most dominant systematics
are focusing on the construction of homogeneous samples of 
low-redshift SNe~Ia \citep[e.g.,][]{2006PASP..118....2H, 2010ApJS..190..418G, 
2011AJ....142..156S}, improvements in photometric calibration 
techniques \citep[e.g.,][]{2010SPIE.7735E..64R, 2015MPLA...3030030S}, 
and an expanded understanding of SN~Ia intrinsic colours and 
dust corrections \citep[e.g.,][]{2010AJ....139..120F, 2011ApJ...731..120M, 
2014ApJ...789...32B}. Additional efforts to further understand 
SN~Ia progenitor systems and their explosion physics are being 
made through, amongst others, focused studies of their 
ultraviolet \citep[UV; e.g.,][]{2016MNRAS.461.1308F} and NIR 
properties \citep[e.g.,][]{2013ApJ...766...72H, 
2015ApJ...806..107D}, as well as studies probing their immediate 
circumstellar environments \citep[e.g.,][]{2011Sci...333..856S, 
2015MNRAS.452.4307P}.

Over the past 30 years, significant observational efforts have 
revealed the existence of multiple subclasses of SNe~Ia. The 
first clear indication of departure from homogeneity arrived with the 
study of the low-luminosity SN~1986G, located in Centaurus A 
\citep[e.g.,][]{1987ApJ...316L..81B, 1987PASP...99..592P, 2016MNRAS.463.1891A}.
Five years later, with the discovery of both the high-luminosity 
SN~1991T \citep{1992ApJ...384L..15F, 1992ApJ...387L..33R, 
1992AJ....103.1632P} and the low-luminosity 
SN~1991bg \citep{1992AJ....104.1543F, 1993AJ....105..301L, 
1996MNRAS.283....1T}, it became obvious that SNe~Ia exhibit 
significant diversity \citep[see, e.g.,][]{1997ARA&A..35..309F}.

At the faint end of the SN~Ia luminosity 
distribution, one finds subtypes including the 
intrinsically red and subluminous SN~1991bg-like SNe~Ia 
\citep{1992AJ....104.1543F, 1993AJ....105..301L, 
1996MNRAS.283....1T} and the so-called {\em transitional}
objects. The light curves of SN~1991bg-like SNe~Ia decline rapidly and 
are characterised by NIR light curves that exhibit 
a single maximum after the optical peak magnitudes are reached. 
In contrast, the transitional subtypes have NIR light curves
exhibiting two maxima, similar to those of normal SNe~Ia, with their first 
maximum being reached before the peak of the optical bands. 
Additionally, they are brighter than the SN~1991bg-like SNe~Ia, despite
having similar $\Delta m_{15}(B)$ values of 
1.7--2.0 mag 
\citep[see, e.g.,][]{2016MNRAS.460.3529A}. 
Examples of transitional SNe~Ia include  
SN~2003hv \citep{2009A&A...505..265L}, 
iPTF~13ebh \citep{2015A&A...578A...9H}, and 
SN~2015bp \citep{2017MNRAS.466.2436S}.

In this paper, we focus on the Type~Ia SN~2007on and 
SN~2011iv, both located in the Fornax cluster member 
NGC~1404. Within the context of the light-curve luminosity vs. width 
relation characterised by the light-curve decline-rate parameter 
$\Delta m_{15}(B)$ \citep{1993ApJ...413L.105P}, these two objects 
are located at the faint end of the SN~Ia luminosity distribution. 
However, with $\Delta m_{15}(B)$ measured values of $\sim 1.7$--2.0 mag, 
along with a number of other photometric and spectroscopic properties 
as discussed below, SN~2007on and SN~2011iv are most akin to 
the transitional SN~Ia iPTF~13ebh \citep{2015A&A...578A...9H}.
As both SN~2007on and SN~2011iv were hosted by the same 
galaxy, they offer a rare opportunity to test the precision of 
transitional SNe~Ia as distance indicators.

Here, comprehensive observations (Sect.~\ref{SEC:OBS}) 
of SN~2007on and SN~2011iv are presented and studied with 
the intent to gain a better understanding on the nature of transitional 
SNe~Ia and their progenitors. In Sect.~\ref{LCanalysis}, a detailed 
analysis of the photometric dataset is presented, including a study 
of their light-curve behaviour, their $B-V$ colour evolution, estimates 
of their host-galaxy reddening, as well as their $^{56}$Ni masses 
derived from constructed bolometric (UV/optical/IR, hereafter 
UVOIR) light curves. In addition, given 
that both objects are very similar and located in the same galaxy, 
they provide an excellent opportunity to test the methods used to 
estimate their distance (Sect.~\ref{SSEC:DIST}). A detailed spectral 
analysis including modelling of the optical maximum-light spectra of 
both SNe~Ia is then presented in Sect.~\ref{SEC:SPA}, followed by 
a discussion in Sect.~\ref{SEC:DISCUSSION}. Our conclusions 
are summarised in Sect.~\ref{SEC:CCL}. 
%
\subsection{Supernovae SN~2007on and SN~2011iv}
\label{SSEC:INTROSNE}
\begin{figure}
\includegraphics[width=15cm]{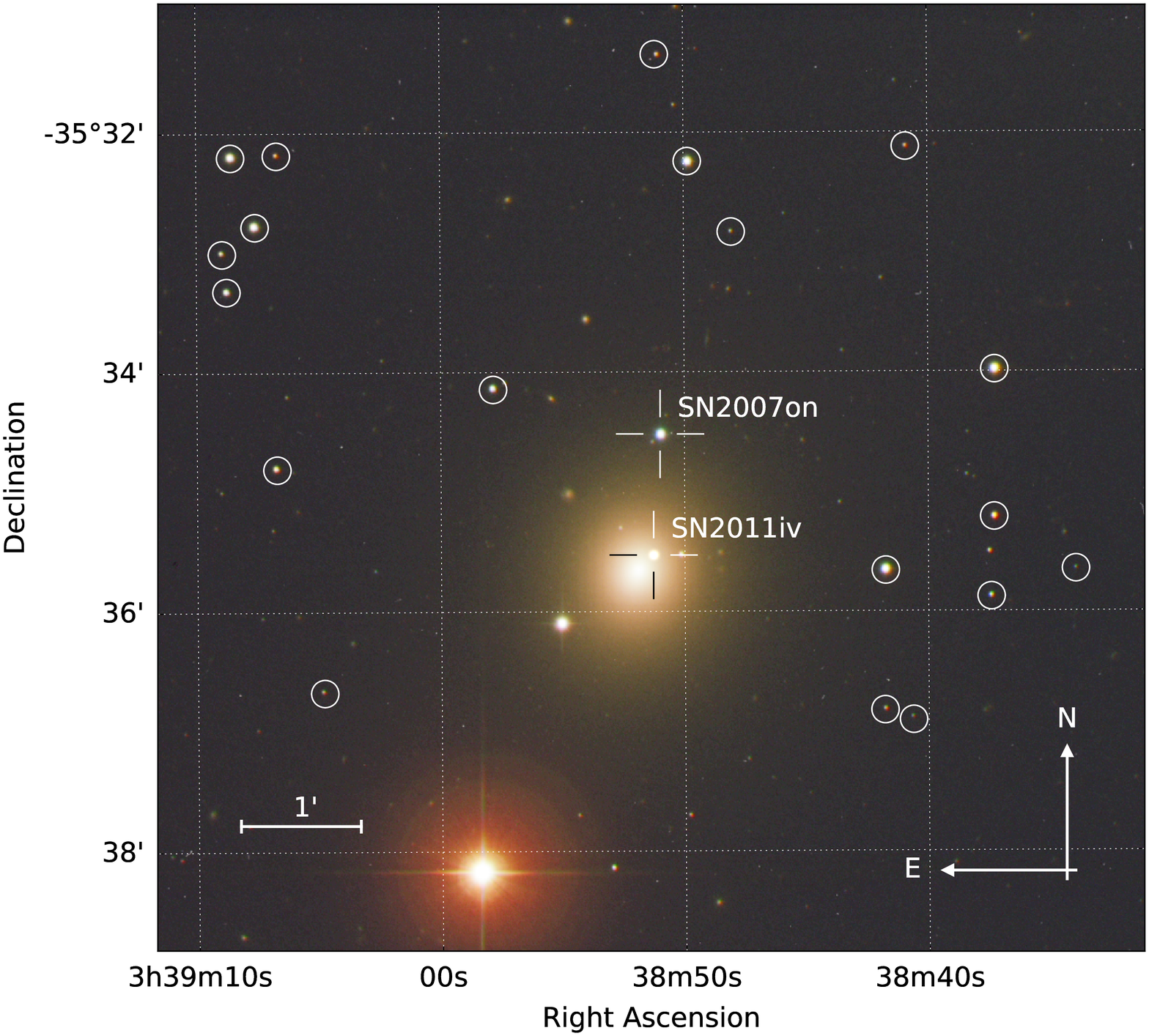}
\vspace{-1.0 cm}
\caption{Composite image of NGC~1404 with the positions of 
SN~2007on, SN~2011iv, and a number of local sequences stars 
indicated.
}
 \label{FIG:FC}
 \end{figure}

%
SN~2007on and SN~2011iv occurred in the Fornax cluster member 
and early-type (E1) elliptical galaxy NGC~1404 \citep{1994AJ....108.2128C}, 
and their location along with a number of local sequence stars is 
presented in Figure~\ref{FIG:FC}. Both  SNe~Ia are northwest 
of the centre of NGC 1404 within the projected bow-shock region of 
the galaxy, which itself is on an infall course directed toward the 
central galaxy of the Fornax cluster, NGC~1399 \citep[e.g.,][]{2005ApJ...621..663M}. 
Interestingly, NGC~1404 has the lowest dust to stellar flux ratio of 
galaxies in the \emph{KINGFISH} survey \citep{2011ApJ...738...89S}, 
suggesting the absence of significant amounts of dust.

SN~2007on was discovered on 5.25 November 2007 UT
\citep{2007CBET.1121....1P} and identified as a subluminous, 
fast-declining SN~Ia \citep{2007ATel.1263....1G, 2007CBET.1131....1M}. 
An X-ray source close to the position of SN~2007on identified in 
archival \emph{Chandra} images was suggested, but not 
unambiguously confirmed, to be the progenitor of SN~2007on 
\citep{2008Natur.451..802V, 2008MNRAS.391..290R}. 

Four years after the discovery of SN~2007on, the Type~Ia 
SN~2011iv was discovered in NGC~1404 on 2.57 December 
2011 UT \citep{2011CBET.2940....1D}. At a projected radius of 
$\lesssim$ 2 kpc from the host's centre, its location is close to the 
effective radius of the galaxy \citep{1994ApJ...436L..75L}, 
unlike SN~2007on, which is at a projected radius of about 8 kpc 
from the centre. Visual-wavelength spectroscopy indicated that 
SN~2011iv was a young, rising SN~Ia \citep{2011CBET.2940....1N, 
2011CBET.2940....3S}. 

There are numerous direct distance measurements obtained with 
various different methods published in the literature and compiled 
on NED\footnote{NED is the NASA/IPAC Extragalactic Database 
\citep{2017AJ....153...37S}.} for NGC~1404.
This includes surface brightness fluctuation (SBF) distances 
ranging from about 15.7 Mpc up to 22.2 Mpc \citep[e.g., ][]{1991ApJ...373L...1T, 
2009ApJ...694..556B, 2002ApJ...564..216L}; as well as among others: 
Fundamental Plane, Planetary Nebula Luminosity Function, and Tully 
Fisher distance estimates (see NED). 
Here we adopt a distance of  $17.9\pm2.9$~Mpc (corresponding to a 
distance modulus of $\mu = 31.27\pm0.20$ mag), which is consistent 
with the  Advanced Camera for Surveys (ACSs) Fornax Cluster Survey 
estimate based on the half-light radii of Globular clusters \citep{2010ApJ...715.1419M}, 
though with a more conservative  uncertainty.

%
\section{Observations}
\label{SEC:OBS}
%
Detailed optical and NIR light curves of SN~2007on obtained by 
the first phase of the Carnegie Supernova Project (CSP-I, 
2004--2009; \citealt{2006PASP..118....2H}) were published by
\citet{2011AJ....142..156S}, while visual-wavelength spectra 
obtained at early and late times are presented by 
\citet{2013ApJ...773...53F} and \citet{2010Natur.466...82M}. 
Here we complement these observations with two previously unpublished 
visual-wavelength spectra at $+$11~d and $+$73~d obtained with 
the ESO-NTT ($+$ EMMI: ESO Multi-Mode Instrument) and two NIR spectra
with the ESO-NTT($+$ SOFI: Son of ISAAC).  
Furthermore, we present recalibrations of the spectra and updated photometry 
of SN~2007on computed using a more accurate measurement of 
the local-sequence photometry, as well a vastly improved reduction 
of the late-phase Gemini-South spectrum of SN~2007on.

Over the course of a second instalment of the CSP, referred to as 
CSP-II (2011--2015), detailed optical and NIR follow-up imaging and 
spectroscopy were obtained for SN~2011iv, extending from $-9$~d to 
$+$260~d relative to the epoch of $B$-band maximum. Combining 
the CSP-II observations with five epochs of UV-optical spectroscopy 
obtained with the UltraViolet Optical Telescope (UVOT) aboard the 
{\em Swift} satellite \citep{2005SSRv..120..165B, 2005SSRv..120...95R} 
and seven epochs of {\em Hubble Space Telescope (HST)} ($+$ STIS: 
Space Telescope Imaging Spectrograph) UV/visual-wavelength 
spectroscopy, and a large number of additional visual-wavelength 
and NIR spectra obtained through various facilities, yields the most 
detailed datasets yet obtained for a transitional SN~Ia.
%
\subsection{Ultraviolet, optical and NIR imaging}
%
UV $uvw2$-, $uvm2$-, and $uvw1$-band imaging of both 
SN~2007on and SN~2011iv were obtained with {\em Swift} ($+$ 
UVOT). Photometry of SN~2007on and SN~2011iv was computed 
following the method described in detail by \citet{2014Ap&SS.354...89B}, 
who use the calibration published by \citet{2011AIPC.1358..373B}. 
The {\em Swift} UVOT images and photometry are also available 
as part of the {\em Swift} Optical Ultraviolet Supernova Archive 
\citep[SOUSA;][]{2014Ap&SS.354...89B}.
In the process of computing definitive photometry, science images 
of SN~2011iv were used for host-galaxy subtraction of 
SN~2007on's  science images, while science images of 
SN~2007on were used for host-galaxy 
subtraction of SN~2011iv's science images. Definitive UVOT 
$uvw2$-, $uvm2$-, and $uvw1$-band photometry is provided in 
Table~\ref{TAB:UVOT}; we note that the photometry of 
SN~2007on presented here supersedes the photometry published 
by \citet{2010ApJ...721.1627M}.

Optical $ugriBV$-band imaging of SN~2007on and SN~2011iv 
was obtained with the Henrietta Swope 1.0 m telescope 
($+$ SITe3 direct CCD camera) located at the Las 
Campanas Observatory (LCO). The NIR $YJH$-band imaging 
of SN~2007on was obtained with the Swope ($+$ RetroCam) and 
the Ir\'{e}n\'{e}e du Pont 2.5 m ($+$ WIRC: Wide Field Infrared Camera) 
telescopes \citep{2011AJ....142..156S}, while in the case of 
SN~2011iv all NIR $YJH$-band imaging was taken with RetroCam 
attached to the Ir\'{e}n\'{e}e du Pont telescope. 

The reduction procedure applied to all imaging data is described in 
detail in the final CSP-I SN~Ia data release (Krisciunas et al., in prep.). 
In brief, point-spread function (PSF) photometry is computed 
differentially to a local sequence of stars in the field of NGC~1404. 
The optical local sequence is calibrated relative to \citet{1992AJ....104..372L} 
($BV$) and \citet{2002AJ....123.2121S} ($ugri$) standard-star 
fields observed over multiple photometric nights. 
The NIR $J$-band and $H$-band local sequences were calibrated 
relative to the \citet{1998AJ....116.2475P} standard stars, while 
the $Y$-band local sequence was calibrated relative to standard 
$Y$-band magnitudes computed using a combination of stellar 
atmosphere models \citep{2003IAUS..210P.A20C} with the $J-K_s$ 
colours of the \citeauthor{1998AJ....116.2475P} standard-star 
catalog \citep{2006PASP..118....2H}.

Absolute optical photometry in the {\em standard} photometric 
system and NIR photometry in the {\em natural} photometric 
system of the local sequences used to compute photometry of 
SN~2007on and SN~2011iv are provided in Table~\ref{tab:opticallocalseq} 
and Table~\ref{tab:nirlocalseq}, respectively. The accompanying 
uncertainties corresponding to the weighted average of the instrumental 
errors obtained over the various nights in which photometric 
standard fields were observed.

Prior to computing photometry of the SNe~Ia, host-galaxy template 
subtraction was performed on each science image. Deep 
host-galaxy template images were obtained with the du Pont telescope 
under excellent seeing conditions well after the SNe~Ia faded 
beyond the detection limit. With galaxy-subtracted science images 
in hand, final optical and NIR photometry of SN~2007on and 
SN~2011iv was computed on the CSP {\em natural} system. The 
optical and NIR photometry of SN~2007on and SN~2011iv is 
provided in Table~\ref{tab:opticalphotometry} and Table~
\ref{tab:nirpot}, respectively.\footnote{Definitive photometry of the 
local sequence stars and the SNe~Ia is also available electronically 
on the CSP webpage \url{http://csp.obs.carnegiescience.edu/
data/}} The associated uncertainty of each photometric data point 
is computed by adding in quadrature of the instrumental error and 
the nightly zero-point error. 
Final {\em Swift} UV and CSP optical and NIR light curves of 
SN~2007on and SN~2011iv are presented in Figure~\ref{FIG:LCCOMP}, 
including best-fit template light curves computed by {\tt SNooPy} 
\citep{2011AJ....141...19B}.
%
\subsection{Ultraviolet, visual-wavelength, and NIR spectroscopic observations}
\label{SSEC:UONSP}
%
Multiple epochs of spectroscopy were obtained of SN~2007on and 
SN~2011iv with a variety of facilities; details summarising these 
observations are provided in Table~\ref{T:SPJ07ON} and 
Table~\ref{T:SPJ11IV}, respectively. 

The spectroscopic time-series of SN~2007on consists of 23 epochs 
ranging from $-$4.0~d to $+$380~d. The  early-phase spectra were 
published by \citet{2013ApJ...773...53F}, while \citet{2010Natur.466...82M} 
presented the two oldest nebular-phase spectra. In addition to these, 
presented here for the first time are two additional epochs of spectra 
obtained with the NTT ($+$ EMMI). These observations were conducted 
between $+$12~d and $+$74~d, and the data were reduced following 
standard procedures within the IRAF\footnote{IRAF is distributed by the 
National Optical Astronomy Observatories, which are operated by the 
Association of Universities for Research  in Astronomy, Inc., under 
cooperative agreement with the US National Science Foundation.} 
environment. Additionally, we present two epochs of NIR spectroscopy 
also obtained with the NTT ($+$ SOFI), and these data 
were reduced as described by \citet{2015A&A...579A..40S}.
 
In the case of SN~2011iv, a total of 23 epochs of visual-wavelength 
spectroscopy was obtained between $-$7~d and $+$276~d.
In addition, twelve epochs of UV spectroscopy were procured, 
yielding one of the most comprehensive UV datasets yet obtained 
for a fast-declining SN~Ia. This includes five epochs ($-$5~d to $+$1~d) 
taken with {\em Swift} ($+$UVOT; wavelength range 0.19--0.68~$\mu$m) 
and seven epochs (ranging from $-0.4$~d to $+$29~d) taken with 
{\it HST} ($+$ STIS; wavelength range 0.18--1.023~$\mu$m;  GO-12592). 
Additionally, 16 epochs of NIR spectroscopy of SN~2011iv were obtained 
between $-$2~d and $+$141~d.  

Spectral data were reduced in the standard manner. {\em Swift} 
spectra were reduced as described by Pan et al. (in prep.), 
{\it HST} data reductions follow the prescription of \citet{2012ApJ...753L...5F}, 
and the ground-based visual-wavelength data were reduced 
following the methods described by \citet{2006PASP..118....2H}. 
NIR spectroscopy  obtained by the Magellan 6.5~m 
Baade telescope ($+$ FIRE; Folded Port Infrared Echellette)
was reduced using the {\tt FIREHORSE} software package developed by Rob 
Simcoe \citep[see][]{2013ApJ...766...72H}, while  data taken with 
the VLT  ($+$ ISAAC: Infrared Spectrometer And Array Camera) 
and the NTT ($+$ SOFI) were reduced as described by 
\citet{2015A&A...579A..40S}. 

The UV spectroscopic time-series of SN~2011iv is plotted in 
Fig.~\ref{FIG:UVSPEC11IV} and the visual-wavelength time-series 
of SN~2007on and SN~2011iv are presented in Fig.~\ref{FIG:OPTISPEC11IV}.
Figure~\ref{FIG:NIRSPEC11IV} displays the NIR spectra of SN~2011iv, 
as well as a spectrum of SN~2007on. Finally, the late-phase visual-wavelength 
spectra of both SNe~Ia are presented in Figure~\ref{FIG:OPSPGT1107}.  
%
%
\section{Photometric Analysis}
\label{LCanalysis}
%
\subsection{Light curves}
\label{SSEC:LCS}
The optical light curves of both objects are well sampled around 
the primary maximum, and high-cadence photometry extends to 
$\sim +80$~d for SN~2007on and $\sim +120$~d for 
SN~2011iv. The NIR light curves of SN~2007on are also densely 
sampled within the first three months of its evolution.  However, the 
NIR light curves of SN~2011iv mainly cover the rise and fall around 
the primary (peak) maximum, with the secondary maximum only 
partially covered within a few days of its peak value. Additionally, 
individual NIR photometric points were obtained between $+$80~d 
and $+$120~d.

\begin{figure*}
\includegraphics[width=18cm]{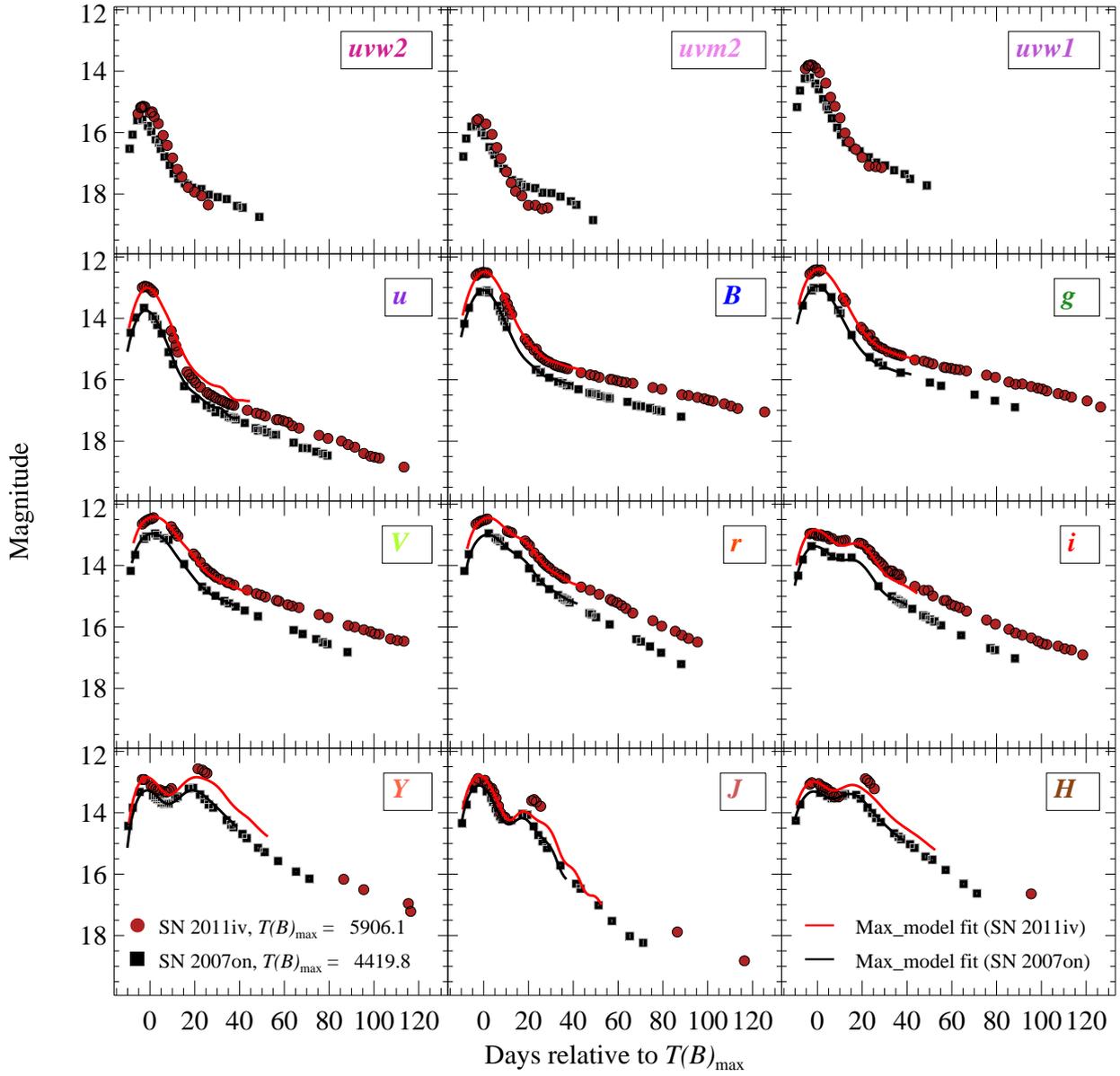}
\caption{UV, optical, and NIR light curves of SN~2007on (squares) 
and SN~2011iv (circles). The black and red solid lines  
represent the best {\tt SNooPy} ``max model'' fits. 
}
 \label{FIG:LCCOMP}
 \end{figure*}

The optical light curves of  SN~2007on and SN~2011iv exhibit, 
compared to normal SNe~Ia, a relatively quick decline from 
maximum light. In the case of the $riYJH$-band light curves, a 
transition to a secondary maximum around 20~d past the primary 
maximum is apparent, which is followed by a linear decline in 
brightness extending to late phases \citep[e.g.,][]{2009A&A...505..265L}. 
For both SNe~Ia, the $H$-band secondary maxima are nearly 
equal to the luminosity of the primary maxima, while the $Y$-band 
secondary maxima exceed the brightness of the primary maxima. 
In Figure~\ref{FIG:LCCOMP}, the light-curve shapes for the two SNe~Ia 
are very similar, yet curiously, SN~2011iv is brighter than SN~2007on 
in all passbands. The observed peak brightness difference 
is found to decline gradually as a function of wavelength from $\sim 0.6$ 
mag in the $B$ band to $\sim 0.35$ mag in the $H$ band. 

Estimates of key light-curve parameters are computed using the 
SN~Ia template light-curve fitting package {\tt SNooPy} 
\citep{2011AJ....141...19B}. {\tt SNooPy} offers several different 
models to fit SN~Ia light curves, and the most appropriate model to 
use depends on the particular question being addressed. To 
estimate the time of maximum brightness, $T_{\mathrm{max}}$, the peak 
magnitude in each passband, $m_{X}$ (where $X= {u, g, r, i, B, V, 
Y, J, H}$ corrected for time dilation, Galactic reddening, and K 
corrections) we use {\tt SNooPy}'s ``max model'' 
\citep[see][Eq. 5]{2010AJ....140.2036S}, adopting 
the colour-stretch parameter, $s_{BV}$.\footnote{Recently 
introduced by \citet{2014ApJ...789...32B}, the colour-stretch 
parameter, $s_{BV}$, is a dimensionless parameter defined to be 
the epoch after maximum light when the $B-V$ colour reaches its 
maximum value, divided by 30 days.} For reasons outlined by
\citet{2014ApJ...789...32B}, $s_{BV}$ is preferred over $\dmfiften$ 
to obtain the aforementioned light-curve parameters in the case 
of fast-declining SNe~Ia. The results are summarised in Table~\ref{T:LCPARAM}.

Figure~\ref{FIG:LCCOMP} compares the resulting best template 
light-curve fits to the optical and NIR light curves. With an $s_{BV}$
value of $0.57\pm0.04$ for SN~2007on and $0.64\pm0.04$ 
for SN~2011iv, both objects are found to evolve photometrically like 
fast-declining SNe~Ia (i.e., $s_{BV} \lesssim 0.7$ and/or $\dmfiften 
\gtrsim 1.7$ mag). This is confirmed from direct Gaussian process 
spline fits to the $B$-band light curves of each object, yielding 
$\dmfiften(B)$ values of $1.96\pm0.01$ mag for SN~2007on and 
$1.77\pm0.01$ mag for SN~2011iv. The measured light-curve 
parameters for SN~2007on and SN~2011iv compare well with 
other works in the literature \citep[e.g.,][]{2011AJ....142..156S, 
2012ApJ...753L...5F, 2014ApJ...789...32B}; although specific light-curve 
parameter values vary, both SNe~Ia are undoubtedly identified 
as subluminous.

Examination of the NIR light curves of both SNe~Ia reveals that 
they peak prior to the epoch of $B$-band maximum. This, together 
with the presence of a secondary maximum, are key photometric 
characteristics of a normal SN~Ia \citep{2009AJ....138.1584K}. In 
contrast, fast-declining SN~1991bg-like SNe~Ia only exhibit a single NIR 
maximum that peaks after the time of $B$-band maximum. In summary, 
the measured $s_{BV}$ values and the spectral characteristics 
discussed below, indicate SN~2007on and SN~2011iv are both 
similar to SN~2003gs \citep{2009AJ....138.1584K}, SN~2003hv 
\citep{2009A&A...505..265L}, and iPTF~13ebh \citep{2015A&A...578A...9H},  
all of which are transitional SNe~Ia.

Figure~\ref{FIG:LWR} displays the luminosity vs. decline-rate relation 
parametrised by $\dmfiften$ (left) and $s_{BV}$ (right) for an 
extended sample of SNe~Ia observed by the CSP-I \citep[e.g.,][]
{2010AJ....139..519C, 2011AJ....142..156S} as well as the  
fast-declining SN~1986G \citep{1987PASP...99..592P} and 
SN~1991bg \citep{1992AJ....104.1543F, 1993AJ....105..301L,
1996MNRAS.283....1T}. Both SN~2007on and SN~2011iv are 
brighter than expected for typical fast-declining SNe~Ia and are 
therefore located closer to the faint end of the luminosity 
decline-rate relation of normal SNe~Ia. 
%
\subsection{$B-V$ and ultraviolet colour evolution}
\label{SSC:BVCOL}
%
Figure~\ref{FIG:BVCOMP} compares the $B-V$ colour evolution of 
SN~2007on and SN~2011iv to that of the normal SN~2006dd and 
the SN~1991bg-like SN~2006mr (both hosted in Fornax~A; see 
Sect.~\ref{SSEC:INTROSNE}), and the transitional SN~Ia 
iPTF13ebh \citep{2015A&A...578A...9H}. The comparison SNe~Ia 
act as representatives for each SN~Ia subgroup, and are chosen 
because of either their close location (member of the Fornax cluster)
or their excellent data coverage. 

While overall the $B-V$ colour evolution of all of these objects 
follows a very similar morphology, there are differences in the $B-V$ 
colour values apparent for the different types of SNe~Ia. Typically, 
at early epochs (before $B$-band maximum), the $B-V$ colours 
have their bluest values. However, as the light curves evolve past 
maximum, the colour values abruptly become redder again, 
successively reaching their maximum over a period of 10~d (in 
the case of SN~2006mr) to 20~d for the other SNe~Ia. Thereafter, 
the $B-V$ colour curves are characterised by a nearly linear decline 
from red to blue colours, again over the period $\sim +30$~d to +90~d. 

Around $B$-band maximum and until the SNe~Ia reach their respective 
maximum $B-V$ value, the colours vary in the sense that more-luminous
SNe~Ia exhibit bluer colours. This effect is likely tied to the 
$^{56}$Ni production, the mixing of $^{56}$Ni, and hence the 
ionisation state of the ejecta \citep{2007ApJ...656..661K}.
Quantifying the $B-V$ colours at maximum light ($+$0~d), we find 
values ranging from $\sim -0.07\pm0.01$ mag for 
SN~2006dd, to between $\sim 0.01\pm0.01$ and 
$0.08\pm0.01$ mag for the transitional objects, to $\sim
 0.73\pm0.01$ mag for the SN~1991bg-like SN~2006mr. As the 
SNe~Ia evolve beyond maximum light, the colour differences  
become even more discrepant (see Fig.~\ref{FIG:BVCOMP}). The 
subluminous SN~2006mr reaches its reddest $B-V$ colour value 
already at about $+$10~d, while the other objects peak later, around 
$+$20~d. 
Beginning around $+$20~d and extending to beyond 
$+$85~d, the $B-V$ colours of SN~2007on are $\sim 0.12 
\pm 0.01$ mag bluer than the colours of SN~2011iv, despite 
SN~2007on being fainter at $B$-band maximum. In Sect.~\ref{modelcomparison} 
we speculate that this difference is due to intrinsic variations 
between the progenitor's central densities, and is not related to any 
effects of dust. 

To further assess the colour evolution of SN~2007on and 
SN~2011iv, we examine their UV colours. We choose to use
the {\em Swift} $uvm2$ band because it gives the highest contrast 
to the optical bands. This is because it has a sharper cutoff at the 
long-wavelength end of its passband as compared with the {\em Swift} 
$uvw1$ and $uvw2$ bands. It is therefore more sensitive to UV 
effects \citep[e.g.,][]{2015ApJ...809...37B}. Displayed in 
Fig.~\ref{FIG:NUVC} are the {\em Swift} $uvm2$--CSP $V$ 
colours (left panel), the CSP $u-B$ colours (middle panel), and the 
CSP $u-V$ colours (right panel) of SN~2007on and SN~2011iv. 
At early epochs, where there are 
concurrent observations beginning around $-$3~d, the UV colours 
($uvm2 - V$, $u-V$) of SN~2007on and SN~2011iv are quite similar, 
suggesting that the two objects have nearly the same conditions in their outer 
layers ($\sim 10^{-3}$--$10^{-2}$~M$_\odot$; see Appendix~\ref{SSC:LECSM}). 
As the SNe~Ia evolve, the UV colours of SN~2011iv become 
redder at epochs past +10~d, resembling the $B-V$ colour 
evolution.
%
\subsection{Host-galaxy reddening}
\label{SSEC:CEEX}
%
According to the NASA/IPAC Extragalactic Database (NED), the 
\citet{2011ApJ...737..103S} recalibration of the \citet{1998ApJ...500..525S} 
dust maps suggests that the Galactic extinction component in the 
direction of NGC~1404 is $A_{V}^{\mathrm{MW}}$ = 0.031 mag. 
Adopting a \citet{1999PASP..111...63F} reddening law characterised 
by $R_{V}$ = 3.1, this corresponds to a negligible 
$E(B-V)_{\mathrm{MW}}$ value of 0.01 mag.

Figure~\ref{FIG:MWDUST} (top panel) displays high-resolution 
visual-wavelength spectra of SN~2007on and SN~2011iv zoomed 
in on the wavelength region where \ion{Na}{i}{~D} absorption is 
expected. The $+$5~d spectrum of SN~2007on \citep{2014MNRAS.443.1849S}
was obtained with the Keck-I telescope ($+$ HIRES: High Resolution 
Echelle Spectrometer; \citet{1994SPIE.2198..362V}), and the $+$55~d spectrum 
of SN~2011iv was obtained with the Magellan Clay telescope ($+$ 
MIKE: Magellan Inamori Kyocera Echelle). The spectra of both 
SNe~Ia exhibit very weak Galactic \ion{Na}{i}{~D} absorption, which 
is consistent with the \citet{1998ApJ...500..525S} dust maps. On the 
other hand, no \ion{Na}{i}{~D} absorption components associated 
with the host galaxy are detected, suggesting minimal to no 
host-galaxy reddening. 
An examination of the UV spectra of SN~2011iv (bottom panel of 
Fig.~\ref{FIG:MWDUST}) reveals several narrow interstellar 
absorption lines including \ion{Fe}{ii}  $\lambda\lambda$2344, 
2374, 2382, 2586, and 2600, and  \ion{Mg}{ii} $\lambda\lambda$2796, 
2803. Close inspection of these absorption features show that
they are not associated with material located at the redshift 
of the host galaxy, but rather originate from material within the 
Milky Way located along the line of sight to NGC~1404.

Estimates of host reddening for SN~2007on and SN~2011iv 
are made by comparing the observed peak colours to the 
intrinsic peak colours as defined by a large sample of 
minimally reddened SNe~Ia \citep{2014ApJ...789...32B}. Using 
{\tt SNooPy}'s ``EBV\_method2'' \citep[see, e.g.,][Eq.~6]{2010AJ....140.2036S}, 
we compute template light-curve fits that imply host-galaxy colour 
excess values of $E(B-V)_{\mathrm{host}}^{\mathrm{07on}} 
= -0.06\pm0.01$ (random) $\pm0.06$ (systematic) mag and 
$E(B-V)_{\mathrm{host}}^{\mathrm{11iv}} = -0.02\pm0.01$ 
(random) $\pm0.06$ (systematic) mag, both consistent with 
minimal to no host reddening.

Host-reddening estimates may also be obtained using the Lira 
relation \citep{1996MsT..........3L}. The Lira relation is based on the 
empirical trend that the $B-V$ colours of normal, 
minimally reddened SNe~Ia evolve uniformly from $+$30~d to 
$+$90~d with a linear slope, and it has been found to also apply to a 
sample of fast-declining SNe~Ia \citep[e.g.,][]{2008MNRAS.385...75T}. 
However, \citet{2014ApJ...789...32B} suggest a parameterisation 
of the Lira law using the colour-stretch parameter, $\sbv$, to account for a 
potentially steeper $B-V$ slope such as exhibited by SN~1991bg-like 
SNe~Ia. Here we adopt the Lira-relation as parameterised by 
\citet{2010AJ....139..120F}, which is represented by a dashed red 
line in Figure~\ref{FIG:BVCOMP}. Comparing the offset of the 
observed $B-V$ colour evolution to that of the Lira relation 
between $+$30~d and $+$90~d, we obtain host-galaxy 
colour-excess values of $E(B-V)_{\rm Lira}^{\rm 07on} 
= -0.025\pm0.01$ (random) $\pm0.04$ (systematic) mag 
and $E(B-V)_{\rm Lira}^{\rm 11iv} = 0.1\pm0.01$ (random) 
$\pm0.04$ (systematic) mag. In what follows, SN~2007on and 
SN~2011iv are assumed to have zero host-galaxy extinction.
%
\subsection{Bolometric light curves and $^{56}$Ni estimates}
\label{SSS:BOLOLC}
%
The exquisite observational datasets of SN~2007on and 
SN~2011iv enable us to construct comprehensive bolometric (UVOIR) light 
curves with the use of the bolometric function contained within 
{\tt SNooPy}. The function has several different ways to construct 
the bolometric light curve. In our case we adopted the so-called SED 
(spectral energy distribution) method. In this method the $uBgVriYJH$-band 
light curves  were first fit with Gaussian process spline functions. 
The resultant spline functions were then used to estimate the colors of the 
SN on the same phases of the \citet{2007ApJ...663.1187H} spectral 
templates\footnote{SNooPy  has an updated set of spectral templates 
extending from optical to NIR wavelengths.}. The spectral templates 
were then multiplied by an appropriate  b-spline function, which ensured 
their synthetic colors match the observed colors. Next, flux bluewards of 
the atmospheric cutoff ($\approx$ 3050 \AA) of the spectral templates was 
estimated by  interpolating to the $uvm2$ flux point.  The total flux was then 
obtained by integrating from the effective wavelength of the $uvm2$ 
passband ($\approx$ 2250~\AA) to the red edge of the $H$ band 
($\approx$ 18,800 \AA). Finally the integrated flux was de-reddened 
and placed on the absolute luminosity scale using the adopted
distance discussed in Sect.~\ref{SSEC:INTROSNE}.

Figure~\ref{FIG:BOLO} (left panel) displays the definitive bolometric 
light curves of SN~2007on and SN~2011iv and the corresponding 
luminosity values are listed in Table~\ref{TAB:UVOIR07on} and 
Table~\ref{TAB:UVOIR11iv}, respectively.  The plot also shows an 
indication of the uncertainty in the bolometric luminosity adopting an 
uncertainty in the adopted distance modulus of $\pm$ 0.20 mag. 
Additionally, for SN~2011iv there are a handful of spectrophotometric 
bolometric points constructed by integrating the SN~Ia flux from 
combined UV, optical, and NIR spectra; these are reported in Table~\ref{TAB:UVOIR11iv2}. 
Given that the NIR spectra include the additional wavelength region 
corresponding to the $K_s$ band, the spectrophotometric bolometric 
points tend to have an ever-increasing amount of luminosity relative 
to maximum as compared to the bolometric points obtained from the 
photometry. The time dependence of the difference between the 
photometric vs. spectrophotometric bolometric points reflects the 
increasing fraction of flux emitted at red wavelengths as the SED 
evolves. Also plotted in the left panel of  Fig.~\ref{FIG:BOLO} is the 
bolometric light curve corresponding to the best-fit DD model of 
SN~2011iv presented below (see Sect.~\ref{modelcomparison} 
and Appendix~\ref{appendix:modelcomparisons}).

The bolometric light curves indicate that SN~2011iv is more luminous 
than SN~2007on at all epochs. Quantitatively, the peak luminosity 
obtained for SN~2007on is 
$L_{\mathrm{bol}}^{\mathrm{max}}$ = $( 5.05 \pm 0.8)\times10^{42}$ erg s$^{-1}$,
and for SN~2011iv it is
$L_{\mathrm{bol}}^{\mathrm{max}}$ = $( 9.03 \pm 0.9)\times10^{42}$ erg s$^{-1}$. 
The peak luminosities are used to estimate the ejected \Ni\ 
abundance through application of Arnett's rule \citep{1982ApJ...253..785A}. 
Following the method described by \citet{2005A&A...431..423S}, 
and adopting the same rise time as employed for modelling the 
maximum-light optical spectra (Sect. \ref{SSS:OPTI40D}, Fig.~
\ref{FIG:OP07MODEL}),  \Ni\ yields of  0.24$\pm$0.05
$\Msun$ and 0.43$\pm$0.06 $\Msun$ are computed for 
SN~2007on and SN~2011iv, respectively.

Given the broad wavelength coverage obtained for SN~2011iv, 
we show in Figure~\ref{FIG:BOLO} (right panel) the fraction of the 
total bolometric flux in the UV, optical, and NIR regimes estimated using observed spectroscopy
Clearly, the contribution from optical wavelengths (3050--8630~\AA) 
dominates the total flux at all phases. At maximum brightness, the fraction 
of flux blueward of the atmospheric cutoff is less than 10\%,
while in the NIR extending from 9000~\AA\ out to 24,900~\AA, it
accounts for about 10\%.  By a month  past maximum the UV fraction 
drops to less than 1\%, while the NIR contribution increases to about 30\%.
Also plotted  is the fraction of flux distributed among the different 
wavelength regimes for the best-fit DD model of SN~2011iv, including 
the fraction of flux in the mid-IR (MIR).
%
\section{Type~Ia supernova distance to NGC~1404}
\label{SSEC:DIST}
%
Equipped with the optical and NIR light curves of SN~2007on and 
SN~2011iv, we can use them to independently ascertain the 
distance to their host galaxy, NGC~1404. We focus our efforts on 
determining the relative {\it difference} between the distances inferred 
from SN~2007on and SN~2011iv, rather than on their absolute distances.

Assuming SN~2007on and SN~2011iv are indeed located in 
NGC~1404, two factors are key for explaining the difference in 
their observed flux:  extinction owing to dust 
and the luminosity vs. decline-rate relation 
\citep[][]{1999AJ....118.1766P}. To investigate this, we construct a 
Markov Chain Monte Carlo (MCMC) simulation that models the distance to 
each object as a function of its observed magnitude. For details of 
the procedure, see \citet[][]{2011AJ....141...19B}. To compute the 
host extinction, we use the results from \citet{2014ApJ...789...32B} 
assuming the Gaussian mixture model as a prior for $R_V$. Similar 
to what was discussed previously (Sect.~\ref{SSEC:CEEX}), 
here we also found that both objects only suffer minimal to no host 
reddening. However, it is important to include the uncertainties and 
their covariances in the analysis. From the resulting MCMC chains, 
we compute the statistics on the {\em difference} between the 
distances of SN~2007on and SN~2011iv for each observed passband. 

Interestingly, the distances derived from the optical bands are more 
discrepant between the two SNe than the distances derived 
using the NIR bands. This is demonstrated in Figure~\ref{FIG:DISTANCES},
which shows the probability distribution of the differences in distance 
moduli computed for the $B$- and $H$-band peak luminosities of our 
two SNe after applying the luminosity vs. colour and luminosity 
vs. decline-rate corrections. The probability distributions of the 
difference in distance moduli is narrower in the $H$ band because 
the corrected NIR luminosities are less sensitive to errors in the 
extinction correction. 
Quantitatively, the $B$-band distance modulus of SN~2007on is
($\mu_{0} - 5 {\rm log_{10}}  \cdot {\rm h}_{73}$) $= 31.37 \pm 
0.024$ (random) mag, where h$_{73}$ = H$_0$/(73 km s$^{-1}$ Mpc$^{-1}$), 
and the $H$-band distance modulus is 
($\mu_{0} - 5 {\rm log_{10}}  \cdot {\rm h}_{73}$) $= 31.31 \pm 
0.030$ (random) mag. For SN~2011iv, on the other hand, the $B$-band 
distance modulus is ($\mu_{0} - 5 {\rm log_{10}} 
\cdot {\rm h}_{73}$)  $= 31.07 \pm 0.021$ mag and the $H$-band 
distance modulus is ($\mu_{0} - 5 {\rm log_{10}} 
\cdot {\rm h}_{73}$) $= 31.11 \pm 0.023$ mag.
These values yield differences of $\Delta \mu = 0.30 \pm 0.02$ 
mag and $\Delta \mu = 0.20 \pm 0.01$ mag in the $B$~band 
and $H$~band, respectively (see Fig.~\ref{FIG:DISTANCES}). These 
comparisons correspond to $\sim 14$\% and $\sim 9$\% 
differences in distance. 
%
\section{Spectral analysis}
\label{SEC:SPA}
%
\subsection{HST ultraviolet spectroscopy}
%
\subsubsection{Observations and spectral comparison}
The UV spectrum of a SN~Ia is shaped by a complex set of
processes, including significant line-blanketing driven by Fe-group 
elements at various ionisation states \citep[see, e.g.,][]{2000A&A...363..705M}.  
The UV spectral region therefore offers an opportunity to study 
the \Ni\ and Fe-group element content located in the outer layers of 
the expanding ejecta \citep[e.g.,][]{1998ApJ...495..617H, 
2000ApJ...530..966L, 2003ApJ...590L..83T, 2008MNRAS.391.1605S, 
2012MNRAS.427..103W}. 

\begin{figure*}
\center
\includegraphics[scale=0.7]{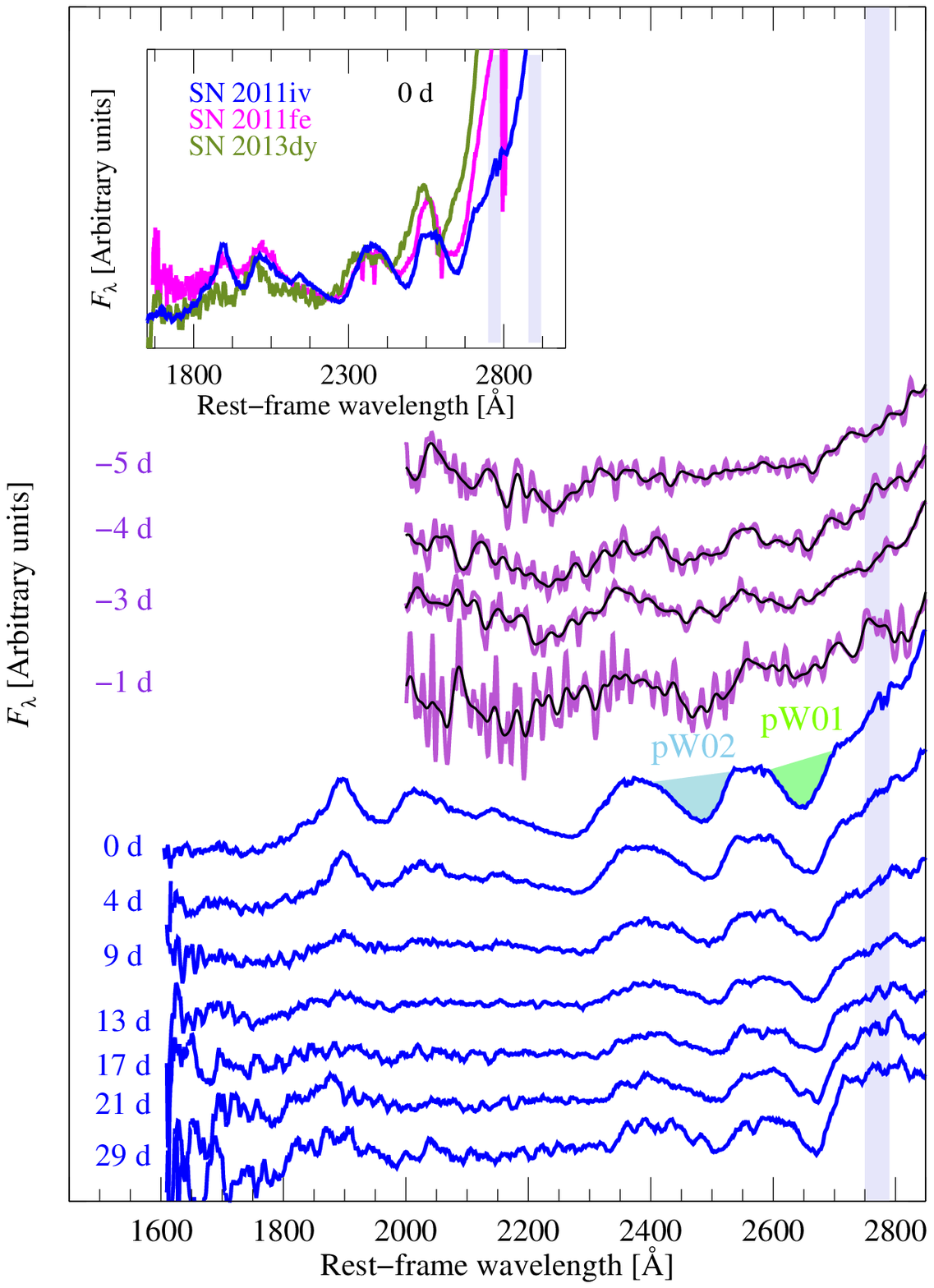}
\includegraphics[scale=0.7]{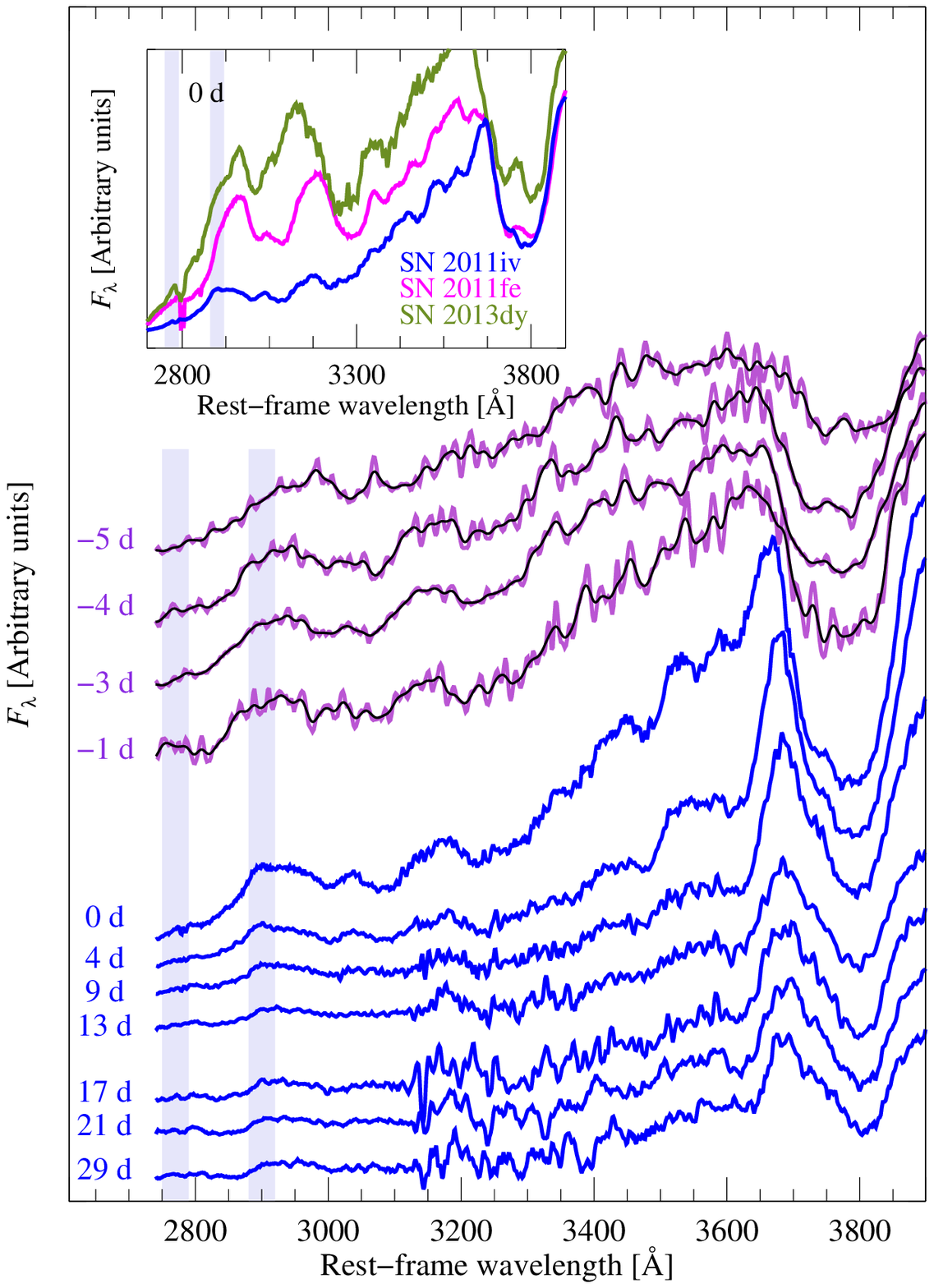}

\caption{Time-series of UV-wavelength spectroscopy of SN~2011iv 
obtained with {\em Swift} (purple) and {\it HST} (blue) over the course of 
nearly a month beginning 5~d before maximum brightness. Smoothed 
versions of the {\em Swift} spectra are presented in black. The insets 
are close-up views of the wavelength ranges 1700--3000 $\AA$ ({\it left}) 
and 2900--3900 $\AA$ ({\it right}), comparing the 0~d spectrum of 
SN~2011iv to the 0~d spectra of SN~2011fe \citep[pink;][]{2014MNRAS.439.1959M} 
and SN~2013dy \citep[green;][]{2015MNRAS.452.4307P}. The light-blue 
vertical areas mark the regions $f_{\lambda}(2770)$ and $f_{\lambda}(2900)$, 
which define the UV ratio $\mathcal{R}_{\mathrm{UV}}$ (see 
Appendix~\ref{SSS:UVRAT}). The green and blue shaded regions indicate 
the area enclosed by the pseudo-equivalent widths defined as pW01 
and pW02, respectively (Sect.~\ref{SSS:UVPEW}).
}
 \label{FIG:UVSPEC11IV}
 \end{figure*}

Given the limited observational coverage to date of the UV portion 
of SNe~Ia, we have undertaken a detailed examination of our UV 
spectroscopic time-series (Sect.~\ref{SSEC:UONSP}). From 
Figure~\ref{FIG:UVSPEC11IV}, it is evident that SN~2011iv is rich 
in prominent P-Cygni features, including those located at 
2300~\AA, 2500~\AA, 2650~\AA, 3000~\AA, and 3200~\AA,
which are typically caused by Mg~II, Fe~II, and other Fe-group elements.
The early-epoch ($+$0~d to $+$9~d) spectra also exhibit a 
conspicuous feature around 1900~\AA. A similar feature typically 
attributed to \ion{Fe}{ii} or \ion{Co}{ii} is also present in the normal 
Type~Ia SN~1992A \citep{1993ApJ...415..589K} and SN~2011fe 
\citep{2014MNRAS.439.1959M}, as well as in the Type~II-P 
SN~1999em \citep[][]{2009ApJ...700.1456B} and the Type~IIb 
SN~2001ig \citep[][]{2015ApJ...803...40B}.

Contained within the inset of Fig.~\ref{FIG:UVSPEC11IV} is a 
comparison between the maximum-light spectrum of SN~2011iv to 
similar-epoch spectra of the normal Type~Ia SN~2011fe 
\citep{2014MNRAS.439.1959M} and SN~2013dy \citep{2015MNRAS.452.4307P}.
SN~2011iv clearly shows notable differences in the UV spectral 
range. As demonstrated in the inset of the left panel, the flux level 
in the range 1600--2600 $\AA$ differs marginally between the 
depicted SNe~Ia. However, the inset in the right panel reveals a 
significantly lower flux level in SN~2011iv at 2600--3600 
$\AA$ as compared to the other objects. Furthermore, 
significant differences in line strengths, shapes, and locations are 
apparent for the most prominent features. 

SN~2011fe and SN~2013dy are chosen for comparison because
they are both normal objects with high-quality data obtained at 
similar wavelengths and epochs. Furthermore, a similar comparison 
as shown in the insets of Figure~3 has been discussed elsewhere 
\citep{2012ApJ...753L...5F,2015MNRAS.452.4307P, 
2016MNRAS.461.1308F}. The origin of the notable flux-level differences
exhibited by SNe~Ia in the UV is a matter of open debate. 
Detailed modelling of SN~Ia spectra points to a rather complex 
interplay between various physical parameters. This includes ejecta
stratification (mixing), metallicity, modification of density structures 
due to pulsations \citep{2004ApJ...607..391G}, spectral line 
formation and ionisation states \citep[e.g.,][]{1998ApJ...495..617H, 
2000ApJ...530..966L, 2014MNRAS.439.1959M, 2008MNRAS.391.1605S}, 
as well as geometric and viewing-angle effects that may alter the 
UV flux level \citep[e.g.,][]{2009MNRAS.398.1809K}.
In the remainder of this section, we  quantify the similarities and 
differences between the UV properties of SN~2011iv and other 
normal SNe~Ia.
%
\subsubsection{UV pseudo-equivalent width}
\label{SSS:UVPEW}

To quantify the UV spectral properties of the two prominent 
absorption features indicated in Fig.~\ref{FIG:UVSPEC11IV} (left 
panel), measurements are made of their pseudo-equivalent width 
(pEW), which is a common method to quantify spectral 
properties of SNe~Ia at optical wavelengths 
\citep[e.g.,][]{2004NewAR..48..623F, 2007A&A...470..411G, 
2013ApJ...773...53F}. Parameters that define the two UV pEW 
indicators considered here were obtained following the prescription 
of \citet{2007A&A...470..411G} and are summarised in 
Table~\ref{T:PSWFL}. In short, the two pEW indicators are 
defined from the blue to the red maximum of an absorption trough.

Based on this definition, Figure~\ref{FIG:UVPWS} shows the resulting
temporal evolution of the derived UV pEWs for SN~2011iv, as 
well as for the comparison SNe (SN~1992A, SN~2011fe, and 
SN~2013dy), chosen as they are representatives of normal SNe~Ia 
with excellent UV data. Interestingly, the pEW values of SN~2011iv 
resemble those of SN~2011fe, while the pEW values of 
SN~1992A evolve similar to those of SN~2013dy. The pEW 
values of pW02 appear to increase until around $+$10~d past 
maximum light, whereafter the pEW steadily declines in a similar 
fashion for the considered SNe~Ia. 
%
\subsubsection{The 3000 $\AA$ feature}
%
Features at longer wavelength (i.e., $\gtrsim 2800$ $\AA$) are 
formed well within the SN ejecta and are products of 
nuclear burning \citep[e.g.,][]{1986ApJ...306L..21B,
1993ApJ...415..589K, 2008MNRAS.391.1605S}.
Conspicuous features in this region are those around 3000 $\AA$ 
and 3250 $\AA$, which are blends of \ion{Fe}{ii} or \ion{Co}{ii} 
\citep[e.g.,][]{1986ApJ...306L..21B}. Comparing the 3000 $\AA$ 
absorption feature of the maximum-light spectrum of SN~2011iv 
(Fig.~\ref{FIG:UVSPEC11IV}) with that of other SNe~Ia reveals 
noticeable differences in its shape. SN~2011iv and SN~2011fe 
exhibit a broad ``W-shaped'' absorption trough, while in the case 
of SN~2013dy this feature is ``V-shaped.'' A V-shaped feature is 
also present in SN~1992A \citep{1993ApJ...415..589K}. We 
measured the pEW of the entire 3000 $\AA$ feature for epochs 
around maximum light in a similar fashion as for the other UV 
features and find values of about 40 $\AA$, in 
agreement with measurements in the literature 
\citep[e.g.,][]{2008ApJ...686..117F}. 
%
\subsubsection{No evidence of SN~Ia ejecta and circumstellar interaction.} 
%
UV observations  have been used in the past to provide constraints 
on interaction between the expanding SN~Ia ejecta and circumstellar 
material (CSM).  In  Appendix~\ref{SSC:LECSM} we discuss our efforts 
to search for evidence of  interaction between the ejecta of SN~2011iv 
and any possibly CSM. In short, our efforts  yield no evidence of  
interaction in the case of SN~2011iv.
%
%
\subsection{Optical spectral comparison of early epochs and spectral synthesis}
\label{SSS:OPTI40D}
\begin{figure*}
\center
\includegraphics[scale=0.65]{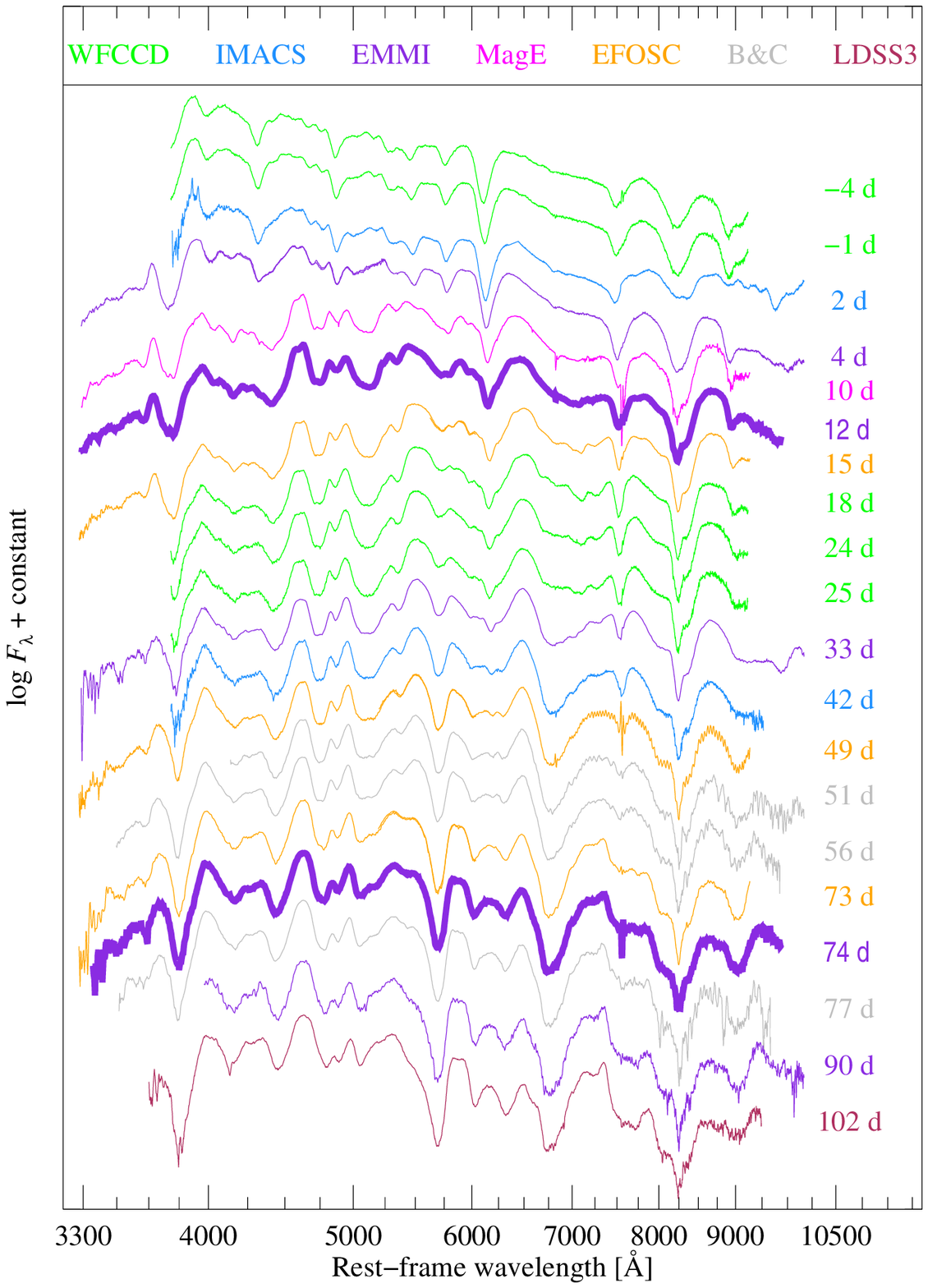}
\includegraphics[scale=0.65]{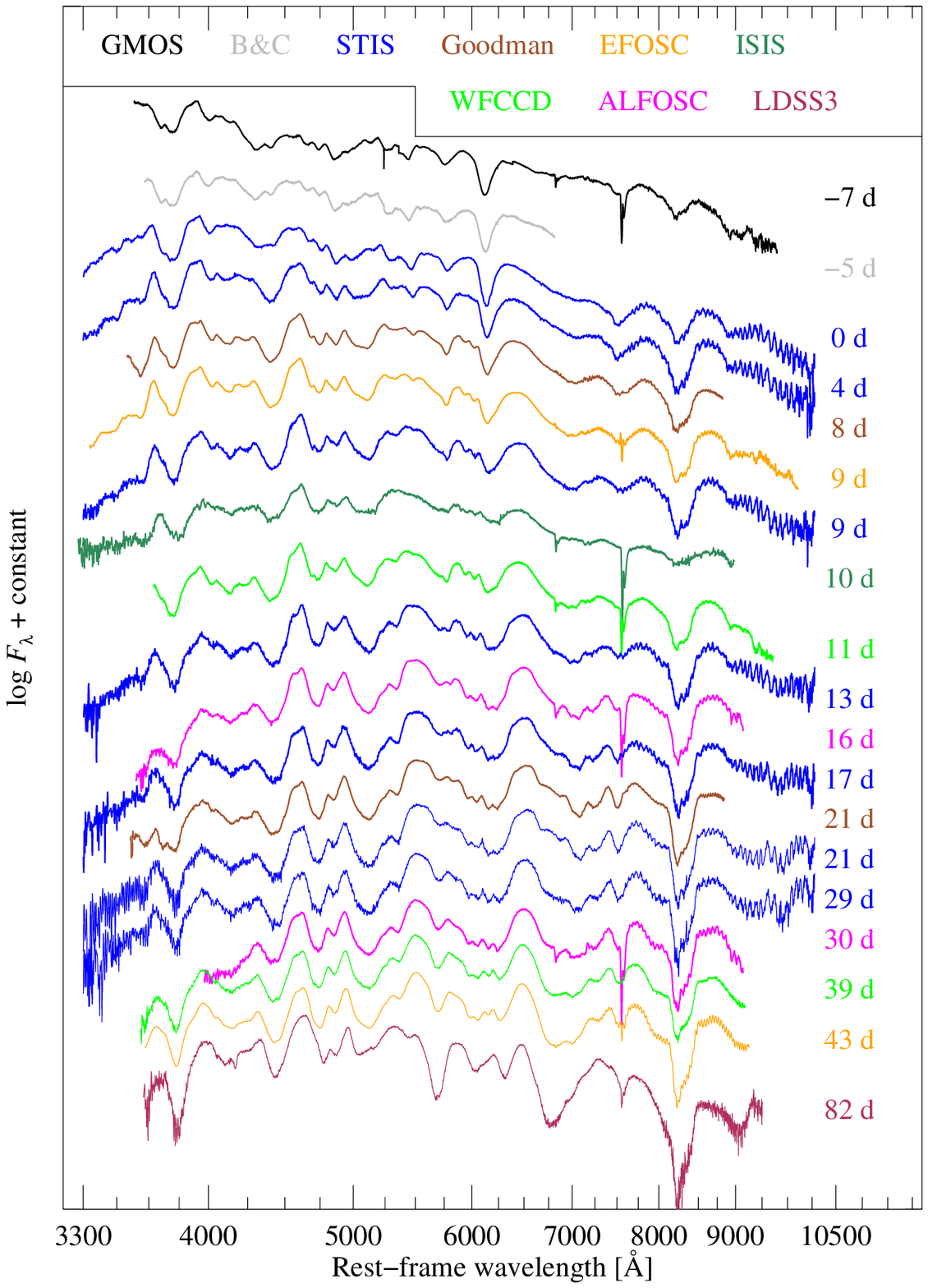}
\caption{Montage of selected visual-wavelength spectra of 
SN~2007on {\em (left)} and SN~2011iv {\em (right)}. The spectra 
are colour coded with respect to the facility used to obtain the 
observations. Previously unpublished NTT ($+$EMMI) spectra of 
SN 2007on are highlighted as thick purple lines. All spectra are 
listed in Tables \ref{T:SPJ07ON} and \ref{T:SPJ11IV}.
}
 \label{FIG:OPTISPEC11IV}
 \end{figure*}

%
Unlike the UV spectral range that is distinguished by significant 
line blending of various Fe-group elements in the outermost layers, 
the visual-wavelength range is characterised by a variety of broad 
P-Cygni profiles from single elements (see, e.g., 
Fig.~\ref{FIG:OPTISPEC11IV}). The study of these features and 
their temporal evolution can provide clues regarding SN~Ia progenitors 
and an avenue to constrain the explosion physics.

Figure~\ref{FIG:OPTISPECMAX} presents a comparison between 
visual-wavelength spectra of SN~2007on and SN~2011iv taken 
around $-$5~d and $+$4~d, along with identifications for 
all of the prominent spectral features. Overall, the spectra are 
similar, and all of the main ions that characterise a SN~Ia are
present. However, as highlighted by the shaded regions in 
Fig.~\ref{FIG:OPTISPECMAX}, notable spectral differences are evident 
at 4300--4500~\AA\ and 4700--5200~\AA. Additionally, 
there are some small differences in the blueshifts of the 
absorption minimum of the \ion{Si}{ii} $\lambda$6355 line.

To facilitate the identification of the various spectral features 
observed in the early epochs of SN~2007on and SN~2011iv, and 
in particular, in the first two shaded regions of Figure~\ref{FIG:OPTISPECMAX}, 
we turn to spectral synthesis modelling using well-established 
techniques applied to the study of numerous SNe~Ia including 
the normal SN~2011fe \citep{2014MNRAS.439.1959M} and 
SN~2014J \citep{2014MNRAS.445.4427A}. Synthetic 
spectra are computed based on  the W7  density profile 
\citep{1984ApJ...286..644N}, allowing for consistency between 
line identification with other models. We note that the majority of 
SN~Ia explosion models produce similar-looking density profiles 
in the regions where the maximum-light spectrum is formed; 
the main differences are in the outermost velocities. Therefore,
the results presented below are independent of whether we use 
a W7 density profile or that associated with our best-fit 
DD explosion models (see Sect.~\ref{modelcomparison}).

Our spectral synthesis calculates the radiation field above a 
blackbody photosphere, with element abundances, luminosity, and 
photospheric velocity varied to produce a best-fit spectrum 
\citep{2000A&A...363..705M}. The code uses the density profile of 
the ``fast deflagration'' single-degenerate W7 model 
\citep{1984ApJ...286..644N,1999ApJS..125..439I}.

Figure~\ref{FIG:OP07MODEL} shows the synthetic spectral models 
computed for the $-$1.0~d spectrum of SN~2007on and the 
spectrum of SN~2011iv taken at maximum light. The model parameters 
for the premaximum spectrum of SN~2007on imply a 
photospheric velocity $v_{\mathrm{ph}} = 9500$ km s$^{-1}$, a 
photospheric blackbody temperature $T_{\mathrm{ph}} = 
10,600$ K, a bolometric luminosity $L_{\mathrm{bol}} = 4.80 
\times 10^{42}$ erg s$^{-1}$, and a rise time to maximum
$t_{\mathrm{rise}} = 17.4$~d. In the case of the synthetic spectrum 
of SN~2011iv, the model parameters are $v_{\mathrm{ph}} = 9500$ 
km~s$^{-1}$, $T_{\mathrm{ph}} = 10,700$ K, $L_{\mathrm{bol}} = 
7.03\times10^{42}$ erg s$^{-1}$, and $t_{\mathrm{rise}} = 17.9$~d. 

The models consist of spectral features attributed to many 
of the same ions found in the spectra of normal SNe~Ia (see 
Fig.~\ref{FIG:OP07MODEL}), and some of the features are 
produced by the blending of several ions. Prominent spectral 
features are attributed to various ions of intermediate-mass 
elements including \ion{Ca}{ii} H\&K, \ion{Si}{ii} 
$\lambda\lambda$3856, 4130, 5972, 6355, \ion{S}{ii} 
$\lambda\lambda$5449, 5623, and the \ion{Ca}{ii} NIR triplet. 
The prevalent feature seen around 7500~\AA\ is 
dominated by a blend of \ion{O}{i} $\lambda$7773, which is a 
characteristic of subluminous SN~1991bg-like SNe~Ia 
\citep[e.g.,][]{2004ApJ...613.1120G, 2008MNRAS.385...75T, 
2016MNRAS.463.1891A}.  

The spectral feature located at 4300--4500 \AA\ 
in SN~2011iv is broader than in SN~2007on (see first shaded 
region in Fig.~\ref{FIG:OPTISPECMAX}). The synthetic spectra 
of SN~2011iv have a more significant contribution of \ion{Si}{iii} 
$\lambda$4553 and \ion{Fe}{iii} $\lambda$4420, whereas in the 
case of SN~2007on this region is dominated by \ion{Mg}{ii} 
$\lambda$4481.

Moving to longer wavelengths, our models indicate that the prominent 
feature in SN~2011iv at 4700--5000~\AA\ is formed from a 
blend of various \ion{Si}{ii} and \ion{Fe}{iii} lines, with the strongest 
contributions from \ion{Si}{ii} $\lambda$5055 and  \ion{Fe}{iii} 
$\lambda$5156, and it contains only a small contribution of 
\ion{Fe}{ii} and \ion{S}{ii}. In the case of SN~2007on, this region in 
the synthetic spectrum is formed by \ion{Si}{ii} $\lambda$5055, 
with some contribution from \ion{S}{ii} $\lambda$5032 and 
\ion{Fe}{ii} $\lambda$5169. The differences between the spectra of 
SN~2007on and SN~2011iv are attributed to the latter being 
intrinsically hotter, which translates into it having ejecta 
characterised by a higher ionisation state. A full detailed spectral 
analysis using the abundance tomography technique will be 
presented in an forthcoming paper (Ashall et al., in prep.).  
%
\subsection{NIR Spectroscopy}
\begin{figure}
\includegraphics[width=\hsize]{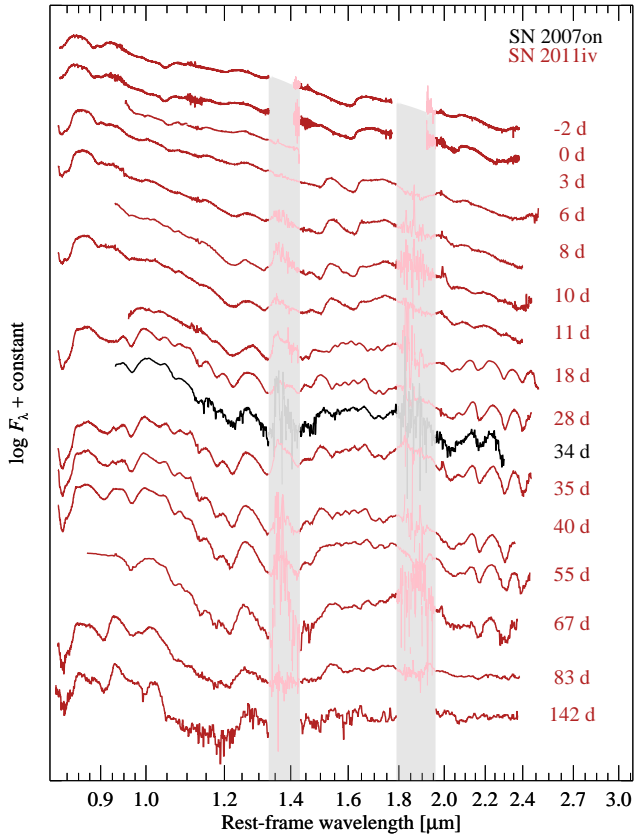}
\vspace{-1.0 cm}
\caption{Selected NIR-wavelength spectra of SN~2007on (black) and 
SN~2011iv (red) taken between $-$2~d and $+$142~d with various 
instruments (see Table~\ref{T:SPJ07ON} and Table~\ref{T:SPJ11IV}). 
The grey vertical bands indicate regions of prevalent telluric absorption.
Some spectra have been smoothed for presentation purposes.
}
 \label{FIG:NIRSPEC11IV}
 \end{figure}

%
At early epochs, the NIR spectra (see Fig.~\ref{FIG:NIRSPEC11IV}) 
of SN~2011iv are rather smooth, exhibiting only a handful of 
features, including the \ion{Mg}{ii} $\lambda$9218 line and an 
absorption dip at $\sim 1.6$ $\mu$m, which is associated with the 
$H$-band break. As the spectrum evolves with time, a multitude of 
features attributed to Fe-group elements (such as \ion{Fe}{II}, 
\ion{Co}{II}, and \ion{Ni}{ii}) emerge and dominate the spectrum, 
particularly at the wavelength regions corresponding to the $H$
and $K_s$ bands. The prominent feature at $\sim 9000$~\AA\ may 
be attributed to a blend of \ion{Mg}{ii} $\lambda\lambda$9218, 
9244, as was identified in the transitional iPTF13ebh 
\citep{2015A&A...578A...9H}. At epochs beyond $+$10~d, the 
latter feature becomes broader and even more prominent, together 
with a feature located around $\sim 9800$~\AA. These absorption 
troughs are likely the result of line blanketing of a large number of 
\ion{Mg}{ii}, \ion{Ca}{ii}, \ion{Co}{ii}, and \ion{Fe}{ii} lines 
\citep{2002ApJ...568..791H}.

\citet{2015A&A...578A...9H} provided a detailed study of the 
$H$-band break observed in SN~2011iv, which is located right 
between the two major telluric regions in the NIR. In summary, 
the profile of the $H$-band break of SN~2011iv evolves differently 
than, say, that of iPTF13ebh, despite both objects being very similar 
otherwise \citep[see][]{2015A&A...578A...9H}. The feature is 
weaker in strength compared to normal SNe~Ia \citep[see]
[Fig.~15]{2015A&A...578A...9H}, though the peak of the $H$-band 
break ratio of SN~2011iv appears to fit into the correlation with the 
light-curve decline rate, $\dmfiften$, and the colour-stretch 
parameter, $\sbv$ \citep{2013ApJ...766...72H}. 

The velocities of the NIR \ion{Mg}{ii} $\lambda$9218 
(Sect.~\ref{SSS:OPLV}, Fig.~\ref{FIG:VELOCITY}) feature 
appear to decrease consistently in SN~2011iv and 
are similar to those observed in the fast-declining 
SN~1991bg and iPTF13ebh. High velocities at early phases also 
characterise the \ion{Mg}{ii} $\lambda$1.0972 $\mu$m~line of 
iPTF13ebh and SN~2011iv \citep[][Fig. 16]{2015A&A...578A...9H}.  

Displayed in Figure~\ref{FIG:NIRSPCOMP1107} are combined 
visual-wavelength and NIR spectra of SN~2007on and SN~2011iv 
taken around a month past maximum brightness, with line identifications 
provided for the most prevalent features. The spectra have been 
calibrated through multiplication by a function that ensures their 
synthetic broad-band magnitudes match the observed broad-band 
magnitudes inferred from the interpolated optical and NIR 
photometry on $+$29~d. The comparison shows that the 
objects are overall quite similar. The spectral features of both 
objects also remain similar at all wavelengths, including the 
\ion{Co}{II} features located between 2.0~$\mu$m and 2.4~$\mu$m.
%
\subsection{Nebular spectroscopy}
\label{SSS:NEBUL}
\begin{figure}
\includegraphics[width=1.0\hsize]{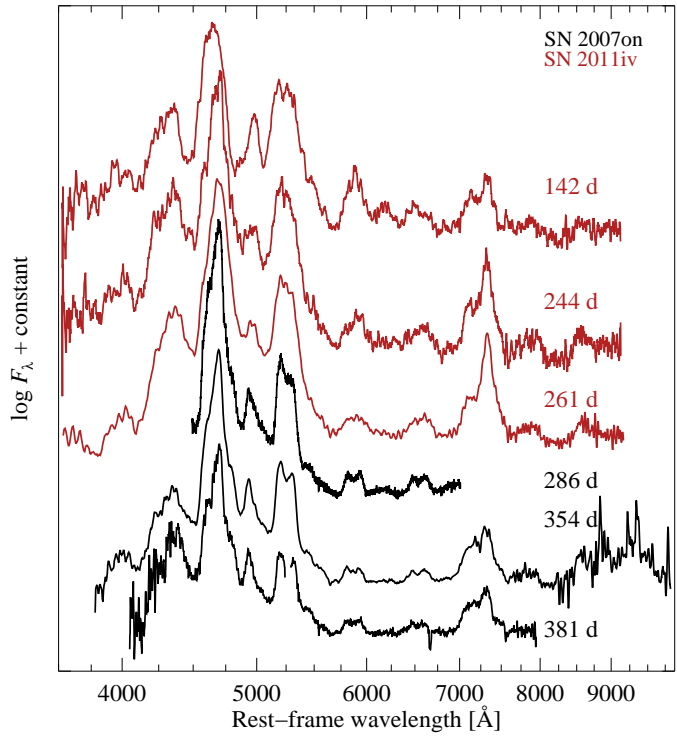}
\caption{Nebular-phase visual-wavelength spectra of 
SN~2007on (black) and SN~2011iv (red).}
 \label{FIG:OPSPGT1107}
 \end{figure}

%
The late-phase visual-wavelength spectra of SN~2007on and 
SN~2011iv provide an opportunity to assess the inner regions of 
SNe~Ia located at the faint end of the luminosity vs. decline-rate 
relation. Figure~\ref{FIG:OPSPGT1107} displays the nebular spectra 
of SN~2007on taken on $+$286~d, $+$353~d, and $+$380~d, as well 
as those of SN~2011iv taken on $+$142~d, $+$244~d, and $+$260~d.
Late-phase spectra of both SNe are characterised by emission 
features formed by numerous blended, mostly forbidden emission 
lines. The strongest emission feature is around 4800 \AA\, 
followed by those in the ranges 4000--4500 $\AA$, 5100--5400 
\AA, and 7000--8000 $\AA$. The 4800 \AA\  feature is dominated 
by \ion{Fe}{iii} with weak contributions of \ion{Fe}{ii}, while 
the 5000--5500 \AA\ feature is dominated by \ion{Fe}{ii} with 
contributions of \ion{Fe}{iii} \citep{1980PhDT.........1A}. Close 
inspection of the spectra of SN~2007on at $+$353~d and $+$380~d 
reveal the existence of double-peak profiles, which were 
first noted by \citet{2015MNRAS.454L..61D}, leading them to suggest 
a bi-modal $^{56}$Ni distribution possibly linked to a non-standard 
explosion scenario. A detailed spectral synthesis study exploring this option 
for SN~2007on and SN~2011iv will be presented in a forthcoming 
publication (Mazzali et al., in preparation).

Preliminary modelling of the nebular spectra has been 
accomplished with a non-local-thermodynamic-equilibrium (NLTE) 
SN nebular code \citep{2007ApJ...661..892M}. 
Spectral synthesis suggests $^{56}$Ni masses of 0.19~M$_{\odot}$ 
produced in SN~2007on and 0.41~M$_{\odot}$ in 
SN~2011iv. These values are within $\sim 23$\% and 5\% of the 
values computed from the peak of the UVOIR light curves of 
SN~2007on and SN~2011iv, respectively.  We note that the 
two different methods to estimate the $^{56}$Ni mass have 
an expected scatter of $\sim 20$\% \citep[see][]{2006A&A...460..793S}.
\section{Discussion}
\label{SEC:DISCUSSION}
\subsection{SN~2007on and SN~2011iv: transitional SNe~Ia}
\label{sec:summary}
\begin{figure*}
\includegraphics[width=18cm]{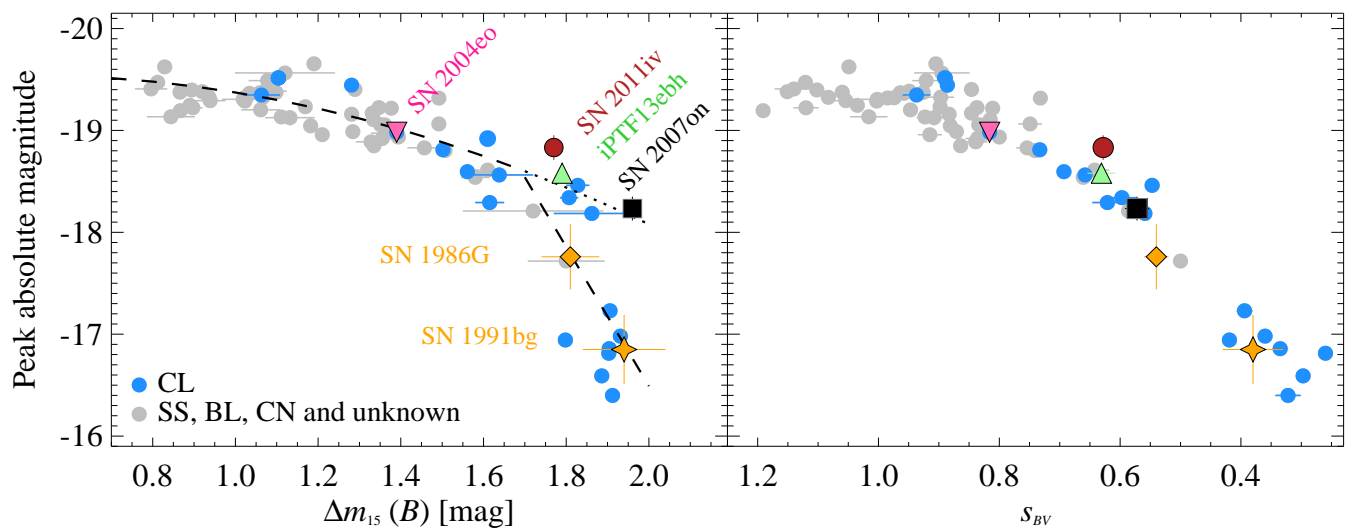}
\caption{Luminosity vs. decline-rate relation populated with a subset 
of CSP-I SNe~Ia \citep{2014ApJ...789...32B} and a few fast 
decliners from the literature. The relation is parameterised by 
$\Delta m_{15}$ (left) and $s_{BV}$ (right). The absolute $B$-band 
magnitudes are extinction corrected as described by 
\citet{2014ApJ...789...32B}, and distances are computed using 
a Hubble constant H$_{0}$ = 73 km s$^{-1}$ Mpc$^{-1}$.
A distance modulus of $\mu = 31.27$ mag is used to place 
SN~2007on (black square) and SN~2011iv (red circle) on the 
luminosity scale. The blue circles correspond to SNe~Ia 
classified as CL (cool) on the \citet{2006PASP..118..560B} 
diagram, whereas the grey circles represent SNe~Ia with either 
different (i.e., SS, CN, BL) or unknown Branch spectral subtype 
classifications.
Additionally, included for comparison are the low-luminosity Type~Ia 
SNe~1986G (yellow diamond) and 1991bg (yellow star), the 
transitional Type~Ia iPTF13ebh (green triangle), and the normal 
Type~Ia SN~2004eo  (pink downward triangle). The dashed curves 
represent the parameterised luminosity vs. decline-rate relation for 
normal SNe~Ia \citep{1999AJ....118.1766P} and for the subluminous 
SNe~Ia \citep{2008MNRAS.385...75T}. The dotted curve represents 
an interpolation of the \citet{1999AJ....118.1766P} relation for 
SNe~Ia with $1.7 < \Delta m_{15}(B) <$ 2.0 mag.
}
 \label{FIG:LWR}
 \end{figure*}

%
We have presented a comprehensive set of photometric and 
spectroscopic observations for the Type~Ia SN~2007on and 
SN~2011iv, both hosted by NGC~1404, and provided a detailed 
analysis of their data. The findings in this paper suggest that 
both SNe~Ia are transitional objects, which means they 
have properties photometrically and spectroscopically 
between those of normal SNe~Ia and subluminous, SN~1991bg-like 
SNe~Ia. Figures~\ref{FIG:LCCOMP} and \ref{FIG:LWR} demonstrate 
that SN~2011iv is brighter than SN~2007on by $\sim 0.6$ mag in the $B$~band 
and 0.35 mag in the $H$~band at maximum light. Although both 
objects are nearly as bright as the faintest normal SNe~Ia, their 
optical decline rates are consistent with fast-declining, 
subluminous SNe~Ia, with $\dmfiften$ typically $\gtrsim 1.7$.
Nonetheless, from a comparison of the colour-stretch parameter, 
$\sbv$ (Fig.~\ref{FIG:LWR}), we find that both SNe~Ia are 
positioned in the luminosity vs. $s_{BV}$ relation between normal 
SNe~Ia ($s_{BV} \gtrsim 0.8$) and SN~1991bg-like objects ($s_{BV} 
\lesssim 0.5$). 

Additionally, the NIR light curves of SN~2007on and SN~2011iv 
are found to peak prior to the time of $B$-band maximum, 
and both exhibit a secondary NIR maximum. These 
characteristics are consistent with normal SNe~Ia and are 
contrary to {\it bona fide} SN~1991bg-like  SNe~Ia that  
exhibit single-peaked NIR light curves, which typically peak $\sim 
2$--5 days {\em after}  $B$-band maximum \citep[][]{2009AJ....138.1584K}. 

To place order among the normal, transitional, and subluminous 
SNe~Ia, presented in Figure~\ref{FIG:OPSNSPCOMP} is a comparison 
of near-maximum spectra  (arranged from top down by decreasing 
peak luminosity) of  the normal Type~Ia SN~2004eo \citep{2007MNRAS.377.1531P}, 
the transitional Type~Ia SNe~2011iv, iPTF13ebh \citep{2015A&A...578A...9H},
and 2007on, and the subluminous Type~Ia SNe~1986G 
\citep{1992A&A...259...63C} and 1991bg \citep{1996MNRAS.283....1T}. 
Examination of the spectral sequence reveals that the normal and 
transitional objects exhibit spectral features related to the transition 
of doubly to singly ionised Fe-group elements, while the 
subluminous objects exhibit \ion{Ti}{ii}. The strength of these 
features and how they vary can be understood to first order by a 
range in photospheric temperatures \citep{1995ApJ...455L.147N}.
Within this framework, \ion{Ti}{ii} features are present in subluminous 
SNe~Ia characterised by cooler photospheres, while \ion{Fe}{ii} features 
appear with increased temperature and luminosity, followed by even 
higher ionisation driven by increased temperatures and the 
emergence of \ion{Fe}{iii}. This is consistent with the luminosity vs. 
decline-rate relation \citep[e.g.,][]{1996ApJ...472L..81H}, and as 
explained below, it is linked to the amount of $^{56}$Ni produced 
during the explosion.
%
\subsection{Model comparison}
\label{modelcomparison}

To gain a theoretically based understanding of the key 
differences between SN~2007on and SN~2011iv we  seek guidance 
from a suite of spherical one-dimensional (1-D)
delayed-detonation (DD) explosion models of Chandrasekhar-mass 
($M_{\rm Ch} \approx~1.4$~M$_\odot$)  carbon-oxygen white dwarfs 
\citep{2002ApJ...568..791H}. Spherical 1-D models are used as current 
3-D hydrodynamical models predict significant mixing throughout the 
envelope which is inconsistent with observations (see Appendix~\ref{appendix:burning}  for discussion). 

In standard DD models, a larger amount of $^{56}$Ni is produced during 
the detonation burning phase compared to the deflagration burning phase 
\citep{1991A&A...245..114K}. During the deflagration phase, 0.25 to 
0.30~$M_\odot$ of carbon-oxygen is required to be burned to lift the 
white dwarf from its gravitational potential. In spherically symmetric explosion 
models, this amount is conveniently parameterised by the transition density 
(hereafter $\rho_{\rm tr}$), which marks the density of the burning material 
when the laminar burning flame transitions from traveling at less than 
the local sound speed (a deflagration)  to faster than the local sound speed 
(a detonation). During the deflagration phase, burning reaches nuclear 
statistical equilibrium (NSE), leading to the production of mostly iron-group 
elements. The abundances of isotopes depends on the level of neutronisation, 
which is specified by the electron to baryon fraction: $Y_e = Y_p / (Y_p + Y_n)$. 
With increasing central density (hereafter $\rho_{\rm c}$), electron capture 
shifts NSE away from the production of radioactive $^{56}$Ni and toward 
the production of  stable Fe-group elements (e.g., $^{58}$Ni). Therefore, 
with increasing $\rho_{\rm c}$, the abundance of stable Fe-group elements 
increases relative to $^{56}$Ni, and inevitably produces a central hole in the 
$^{56}$Ni distribution.\footnote{Central density plays an influential
role in the production of radioactive $^{56}$Ni; increased values 
lead to electron-capture rates that shift the synthesis of elements in nuclear 
statistical equilibrium (NSE) away from $^{56}$Ni and toward 
stable iron-group elements.  In high-metallicity white dwarfs, settling of
$^{22}$Ne in the core  can also shift the NSE abundances away 
from $^{56}$Ni in the central regions, but this requires very long 
evolutionary times on the order of $5+$ billion years \citep{2011A&A...526A..26B}. 
Since the amount of $^{22}$Ne is relatively limited, 
the size of the $^{56}$Ni hole is expected to account for 
$\leq 0.05$~M$_\odot$ \citep{2001ApJ...549L.219B}.
}

During the deflagration phase, depending on $\rho_{\rm c}$,  
$^{56}$Ni  production can range from very little up to 
$\sim 0.3$ M$_\odot$ \citep[e.g.,][]{1984ApJ...286..644N, 
2000ApJ...536..934B, 2002ApJ...568..791H, 2015ApJ...806..107D}. 
During the detonation phase, depending on the value of 
$\rho_{\rm tr}$, $^{56}$Ni production can range from very 
little to  $\sim 0.6$ M$_\odot$ \citep{2002ApJ...568..791H}. 
Therefore, the total production of $^{56}$Ni in DD models can 
range from very little  up to $\sim 0.9$~M$_\odot$, with 
a varying distribution of $^{56}$Ni  within the expanding ejecta.
While in normal SNe~Ia the total $^{56}$Ni mass is 
dominated by contributions from both the deflagration and 
detonation phases, in subluminous SNe~Ia the $^{56}$Ni is 
produced primarily during the deflagration phase, leading to a 
$^{56}$Ni distribution more centrally condensed and 
influenced by $\rho_{\rm c}$. 

Within this  well-established framework, the higher peak luminosity 
(and hence slower decline rate) of SN~2011iv compared to 
SN~2007on is attributed to having produced more $^{56}$Ni, and 
this in turn produced the bluer colours exhibited by SN~2011iv at early 
times (see Fig.~\ref{FIG:BVCOMP}).
 The dependence of temperature 
and luminosity on the $^{56}$Ni mass was described decades ago 
by \citet{1982ApJ...253..785A}, and this relationship also drives  the 
spectroscopic sequence plotted in Fig.~\ref{FIG:OPSNSPCOMP} 
\citep[see also][]{1995ApJ...455L.147N}. Specifically, objects with 
smaller amounts of $^{56}$Ni  will have lower-temperature photospheres 
leading to the presence of \ion{Ti}{ii} lines, while higher $^{56}$Ni mass 
objects will have  hotter photospheres and therefore higher-ionisation 
conditions leading to the presence of  \ion{Fe}{ii} and/or  \ion{Fe}{iii} 
features.

As demonstrated in Fig.~\ref{FIG:BVCOMP}, despite SN~2011iv 
being brighter and bluer than SN~2007on at early epochs, between 
$+20$~d and $+85$~d SN~2011iv appears $\sim 0.12\pm0.01$ mag 
redder than SN~2007on, and this behaviour is confirmed by  the UV vs. 
optical  colour  (see Fig.~\ref{FIG:NUVC}). This rather perplexing 
behaviour can be explained by the progenitor white dwarf of SN~2011iv 
having a larger $\rho_{\rm c}$ compared to SN~2007on. As discussed 
above, increasing $\rho_{\rm c}$ leads to the production of more 
stable Fe-group elements at the expense of radioactive $^{56}$Ni. Less
$^{56}$Ni is produced in the deflagration phase owing to the higher
value of $\rho_{c}$. In turn, this leads to less centrally condensed 
$^{56}$Ni, less heating of the central regions of the ejecta, and 
(consequently) redder colours at late times. 

To demonstrate the influence  $\rho_{\rm c}$ has on the $B-V$
colour evolution, plotted in the inset of Figure~\ref{FIG:BVCOMP} is
the $B-V$ colour evolution of SN~2007on and SN~2011iv compared to
that corresponding to the best-fit DD models (see
Appendix~\ref{appendix:modelcomparisons}).  At early times the
models match the colours quite well, while at later times the models
are  steeper than  observed in SN~2007on and SN~2011iv. This is
largely due to forbidden lines become increasingly important, and
the processes associated with these transitions are  difficult to
model because of a lack of atomic data  \citep[see, e.g.,][]{1995ApJ...443...89H, 
2014MNRAS.441.3249D, 2015MNRAS.454.2549B, 
2015ApJ...798...93T,brian_14JIR14,friesen11fe_17}. Nonetheless the 
best-fit DD models demonstrate that with increased $\rho_{\rm c}$ one 
obtains redder $B-V$ colours along the Lira relation, while models with 
reduced  $\rho_{\rm c}$ exhibit bluer $B-V$ colours. 
Assuming a canonical value of $\rho_{\rm c}$ = 2 $\times$ 
10$^{9}$~g~cm$^{-3}$ for SN~2011iv, a $B-V$ color offset of 0.12 mag 
between models during the phases coincident with the Lira relation corresponds
to a model with $\rho_{\rm c}$ = 1 $\times$ 10$^{9}$ g~cm$^{-3}$ for SN~2007on. 
As shown in \citet{2017ApJ...846...58H}, the offset is relatively stable 
over the period of the Lira relation for models of different brightness.
We note that a similar result was also found  from the detailed analysis 
of the low-luminosity SN~1986G \citep{2016MNRAS.463.1891A}.
%
\subsection{Variations in luminosity}
%
Tentative evidence exists that SNe~Ia residing in early-type 
galaxies exhibit smaller Hubble scatter than SNe~Ia hosted in 
late-type galaxies \citep{2003MNRAS.340.1057S}. 
This is thought to be driven by (i) the progenitor stars being older 
and spanning a smaller mass range, and (ii) dust extinction 
being less significant in early-type hosts compared with 
late-type hosts. Some high-redshift experiments have therefore 
been designed to target early-type galaxies in distant clusters.

With SN~2007on and SN~20011iv being both located in the same 
galaxy, they offer a rare opportunity to test the assertion that SNe~Ia 
located in early-type hosts provide minimal dispersion in their peak 
luminosities. However, as we have shown in 
Sect.~\ref{SSEC:DIST}, even after correcting for colour stretch and 
colour (extinction), both objects exhibit significantly different peak 
absolute magnitudes. This is in contrast to the three normal 
($\Delta m_{15}(B) \approx 1.1$--1.2 mag) SNe~Ia hosted in the 
early-type galaxy Fornax A, whose distances were found to be consistent 
at the 3\% level \citep{2010AJ....140.2036S}. Therefore, the 
significant discrepancy between the distances of SN~2007on and 
SN~2011iv serves as a cautionary tale for the use of transitional 
SNe~Ia located in early-type hosts in the quest to measure 
cosmological parameters. The implications of this finding are 
important in cosmology; for example, more than half of the 
SNe~Ia used by \citet{2012ApJ...746...85S} to constrain the 
high-redshift ($z > 1$) end of their Hubble diagram are fast-declining 
SNe (i.e., objects best described by the SALT parameter 
$x{_1}  < -1$, or equivalently $\Delta m_{15}(B) > 1.4$ mag).
 
As described above, within the context of the DD $M_{\rm Ch}$ 
models employed in this study, the $^{56}$Ni production depends 
both on $\rho_{\rm c}$ and $\rho_{\rm tr}$. While $\rho_{\rm tr}$  
is the primary driver of the luminosity vs. decline-rate relation, 
$\rho_{\rm c}$ plays an important role for the faster-declining 
SNe~Ia which produce smaller amounts of radioactive $^{56}$Ni as 
compared to normal-luminosity SNe~Ia. In SNe with the 
same $\Delta m_{15}(B)$ (so-called ``twins''), this secondary parameter 
can lead to changes in the peak brightness of 0.05 mag for normal-brightness 
SNe~Ia, and by up to 0.7 mag in the case of twin subluminous SNe~Ia 
(Hoeflich et al. 2017).

\citet{2010ApJ...710..444H} studied the variation in brightness 
between the peak and the tail (roughly the brightness at day $+$40) 
caused by variations in progenitor mass, metallicity, accretion rate, 
$\dot M$, and central density.  The effects of variations in $\rho_c$ 
are further elucidated in Figure 7 of Hoeflich et al. (2017), 
which shows the variation of heating from gamma-ray deposition 
as a function of central concentration of radioactive nickel. This 
more central heating in lower-$\rho_c$ SNe leads
to a hotter central region and therefore bluer colours, with a
variation in $B-V$ of up to 0.2 mag. SNe of comparable central
densities have similar Lira relations. The differences in SN~2007on 
and SN~2011iv lead to expected variations in $\rho_c$ of
a factor of $\sim 2$ (see Sect.~\ref{modelcomparison}). For this 
density variation, we expect SN~2007on would be brighter at 
maximum by about 0.18 mag than a model with the same $\rho_c$ 
of SN~2011iv. These modifications bring the distance
of both SNe to within $\sim 0.1$ mag. In light of these 
results, our findings suggest that observations extended to at 
least $+$40 days are required to break the degeneracy between 
$\rho_{\rm c}$ and the luminosity vs. decline-rate relation in fast-declining 
SNe~Ia. Consequently, future SN~Ia cosmological experiments 
should consider obtaining photometry at phases coincident with the 
Lira relation.
  
\section{Summary and Conclusions}
\label{SEC:CCL}
%
We have presented a comprehensive set of photometric and 
spectroscopic observations --- spanning from UV through NIR wavelengths --- 
of the transitional Type~Ia SNe~2007on and 2011iv. The detailed 
observational dataset combined with modeled comparisons, and the 
fact  both objects were  located in the same host galaxy NGC~1404, 
allowed us to obtain insights on their progenitors and their ability to 
serve as distance indicators. The main findings of this study are
as follows.

\begin{itemize}

\item The transitional Type~Ia SNe~2007on and 2011iv exhibit 
spectral and light-curve properties consistent with being an extension 
of the normal SN~Ia population, and are not consistent with the 
observed properties of subluminous SN~1991bg-like SNe~Ia. 

\item  The difference in their peak luminosities is caused by differences 
in their $^{56}$Ni production. This drives the differences in their $B-V$ 
colours around maximum brightness, as well as subtle differences in their 
spectroscopic properties.

\item The $B-V$ colour evolution of SN~2011iv is found to transition
from being bluer to redder than SN~2007on between maximum brightness and 
several weeks later. We suggest that this behaviour is linked to the progenitor 
of SN~2011iv having  a higher $\rho_{\rm c}$ than the progenitor of 
SN~2007on. With a higher $\rho_{\rm c}$, SN~2011iv produced more 
stable $^{58}$Ni in the centre of its ejecta, leading to the formation of a 
central hole in the  distribution of $^{56}$Ni. This ultimately leads to less 
energy deposition in the central region of the ejecta and hence the redder 
colours observed in SN~2011iv as compared with SN~2007on along the 
Lira relation. The colour difference of $\sim 0.12$ mag suggests that
$\rho_{\rm c}$ in SN~2011iv was a factor of two larger than in SN~2007on.

\item  An analysis of the $B$- and $H$-band distance estimates of SN~2007on 
and SN~2011iv reveals relative differences between the two objects of 
$\sim 14$\% and $\sim 9$\%, respectively. These differences serve as a 
warning in the use of transitional SNe~Ia in future efforts to measure 
cosmological parameters. New observational campaigns centred around 
low-luminosity SNe~Ia should place an effort on obtaining data extending 
out to 2--3 months past maximum light, in order to account for luminosity 
discrepancies related to $\rho_{\rm c}$ differences in their progenitors.

\end{itemize}
%
 \begin{acknowledgements}
We thank J. Silverman for useful discussions pertaining to the {\it HST} 
observations and E. Newton for obtaining some NIR spectroscopic 
observations. Supernova research at Aarhus University is support in 
part by a Sapere Aude Level 2 grant funded by the Danish Agency for 
Science and Technology and Innovation, and the Instrument Center for 
Danish Astrophysics (IDA). M. Stritzinger is also supported by a 
research grant (13261) from VILLUM FONDEN. The CSP-I is supported 
by the US National Science Foundation (NSF) under grants AST-0306969, 
AST-0607438, AST-1008343, AST-1613426, AST-1613455, and AST-1613472. 
A portion of the work presented here was done at the Aspen Center for 
Physics, which is supported by US NSF grant PHY-1066293.
G. Pignata acknowledges support provided by the Millennium Institute of 
Astrophysics (MAS) through grant IC120009 of the Programa Iniciativa 
Cientifica Milenio delMinisterio de Economia, Fomento y Turismo de Chile. 
N. Elias de la Rosa acknowledges financial support by the 1994 PRIN-INAF 
2014 (project ``Transient Universe: unveiling new types of stellar explosions 
with PESSTO'') and by MIUR PRIN 2010--2011, ``The dark Universe and the 
cosmic evolution of baryons: from current surveys to Euclid.''
S. Benetti is partially supported by the PRIN-INAF 2014 project ``Transient 
Universe: unveiling new types of stellar explosions with PESSTO.''
P. Hoeflich acknowledges financial support by the grant 1715133 by the  
National  Science Foundation entitled ``Signatures of Type Ia Supernovae 
Explosions and their Cosmological Implications." 
A. V. Filippenko is grateful for financial assistance from US NSF grant
AST-1211916, the TABASGO Foundation, and the Christopher R. Redlich 
Fund; he also acknowledges {\it HST} grants GO-13286, GO-13646, and 
AR-14295 from STScI, which is operated by AURA under NASA contract NAS 5-26555.
Filippenko's work was conducted in part at the Aspen Center for Physics, which is 
supported by NSF grant PHY-1607611; he thanks the Center for its hospitality 
during the neutron stars workshop in June and July 2017.
R. J. Foley is supported by
NASA under Contract No.\ NNG16PJ34C issued through the {\it WFIRST} Science 
Investigation Teams Programme.  The UCSC group is supported in part by 
NASA grant 14-WPS14-0048, US NSF grant AST-1518052, and from fellowships 
from the Alfred P.\ Sloan Foundation and the David and Lucile Packard Foundation.
We  thank the STScI staff for accommodating our target-of-opportunity programs. 
A.\ Armstrong, R.\ Bohlin, S.\ Holland, S.\ Meyett, D.\ Sahnow, P.\ Sonnentrucker, 
and D.\ Taylor were critical for the execution of these programs. Finally, 
we are grateful to N.\ Gehrels and the {\it Swift} team for executing 
our programme promptly.
\end{acknowledgements}
%
\bibliographystyle{aa}

\bibliography{oneA}
 \clearpage

\clearpage
%
\clearpage
 \clearpage
\clearpage
\clearpage
\clearpage
\begin{figure*}
\includegraphics[width=1.0\hsize]{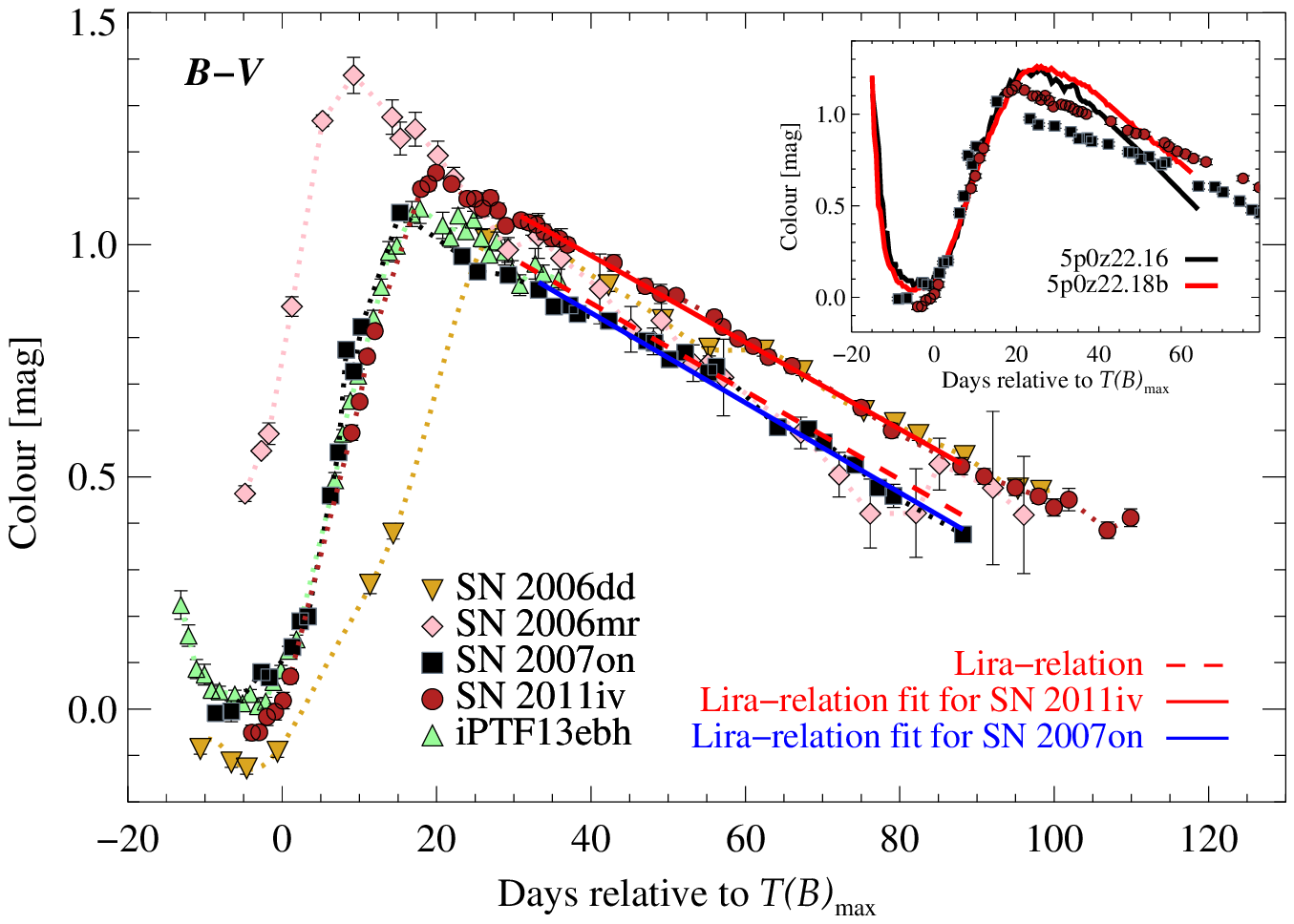}
\caption{Temporal evolution of the observed $B-V$ pseudo-colour. 
The filled red circles, black squares, light-green triangles, 
pink diamonds, and golden downward triangles represent the 
data of (respectively) SN~2011iv, SN~2007on, iPTF13ebh, SN~2006mr, 
and SN~2006dd. The data of the normal SN~2006dd and 
the subluminous SN~2006mr are from \citet{2010AJ....140.2036S} and 
\citet{2010AJ....139..519C}, respectively. All colours have been 
corrected for Galactic reddening. The Lira relation from 
\citet{2010AJ....139..120F} is indicated as a red dashed line, and 
Lira-relation fits to the data are shown as red (SN~2011iv) and 
blue (SN~2007on) solid lines. The top-right inset contains the 
$B-V$ colour evolution of SN~2007on and SN~2011iv compared to 
their corresponding modeled $B-V$ colour evolution predicted by the 
best-fit DD models presented in  Appendix~\ref{appendix:modelcomparisons}.
}
 \label{FIG:BVCOMP}
 \end{figure*} 
%
\clearpage
 \begin{figure*}
\includegraphics[width=1.0\hsize]{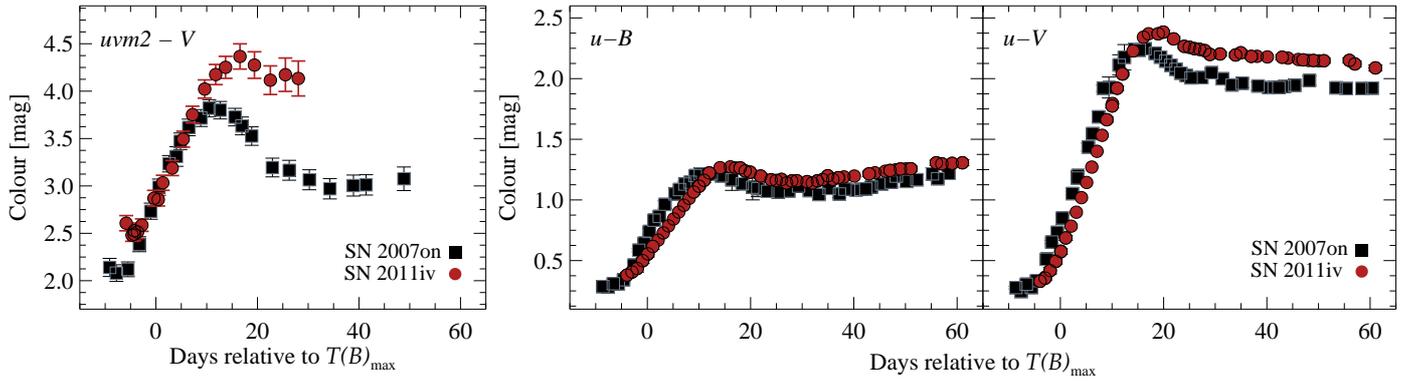}
\caption{ Near-UV colours of SN~2007on (black squares) 
and SN~2011iv (red circles). Presented are the {\em Swift} 
$uvm2$ -- CSP $V$ colours  (left panel), followed by CSP 
$u-B$ and $u-V$ colours in the middle and right panels, 
respectively.
}
 \label{FIG:NUVC}
 \end{figure*}
%
\clearpage
 \begin{figure}
\begin{center}
\includegraphics[width=\hsize]{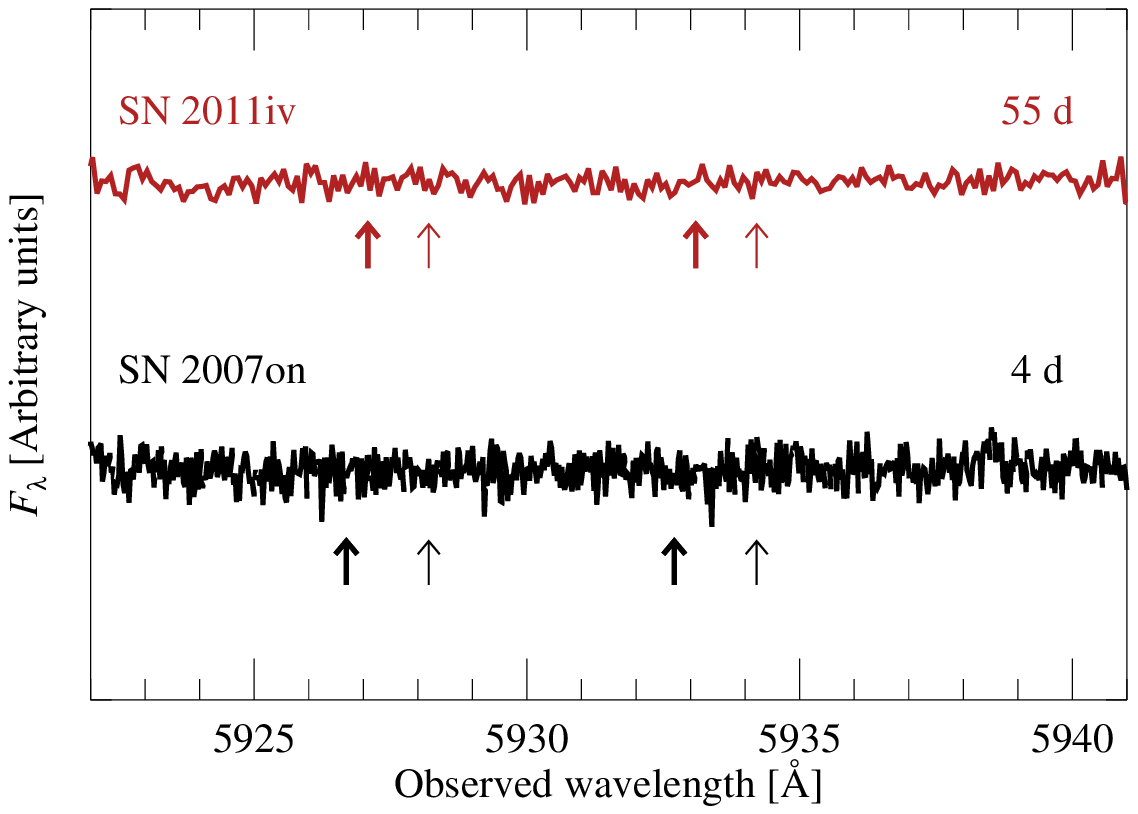}
\includegraphics[width=\hsize]{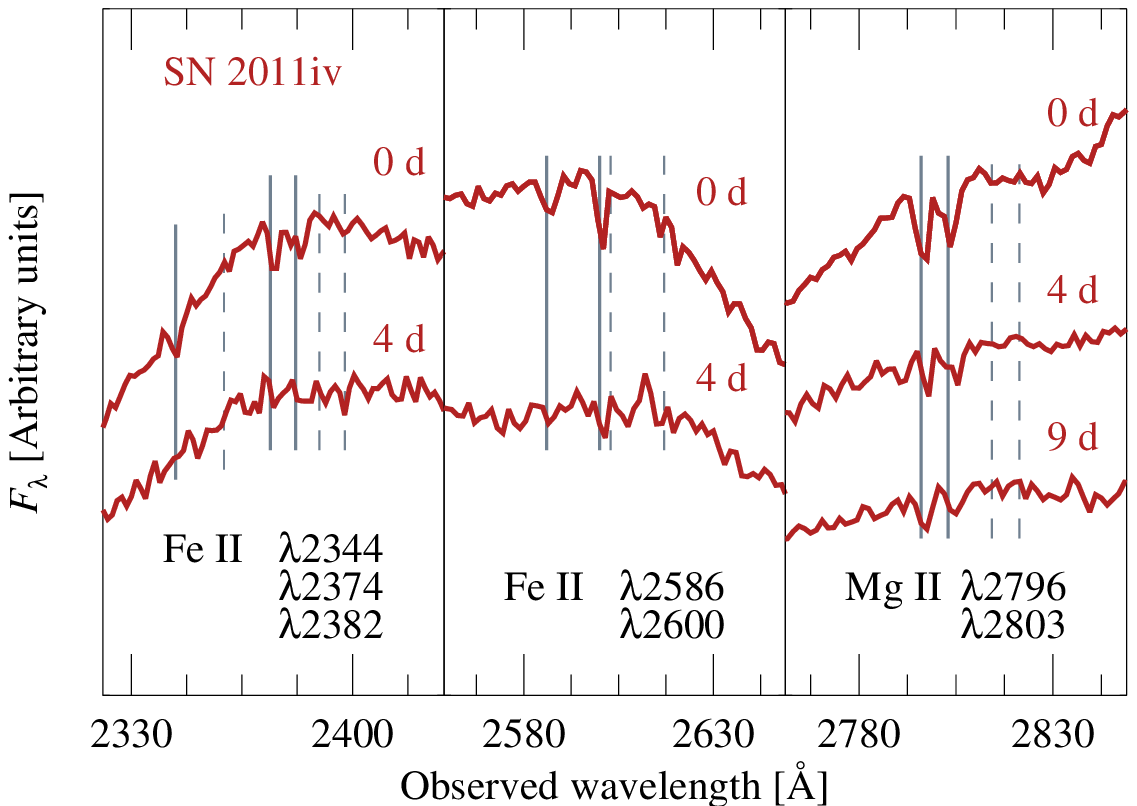}
\end{center}
\caption{Host and interstellar absorption. 
{\em(top)} High-resolution visual-wavelength spectra of 
SN~2011iv (top) and SN~2007on (bottom) 
\citep{2014MNRAS.443.1849S}, zoomed in at the expected 
location of the host-galaxy \ion{Na}{i}{~D} absorption. The thin 
arrows indicate the positions of the \ion{Na}{i}{~D} lines at the 
heliocentric velocity of NGC~1404, while the thick arrows indicate 
the expected positions based on the rotation curve of the host 
\citep{1998A&AS..133..325G}. In both cases, no \ion{Na}{i}{~D} 
lines are discernible, suggesting minimal to no host-galaxy 
reddening. {\em (bottom)} Interstellar Fe and Mg absorption features 
in UV spectra of SN~2011iv. The spectra around maximum light of 
SN~2011iv exhibit narrow absorption features of   
\ion{Fe}{ii} $\lambda$2344,
\ion{Fe}{ii} $\lambda$2374, and
\ion{Fe}{ii} $\lambda$2382 (left panel),
\ion{Fe}{ii} $\lambda \lambda$2586, 2600 (middle panel), and
\ion{Mg}{ii} $\lambda \lambda$2796, 2803 (right panel)
at the position of the Milky Way (solid grey lines). The dashed grey 
lines indicate the expected position of the absorption features, if 
they originate from the host galaxy. 
}
 \label{FIG:MWDUST}
 \end{figure}
\clearpage
\begin{figure*}
\includegraphics[width=0.5\hsize]{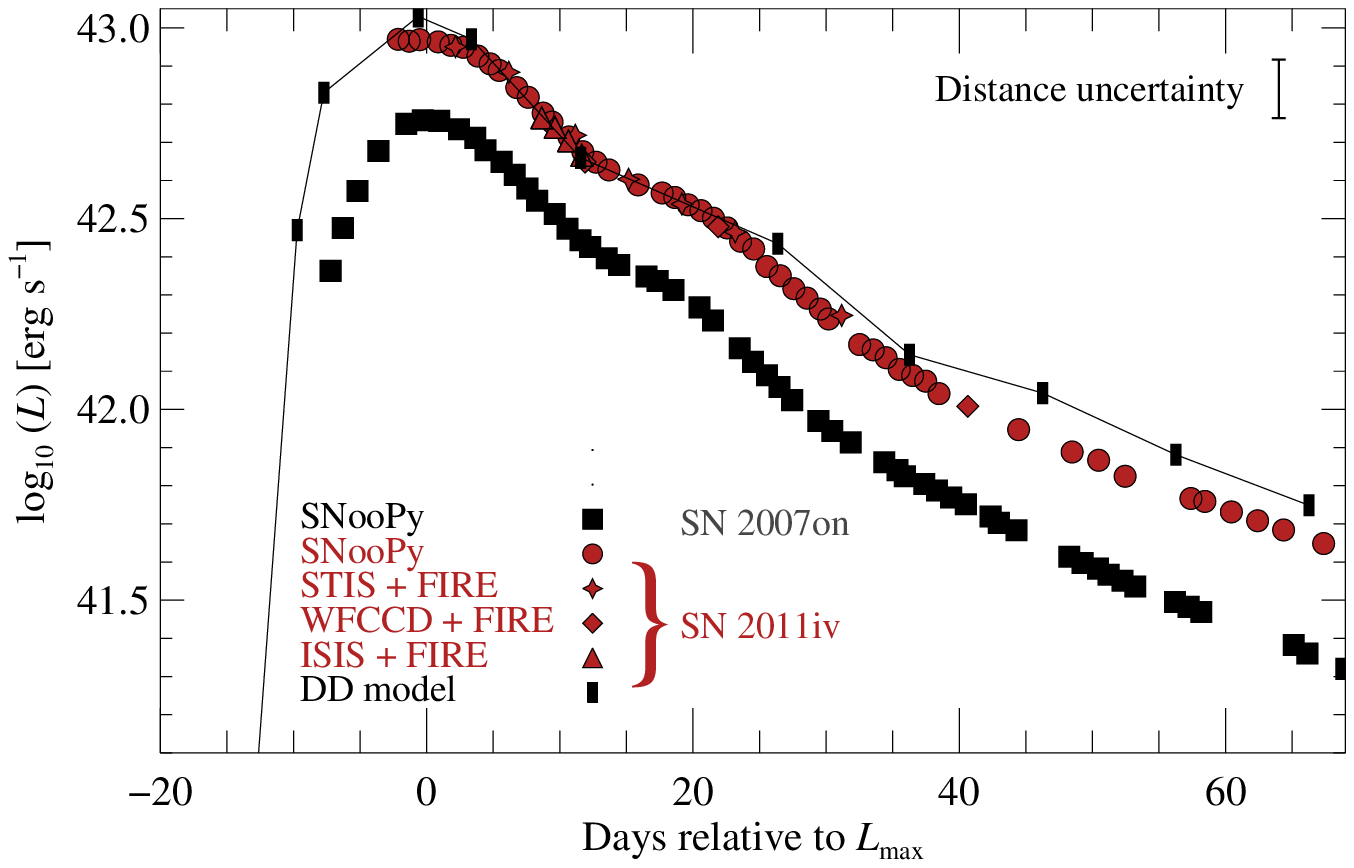}
\includegraphics[width=0.5\hsize]{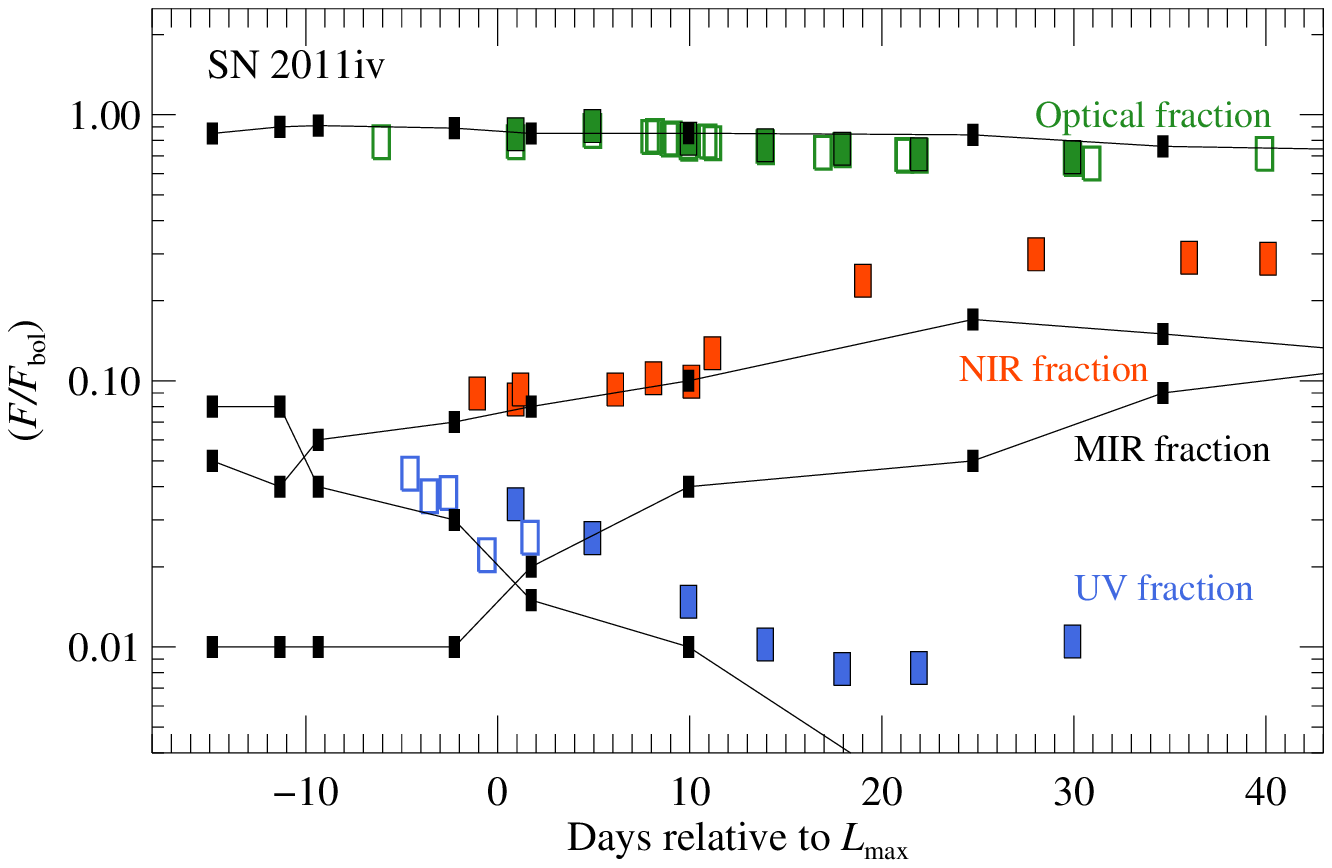}
\caption{
({\it left}) Bolometric light curves of SN~2007on (black) and 
SN~2011iv (red) constructed from observations constraining the flux  
extending from $uvm2$ to the $H$ band ($\sim$ 2250--18,800~\AA; UVOIR).
Complementing the majority of SN~2011iv measurements are a 
handful of additional bolometric points covering the wavelength 
range extending from the UV (1620~\AA) to $K_s$ (24,900~\AA).
The error bar on the bolometric luminosity is presented in the top right of the plot, 
and assumes an error in  the adopted distance modulus of $\pm$0.2 mag.
Also shown is the bolometric light curve (line) of model 5p0z22.18b 
presented in Sect.~\ref{modelcomparison}.
({\it right}) UV (blue), optical (green), and NIR (red) flux fractions of 
the total bolometric light curve of SN~2011iv computed from observed spectroscopy.
The open green symbols illustrate reduced 
optical flux fractions from 3480~\AA\ to 8630~\AA, while the solid green 
symbols represent the full range from the atmospheric cutoff at 
3050~\AA\ to 8630~\AA. The UV fraction extends from either 1620~\AA\  
(STIS, solid blue symbols) or 1900~\AA\  (UVOT, open blue symbols) 
to 3050~\AA, and  the NIR fraction extends from 9000~\AA\ to 24,900~\AA. 
Shown as connected filled squares is the fraction of flux distributed 
among the different wavelength regimes for model 5p0z22.18b. This includes 
the predicted MIR fraction (24,900~\AA\ to 1,000,000~\AA), which drives 
the difference between the observed and modeled UVOIR light curves 
beginning around a month past maximum brightness. 
}
 \label{FIG:BOLO}
 \end{figure*} 
%
\clearpage
\begin{figure}
\includegraphics[width=1.2\hsize]{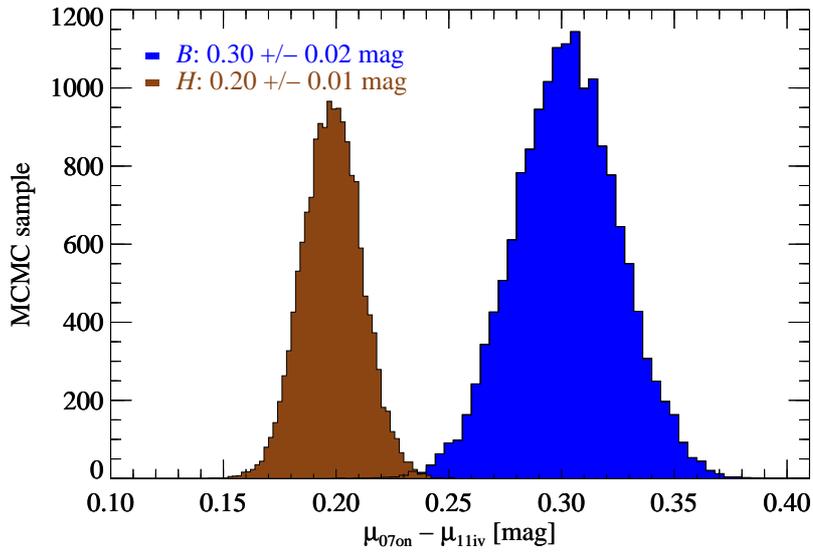}
\caption{Probability distribution of the differences in dereddened distance 
computed for SN~2007on and SN~2011iv from their peak 
$B$- and $H$-band magnitudes, after having applied corrections 
for the luminosity vs. colour and luminosity vs. decline-rate 
relations. The $H$-band distribution is narrower because it is less 
sensitive to errors in the extinction correction.
}
\label{FIG:DISTANCES}
\end{figure} 
%
\clearpage
\begin{figure}
\includegraphics[width=1.0\hsize]{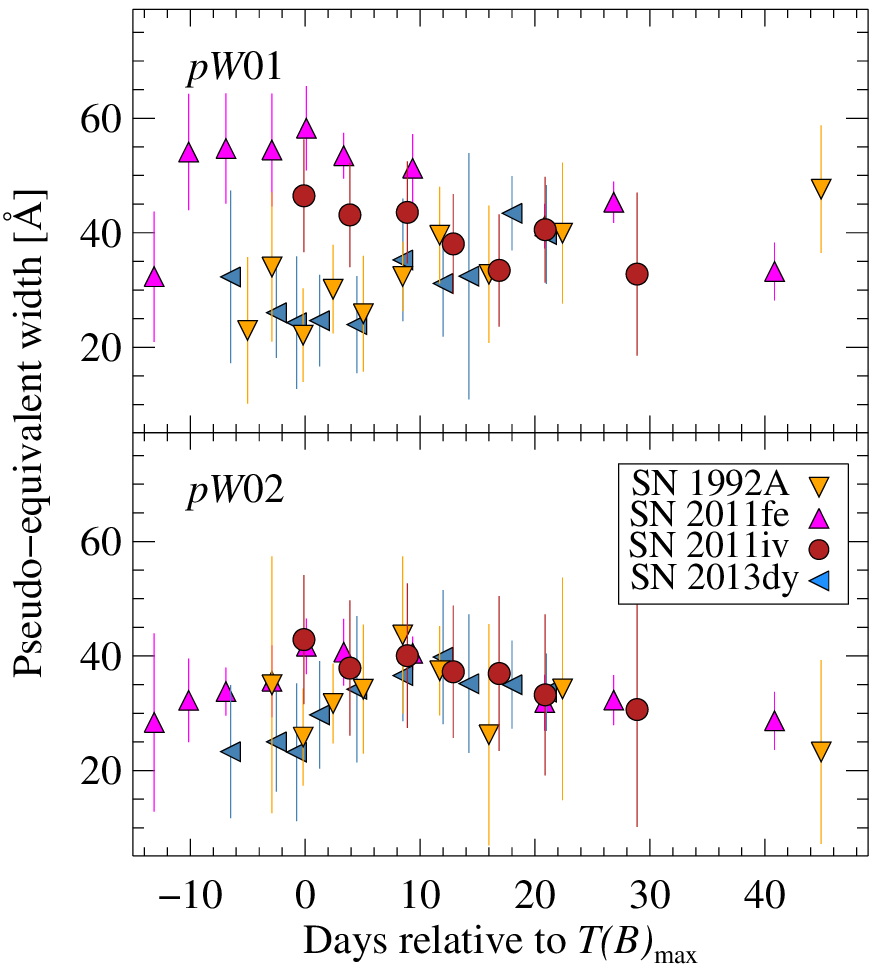}
\caption{The pEW of the prominent $UV$ spectral absorption 
features designated pW01 and pW02. The pEW 
measurements are presented for SN~1992A 
\citep[][yellow downward triangles]{1993ApJ...415..589K}, 
SN~2011fe \citep[][pink upward triangles]{2014MNRAS.439.1959M}, 
SN~2011iv (red circles), and SN~2013dy 
\citep[][blue left-sided triangles]{2015MNRAS.452.4307P}.
}
\label{FIG:UVPWS}
\end{figure} 
%
\clearpage 
\begin{figure*}
\includegraphics[width=1.0\hsize]{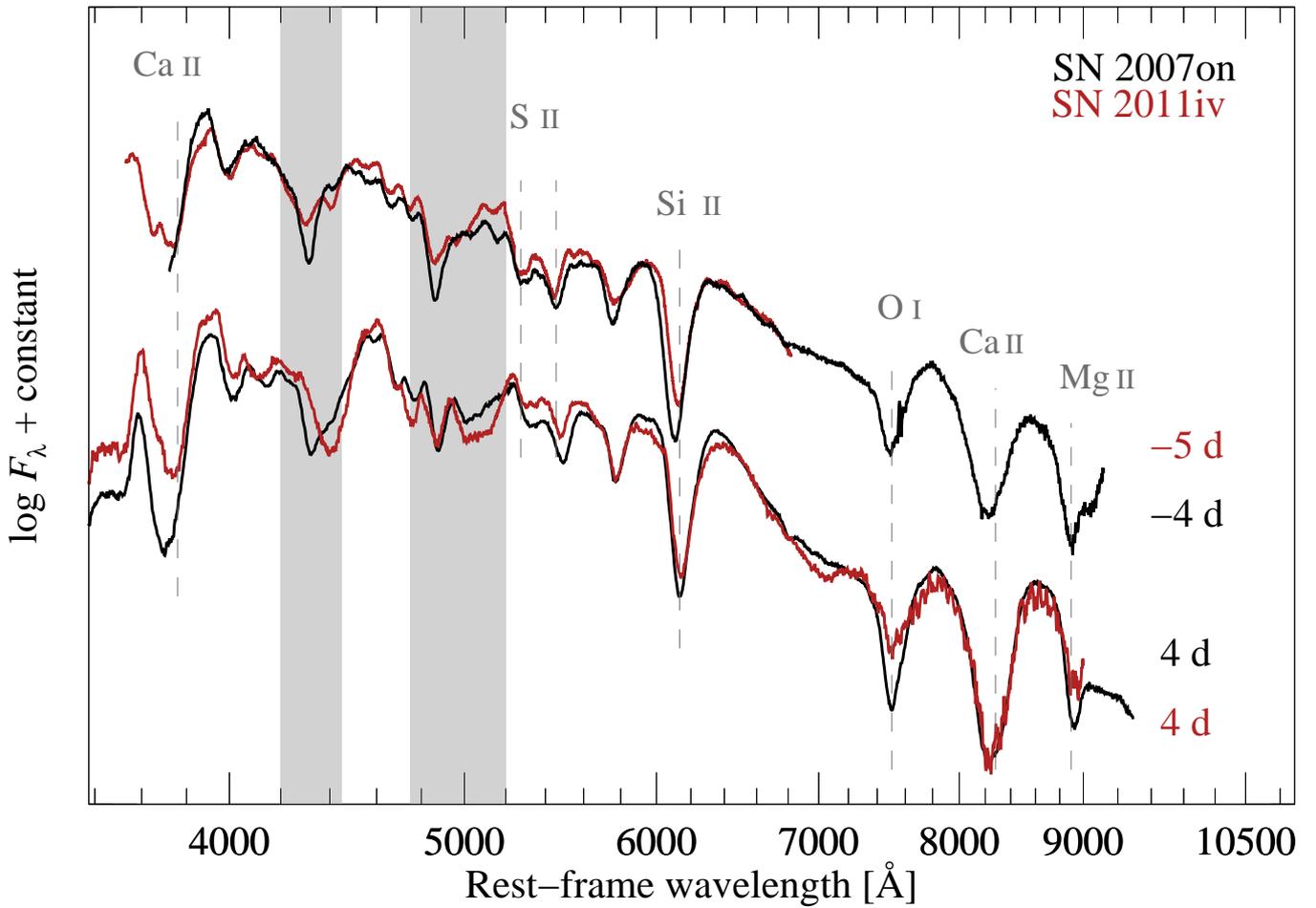}
\caption{Comparison of visual-wavelength spectra of 
SN~2007on and SN~2011iv taken around maximum light. As 
described in Sect.~\ref{SSS:OPTI40D}, the grey vertical bands 
highlight prominent features around 4200--4400~\AA\ and 
4800--5200~\AA, where the two objects show clear disagreement.
}
\label{FIG:OPTISPECMAX}
\end{figure*}

\clearpage
 \begin{figure}
\begin{center}$
\begin{array}{cc}
\includegraphics[width=1.5\hsize]{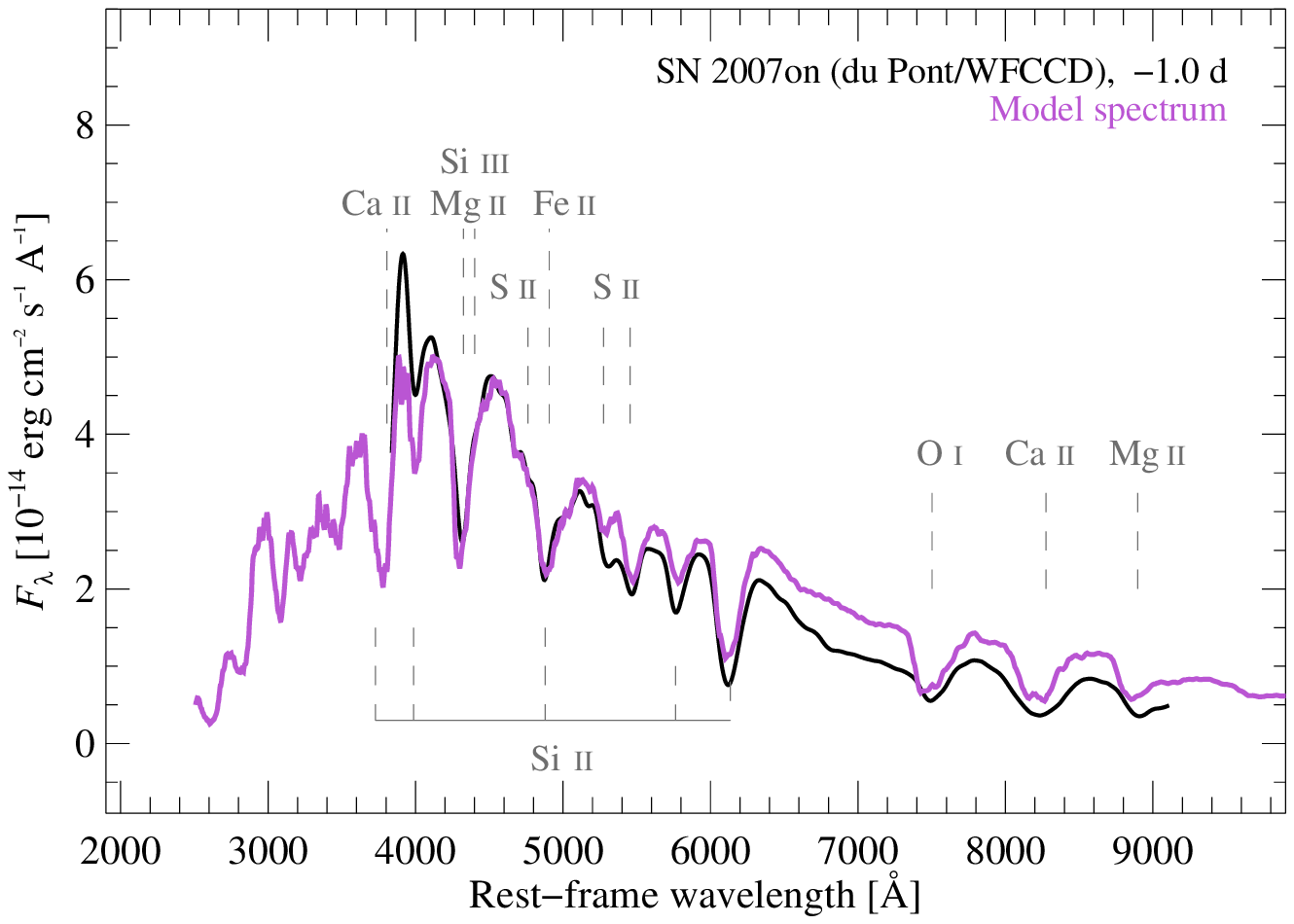}\\
\includegraphics[width=1.5\hsize]{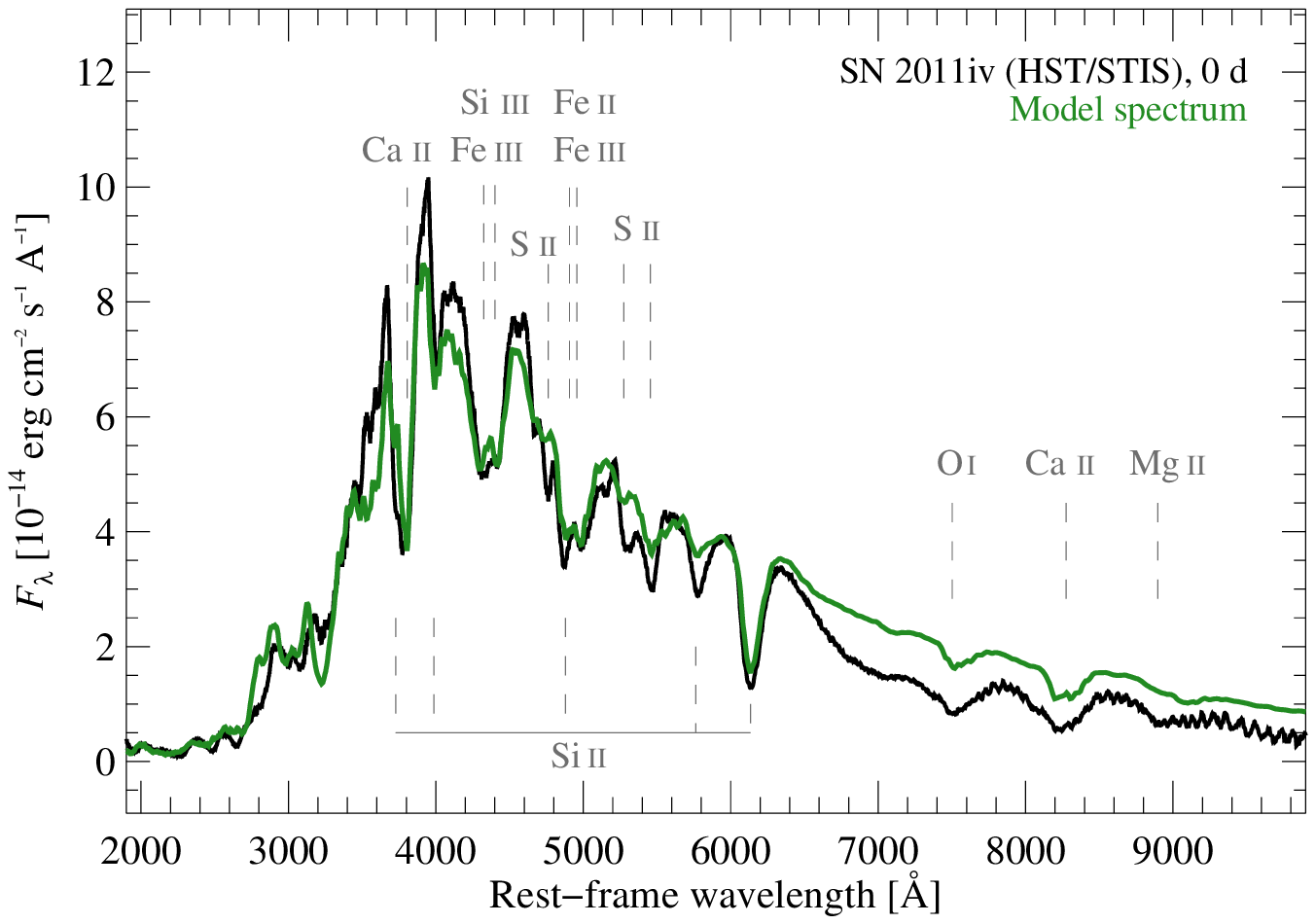}
\end{array}$
\end{center}
\caption{Near-maximum-light visual-wavelength spectra (black) 
of SN~2007on ({\it top}) and SN~2011iv ({\it bottom}) compared to 
our best-fit modelled spectrum.
}
\label{FIG:OP07MODEL}
\end{figure}
%
\clearpage
\begin{figure*}
\includegraphics[width=1.0\hsize]{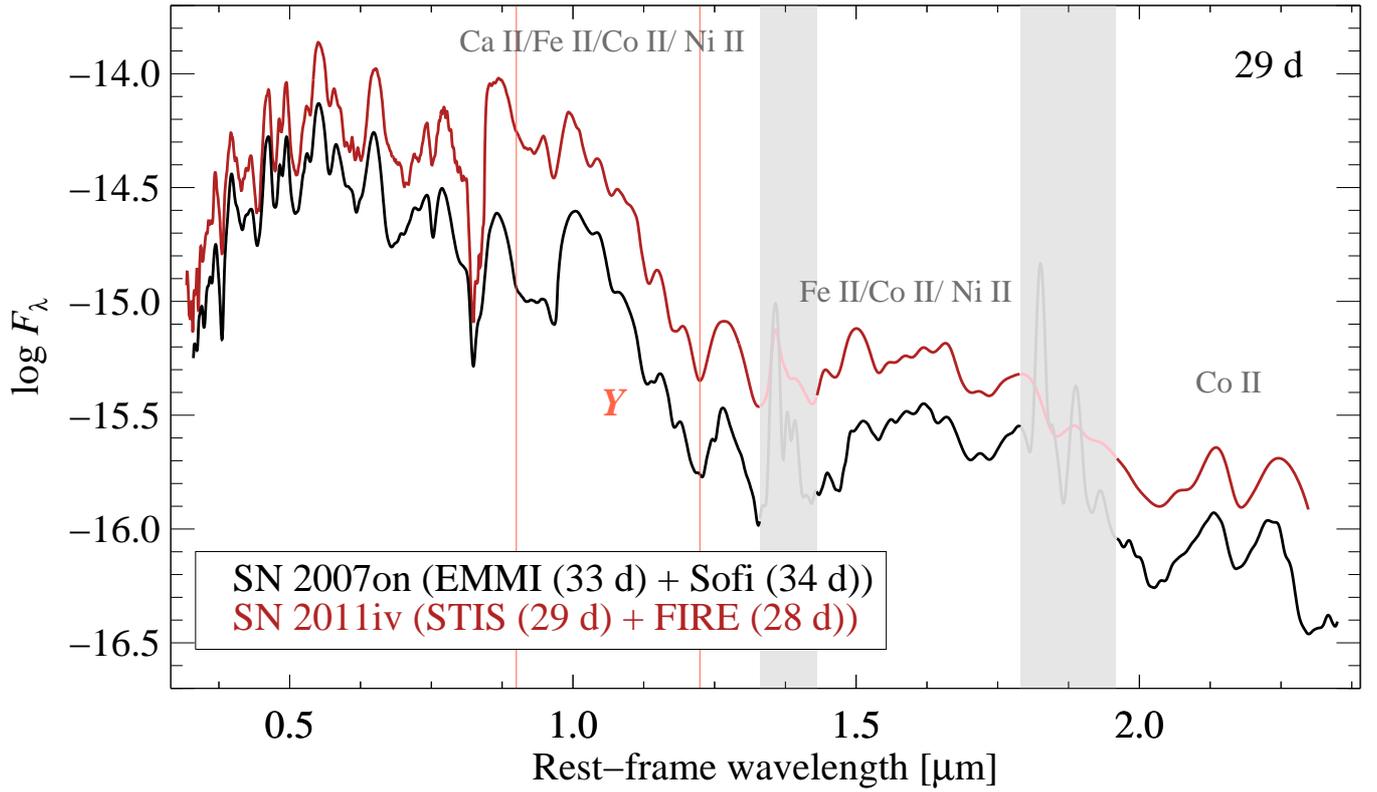}
\caption{Comparison of visual-wavelength and NIR spectra of 
SN~2007on and SN~2011iv taken around a month past maximum light. 
Each spectrum has been calibrated to match the broad-band 
photometry on $+$29~d and smoothed for presentation purposes.
}
 \label{FIG:NIRSPCOMP1107}
 \end{figure*}
%
\clearpage
 \begin{figure*}
\includegraphics[width=0.6\hsize]{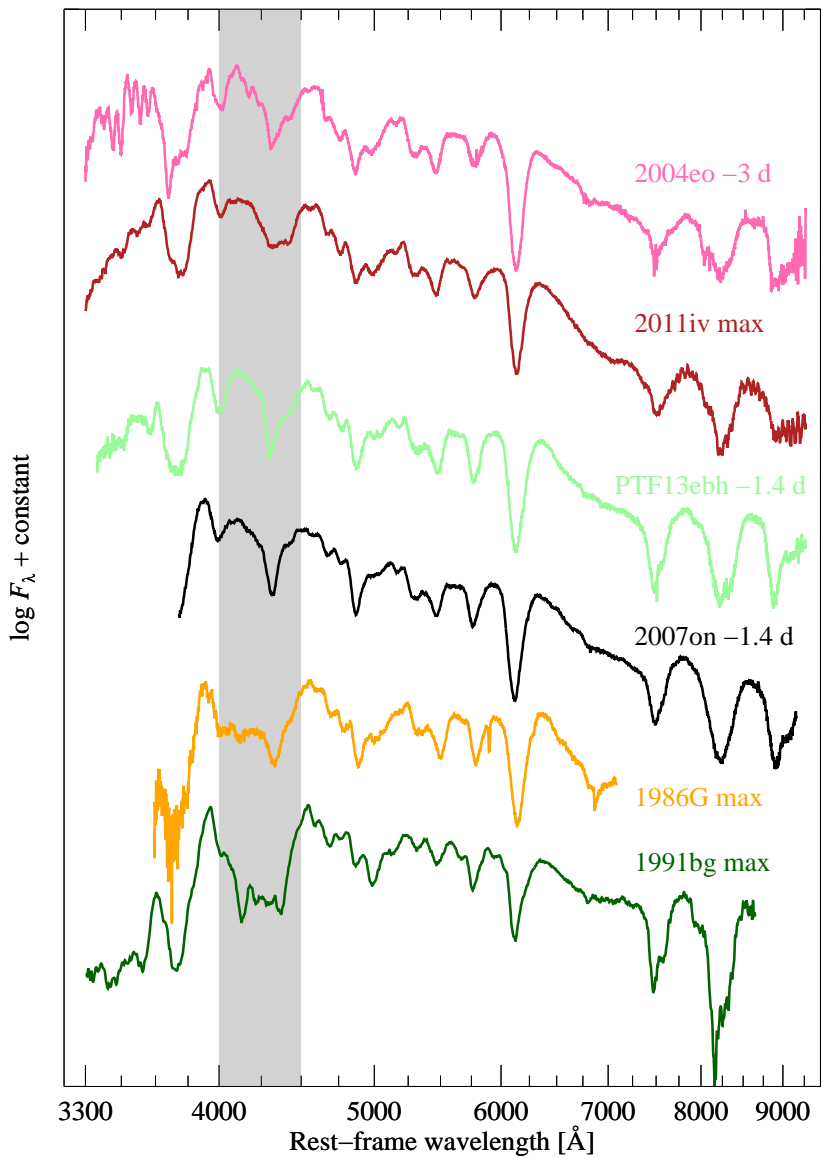}
\caption{Visual-wavelength spectral comparisons around 
maximum brightness (arranged from top to bottom by decreasing 
peak luminosity) of the normal SN~2004eo 
\citep{2007MNRAS.377.1531P}, SN~2011iv, iPTF13ebh 
\citep{2015A&A...578A...9H}, SN~2007on, SN~1986G 
\citep{1992A&A...259...63C}, and SN~1991bg 
\citep{1996MNRAS.283....1T}.  
The spectra of SN~1986G were dereddened using the reddening 
parameters of $A_V = 2.03$ mag and $R_{V} = 2.57$,  
while spectra of SN~1991bg were dereddened using the 
parameters $A_V = 0.22$ mag and $R{_V} = 3.1$. The grey 
shaded region highlights the 4150~\AA\ \ion{Ti}{II} feature that    
is weak in SN~1986G, prevalent in SN~1991bg, and not present 
in the other comparison objects (see text for discussion). 
}
\label{FIG:OPSNSPCOMP}
 \end{figure*}
%
\clearpage
\input{UVOT.tex}

\clearpage 
\input{OPTICALLOCALSEQ.tex}

\clearpage
\input{NIRLOCALSEQ.tex}

\clearpage
\input{FINALOPTICALPHOT.tex}

\clearpage
\input{FINALNIRPHOT.tex}

\clearpage
\input{TAB_SPJ07on.tex}

\clearpage
\input{TAB_SPJ11iv.tex}

%
\clearpage
\input{TAB_LCPARAM}

\clearpage
\input{UVpEW.tex}

\clearpage
\input{BOLOUVOIR_07.tex}
\clearpage
\input{BOLOUVOIR_11iv.tex}
\clearpage
\input{BOLOUVOIR_SPECS_11iv.tex}

\clearpage

%
\begin{appendix}
%
\section{The UV flux ratio}
\label{SSS:UVRAT}
%
To calculate the UV flux ratio, $\mathcal{R}_{\mathrm{UV}}$,
we adopt the definition of \citet{2008ApJ...686..117F} as 
$\mathcal{R}_{\mathrm{UV}} = f_{\lambda}(2770~\AA) / f_{\lambda}(2900~\AA)$.
This is done for both SN~2011iv and other SNe~Ia with available 
UV spectra including SN~1980N, SN~1986G, SN~1992A 
\citep{2008ApJ...686..117F, 1993ApJ...415..589K}, SN~2009ig 
\citep{2012ApJ...744...38F}, SN~2011fe \citep{2014MNRAS.439.1959M},
and SN~2013dy \citep{2015MNRAS.452.4307P}. To compute 
$\mathcal{R}_{\mathrm{UV}}$, all of the spectra were 
corrected for extinction using \texttt{IDL} routines equipped with the 
standard extinction-curve parameterisation of \citet{1999PASP..111...63F}. 
Milky Way reddening values and the host-galaxy recessional velocities are 
taken from NED. For SN~1986G, we adopt host-galaxy 
reddening parameters of $A_V = 2.03$ mag and $R_{V} = 2.57$ 
\citep{2013ApJ...779...38P}.

\begin{figure}
\includegraphics[width=1.7\hsize]{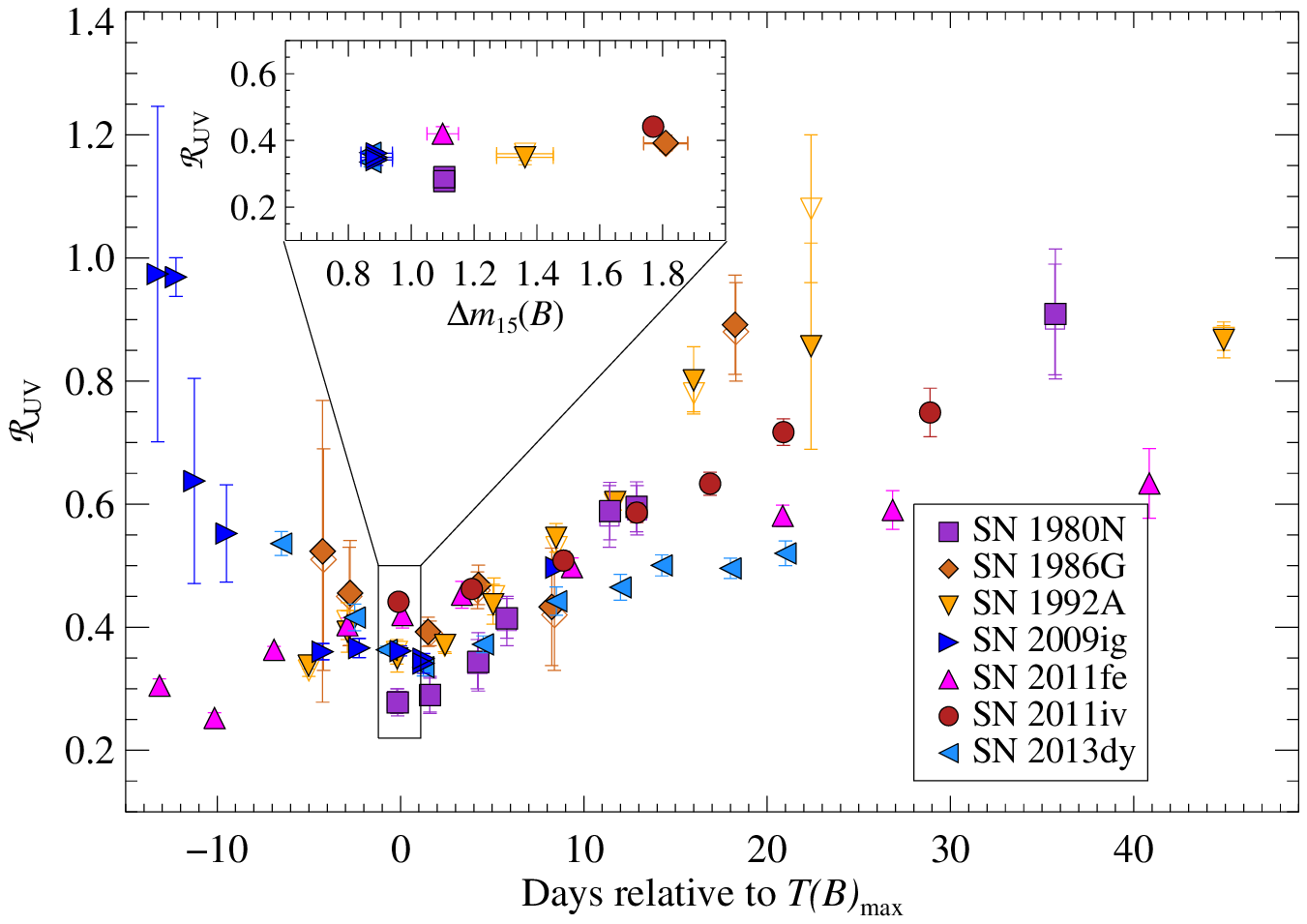}
\caption{Temporal evolution of the UV ratio, 
$\mathcal{R}_{\mathrm{UV}}$. The $\mathcal{R}_{\mathrm{UV}}$ 
values for SN~2011iv between 0~d and $+$29~d are shown as red 
filled circles. For comparison, we calculated the 
$\mathcal{R}_{\mathrm{UV}}$ for SN~1980N and SN~1986G 
\citep{2008ApJ...686..117F}, SN~1992A \citep{1993ApJ...415..589K}, 
SN~2009ig \citep{2012ApJ...744...38F}, and for SN~2011fe 
and SN~2013dy \citep{2015MNRAS.452.4307P}.
Results are shown as solid orange, purple, yellow, and dark-blue 
symbols, respectively. Additionally, the derived 
$\mathcal{R}_{\mathrm{UV}}$ from \citet{2008ApJ...686..117F} for 
SN~1986G, SN~1980N, and SN~1992A are presented as open 
symbols consisting of the same colour and shape. The small 
differences arise from different assumptions about redshift and 
host-galaxy reddening, as described in Appendix~\ref{SSS:UVRAT}. 
}
 \label{FIG:UVRAT}
 \end{figure}

Figure~\ref{FIG:UVRAT} displays the temporal evolution of 
$\mathcal{R}_{\mathrm{UV}}$ for the selected sample of SNe~Ia,
which are characterised by different values of $\dmfiften$ and 
host-galaxy properties. Seemingly, $\mathcal{R}_{\mathrm{UV}}$ 
does not follow a simple trend with epoch, though it appears that 
for all SNe~Ia $\mathcal{R}_{\mathrm{UV}}$ increases between 0~d 
and $+$40~d. However, prior to $B$-band maximum brightness, 
$\mathcal{R}_{\mathrm{UV}}$ either increases or decreases over 
time for the SN~Ia sample. The inset of Fig.~\ref{FIG:UVRAT} 
provides a close-up view of the epoch around $B$-band maximum 
($\pm 1$~d) and indicates that $\mathcal{R}_{\mathrm{UV}}$ of the 
selected SNe~Ia does not follow a clear trend with $\dmfiften$.

Additionally, also compared in Fig.~\ref{FIG:UVRAT} are 
$\mathcal{R}_{\mathrm{UV}}$ values of SN~1980N, SN~1992A,  
SN~1986G, and those of \citet{2008ApJ...686..117F} computed 
using the same set of data. While the results overall agree, there 
are small differences owing to different assumed values of reddening 
and/or host-galaxy recession velocity.  As also discussed in 
the literature \citep[e.g.,][]{2012MNRAS.427..103W}, 
$\mathcal{R}_{\mathrm{UV}}$ is not a broad-band colour 
measurement, and it may therefore be sensitive to various
issues pertaining to both the quality of the data and 
data analysis, but also environmental and/or intrinsic differences.
%
\clearpage
 
 \clearpage
\section{Analysis of visual-wavelength spectroscopy}

\subsection{Optical pseudo-equivalent width}
\label{SSS:OPEW:wq}
%
In analogy to the UV spectral region, here we quantify the properties 
of the most prominent optical absorption features by measuring the 
pEW. We assess the pEW of eight features in a similar way as outlined 
in Sect. \ref{SSS:UVPEW}, adopting the definition and naming 
convention as described by \citet[][see their Table 4]{2013ApJ...773...53F}. 
In short, the specific pEWs are as follows:
pW1 (\ion{Ca}{ii} H\&K), 
pW2 (\ion{Si}{ii} $\lambda$4130),  
pW3 (\ion{Mg}{ii}), 
pW4 (\ion{Fe}{ii}),
pW5 (\ion{S}{ii} W), 
pW6 (\ion{Si}{ii} $\lambda$5972), 
pW7 (\ion{Si}{ii} $\lambda$6355), 
and pW8 (\ion{Ca}{ii} NIR triplet). 
We note that for SN~2007on and SN~2011iv, the pW3 
(\ion{Mg}{ii}) is a blend dominated by either \ion{Mg}{ii} 
$\lambda$4481 or \ion{Si}{iii} $\lambda$4560, and pW4 
(\ion{Fe}{ii}) is shaped by \ion{S}{ii} $\lambda$5032, 
\ion{Si}{ii} $\lambda$5055, and \ion{Fe}{ii} $\lambda$5083.  

\begin{figure}
\includegraphics[width=1.3\hsize]{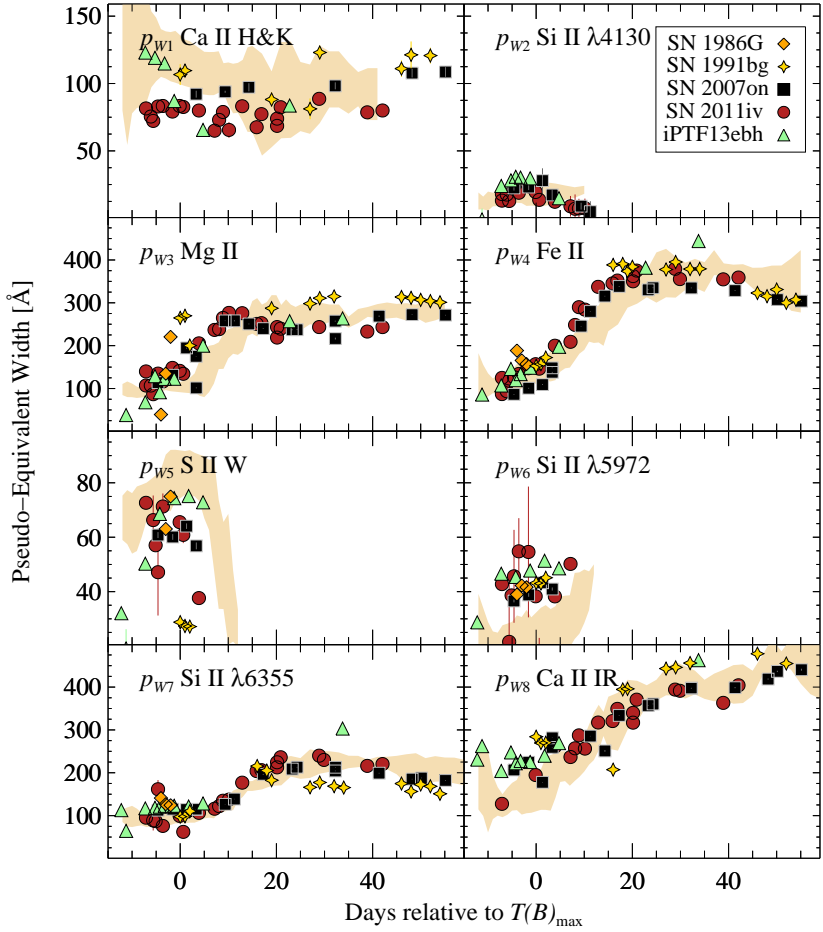}
\caption{Temporal evolution of eight pEW spectral indicators.
Black squares and red dots correspond to the pEW values of 
SN~2007on and SN~2011iv, respectively. Additionally, presented 
are pEW values determined from spectra of the subluminous 
Type~Ia SN~1986G (orange diamond), SN~1991bg (yellow star), 
and the transitional SN~Ia iPTF13ebh \citep[][green triangles]
{2015A&A...578A...9H}. The yellow shaded areas corresponds to 
the average pEW values measured from a large set of normal 
SNe~Ia \citep{2013ApJ...773...53F}.  
}
 \label{FIG:PSWEVOL}
 \end{figure}

Figure~\ref{FIG:PSWEVOL} presents the temporal evolution of the 
pEW values for all eight features as measured from the 
spectroscopic sequences of SN~2007on and SN~2011iv (see 
Tables \ref{T:SPJ07ON} and \ref{T:SPJ11IV}). Additionally, 
Fig.~\ref{FIG:PSWEVOL} displays the pEW values of SN~1986G, 
SN~1991bg, and iPTF13ebh, as well  as the average pEW 
values (and their associated 1$\sigma$ dispersion) as measured 
from the CSP-I SN~Ia  sample \citep{2013ApJ...773...53F}. 

Inspection of Figure~\ref{FIG:PSWEVOL} reveals that fast-declining 
SNe~Ia show very similar pEW values across all epochs, 
although  the pEW values of pW4 are somewhat lower in 
SN~2007on. The measured pEW values of the fast-declining 
SNe~Ia are found to be largely similar with the average pEW 
values of the normal SNe~Ia (yellow shaded region), except in the 
case of pW1 (\ion{Ca}{ii} H\&K) and pW5 (\ion{S}{ii} W), 
which are both consistently below the mean sample values.
On the other hand, the pW6 (\ion{Si}{ii} $\lambda$5972) pEW 
values are larger than what is found in normal SNe~Ia, and this is 
also consistent with observations of other fast-declining SNe~Ia 
\citep{2006MNRAS.370..299H}. This phenomenon is a 
consequence of the increasing \ion{Si}{iii} to \ion{Si}{ii}  
recombination rate that is caused by a decrease in temperature, 
leading to a more prevalent \ion{Si}{ii} $\lambda$5972  feature 
\citep[see][]{2008MNRAS.389.1087H}. Finally, the measured 
pEW values of pW2 (\ion{Si}{ii} $\lambda$4130) and pW7 
(\ion{Si}{ii}  $\lambda$6355) are found to be in agreement with the 
average pEW values of the normal SNe~Ia, while for pW3 
(\ion{Mg}{ii}) and pW4 (\ion{Fe}{ii}) the fast-declining objects 
have larger pEW values until about $+$10~d to $+$20~d.

To categorise the spectroscopic diversity of SNe~Ia, 
\citet{2006PASP..118..560B} introduced a classification scheme 
based on the pEW values of pW6 (\ion{Si}{ii} $\lambda$5972) 
and pW7 (\ion{Si}{ii} $\lambda$6355) at maximum light. There 
are four subclasses defined: the ``cool'' (CL), the ``broad 
line'' (BL), the ``shallow silicon'' (SS), and the ``core normal'' (CN). 
Adopting the definition as outlined by  \citet{2012AJ....143..126B} 
and \citet{2013ApJ...773...53F},  SN~2007on and SN~2011iv are 
both clearly CL SN~Ia.

\subsection{Optical line velocity}
\label{SSS:OPLV}
%
Here we present measurements of the Doppler velocity at 
maximum absorption of the most prominent ions. This  provides an 
estimate of the velocity distribution of the various line-forming 
regions of the ejecta. Measurements were obtained by fitting a 
Gaussian function to the absorption profiles using a custom-made 
\texttt{IDL} program.

\begin{figure}
\includegraphics[width=1.1\hsize]{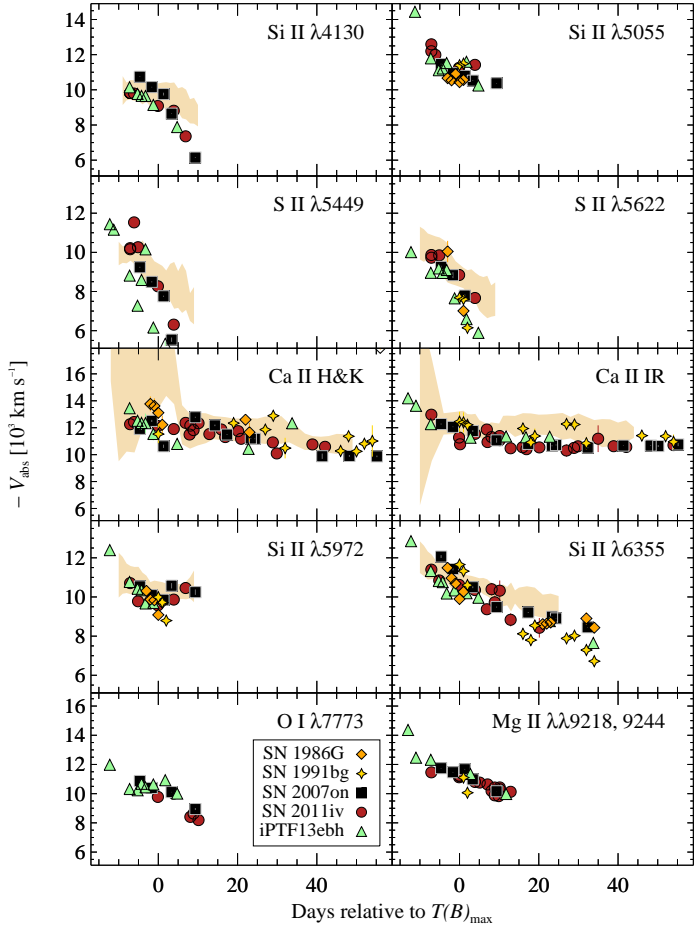}
\caption{Temporal evolution of the Doppler velocity of prominent 
ions. Velocity measurements of SN~2007on (black squares) 
and SN~2011iv (red circles) are compared to measurements
of SN~1986G, SN~1991bg, and iPTF~13ebh (colour coded
as in Fig.~\ref{FIG:PSWEVOL}). The yellow shaded areas 
correspond to the average velocities measured from a large 
set of normal SNe~Ia \citep{2013ApJ...773...53F}.
}
 \label{FIG:VELOCITY}
 \end{figure}

Figure~\ref{FIG:VELOCITY} displays the Doppler velocity 
measurements of SN~1986G, SN~1991bg, SN~2007on, 
SN~2011iv, and iPTF13ebh for ten different spectral line 
features. The uncertainty in the measurements is roughly
500 km~s$^{-1}$. Additionally, the average velocity values (and 
their associated 1$\sigma$ dispersion) of seven absorption 
features obtained from the CSP-I SN~Ia sample  \citep{2013ApJ...773...53F} 
are shown in Figure~\ref{FIG:VELOCITY}. Overall, the velocities 
of the various spectral features in the fast-declining SNe~Ia 
are found to be marginally less than those of normal SNe~Ia. 
Comparison of velocities and their temporal evolution among 
the fast-declining SNe~Ia indicated no significant differences.

\clearpage
\clearpage
\clearpage
\section{Nuclear burning fronts in multi-dimensional objects}
\label{appendix:burning}
The findings presented in Sect.~\ref{modelcomparison} are based
on comparisons of the light-curve properties of SN~2007on and 
SN~2011iv to a suite of spherical 1-D DD explosion models though
the physics and many aspects of nuclear burning fronts are inherently 
multi-dimentional be it the deflagration phase, the transition
between deflagration to detonation, and the detonation phase.

During the last decades, significant progress has been made
  toward a better understanding of the physics of deflagration
  hydrodynamical flames in SNe~Ia.  On a microscopic scale, a
  deflagration propagates due to heat conduction by electrons. Though
  the laminar flame speed is well known, the fronts are hydrodymamical
  unstable which requires multi-dimensional simulations adopting
  various sub-grid scale model
  \citep{n95,rein99,lisewski00,k01,g02,R02,g02,g04,2014MNRAS.438.1762F,zeldovich70}.
  All groups found that Rayleigh-Taylor instabilities (RT) govern the
  morphology of the fronts in the regime of linear instabilities, and
  other instabilities (e.g. Kelvin-Helmholtz) become
  important. Differences in the simulations occur due to the adoption
  of various sub-grid models and due the sensitivity with respect to the initial
  conditions of the WD at the time of the thermonuclear runaway and
  multi-spot ignitions \citep{n96,calder03,plewa04,LAH05}. All
  simulations to date predict significant large-scale mixing of burned
  and unburned material in the WD already during the deflagration
  phase.

Detailed studies of well-observed SNe~Ia suggest there is
 a process at work which partially suppresses RT instabilities during
the deflagration burning phase. For example, direct imaging of the
supernova remnant (SNR) S~Andromedae suggests a large Ca-free
core, indicative of high-density burning and limited mixing
\citep{2007ApJ...658..396F, 2015ApJ...804..140F, 2016sros.confE.115F}.
Additional lines of evidence of partially suppressed RT mixing are 
significantly degraded spectral fits by models  having an injection of 
radioactive material into the Si/S layers \citep{2002ApJ...568..791H}, 
pot-bellied nebular line profiles \citep{2004ApJ...617.1258H, 
2006ApJ...652L.101M, 2003Sci...299...77G, 2010Natur.466...82M, 
2015A&A...573A...2S, 2015ApJ...806..107D}, and broad [\ion{Co}{iii}], [\ion{Fe}{ii}], and
[\ion{Ar}{ii}] spectral features seen in late-time MIR spectra, 
which at the same time exhibit narrow $^{58}$Ni lines \citep{2004ApJ...607..391G,
2015ApJ...798...93T, 2015ApJ...804..140F}. The underlying physics
driving the suppression of RT mixing is currently unknown, though
effects associated with the presence of prevalent magnetic fields
could be a possibility which can suppress large scale instabilties 
\citep{2004ApJ...617.1258H, 2014ApJ...795...84P, 
2014ApJ...794...87R, 2015ApJ...806..107D, 2016AAS...22723716H}, or
it may be related to the initial condition of the WD such as rapid rotation 
\citep{uenishi03,yoon04b,h06}. In this study, we adopt spherical explosion 
models as they naturally partially suppress RT mixing during the deflagration 
burning phase  \citep{2001NuPhA.688...21D}.

Hydrodynamical instabilities and the interplay between hydrodynamics 
and nuclear burning also plays a role in the transition from the deflagration to 
detonation burning phases.  However, here the processes at play are not associated 
with large scale hydrodynamical instabilities, but rather to small scale mixing 
of burned and unburned material at the interface of a compressional shock 
wave following the so-called Zel`dovich mechanism \citep[e.g.,][]{1995ApJ...452..779N, 
1997ApJ...478..678K}. Other mechanisms suggested invoke
crossing shock waves produced in the highly turbulent medium \citep{livne98} 
and mixing by shear flows
in rapidly, differentially rotating WDs \citep{hetal03,uenishi03,yoon04a,yoon04b}.
The mechanism(s) for small scale mixing is still under debate and in all simulations
the transition from the deflagration to detonation phase is initiated by microscopic
mixing within the adopted parameterisation  \citep[see, e.g.,][]{1995ApJ...452...62L,
1995ApJ...452..779N, 1997ApJ...478..678K,gamezo04, 2013A&A...559A.117C}. 

\section{Model comparisons}
\label{appendix:modelcomparisons}

We now present a comparison between predicted observables of the 
models and SNe~2007on and 2011iv. As a baseline we use models  originating 
from a 5 M$_\odot$ main sequence star with solar metallicity \citep{2002ApJ...568..791H}.   
The best-fit model parameters are determined from the interpolations of an 
extend grid of  DD models with $\rho_c = 0.5...6\times10^{9}$ g~cm$^{-3}$  
and $\rho_{tr} = 0.5... 2.7\times10^7$~g~cm$^{-3}$ using the 
$\Delta m_{15}-CMAGIC$ optimization for determining the model 
parameters \citep{2017ApJ...846...58H}. 

Comparison of key observables to  predictions of these  models 
(computed using the CSP-I passbands) are determined  using the absolute 
$B$- and $V$-band light curves, the $B-V$ colors and the $B$- and $V$-band 
decline-rate parameters. We note that the $B-V$ color offset noted in 
Sect.~\ref{SSC:BVCOL} is used to calculate the relative difference in the 
$\rho_{c}$ between SNe~2007on and 2011iv indicates a factor of two 
differences in $\rho_{c}$, assuming  the conical  value of $\rho_c = 2.0\times10^{9}$ g~cm$^{-3}$ for SN~2011iv.

The best-fit model for  SN~2007on is characterized by $\rho_{\rm c} = 1.0\times10^{9}$ 
g~cm$^{-3}$ and $\rho_{\rm tr} = 1.6\times10^{7}$~g~cm$^{-3}$ and 
corresponds to model 5p0z22.16.1E9 (see Hoeflich et al. 2017). Model 5p0z22.16.1E9 
produces 0.008~M$_\odot$ of stable $^{58}$Ni and 0.32~M$_\odot$ 
of radioactive $^{56}$Ni mass of which 0.13~M$_\odot$ is produced  
during the deflagration burning phase and 0.19~M$_\odot$ during the detonation 
burning phase. The difference between the model observables  [$M_{B}$, 
$B-V$ at $T(B)_{\rm max}$, $\Delta m_{15}(B)$, $\Delta m_{15}(V)$] 
of [$-18.29$ mag, $0.11$ mag, $1.69$ mag, $1.16$ mag] and those corresponding to 
 SN~2007on of [$-18.23 \pm 0.08$ mag, $0.08 \pm 
0.01$ mag, $1.96 \pm 0.01$ mag, $1.14 \pm 0.01$ mag] are
[$-0.06$ mag, $0.03$ mag, $-0.27$ mag, $0.02$ mag], respectively. 

The best-fit model for SN~2011iv is characterized by  
$\rho_{\rm c} =  2.0\times10^{9}$ g~cm$^{-3}$ and 
$\rho_{\rm tr} = 1.7\times10^{7}$~g~cm$^{-3}$ and corresponds 
to model 5p0z22.18b, which is an interpolation between models 
5p0z22.16 and 5p0z22.18 (Hoeflich et al. 2002). Model  5p0z22.18b 
produces 0.02~M$_\odot$ of stable $^{58}$Ni  and 0.37~M$_\odot$ 
of radioactive $^{56}$Ni of which  0.16~M$_\odot$  was produced 
during the deflagration burning phase and 0.21 M$_\odot$  
during the detonation burning phase. The difference between the 
model observables  [$M_{B}$, $B-V$ at $T(B)_{\rm max}$, 
$\Delta m_{15}(B)$, $\Delta m_{15}(V)$] of [$-18.64$ mag, $0.09$ mag, $1.67$ mag, 
$1.08$ mag] and those corresponding to  SN~2011iv of [$-18.83 \pm 0.08$ mag, $0.01 
\pm 0.01$ mag, $1.77 \pm 0.01$ mag, $1.08 \pm 0.01$ mag] are 
[$-0.19$ mag, $+0.08$ mag, $-0.10$ mag, 0.0 mag], respectively.
Note, as demonstrated in Fig.~\ref{FIG:BOLO}, that the UVOIR 
light curve computed for 5p0z22.18b is found to fit the UVOIR light 
curve SN~2011iv fairly well.

As an example of the model calculations we show in Fig. D.1  
the best-fit  model for SN~2007on (i.e., 5p0z22.16.1E9) its density  
and velocity structures as a function of mass  (left panel), its final 
abundance tomography (middle panel), and the effect of varying 
$\rho_{c}$ on the final distribution of $^{56}$Ni (right panel). 
The middle panel of Fig.~D.1  demonstrates that the abundance in 
the inner region is dominated by electron capture elements, and the 
size of this region is determined by $\rho_{c}$ (see right panel).
We note that the outer abundance tomography is largely determined 
by the detonation burning phase and therefore on $\rho_{tr}$ (Diamond et al. 2015).
This justifies the procedure to interpolate in $\rho_{c}$  and $\rho_{tr}$.

The interpolated best-fit model for SN~2011iv (i.e., model 5p0z22.18b) 
has an overall similar structure as that shown in Fig.~D.1  for model 5p0z22.16.1E9, 
but its intermediate mass elements are shifted to a higher velocity by  
$\approx$ 1000 km~s$^{-1}$ and the electron-capture elements  extend to 3200 km~s$^{-1}$ (see right panel).

\clearpage
\begin{figure*}
\begin{center}$
\begin{array}{ccc}
\includegraphics[width=0.30\textwidth]{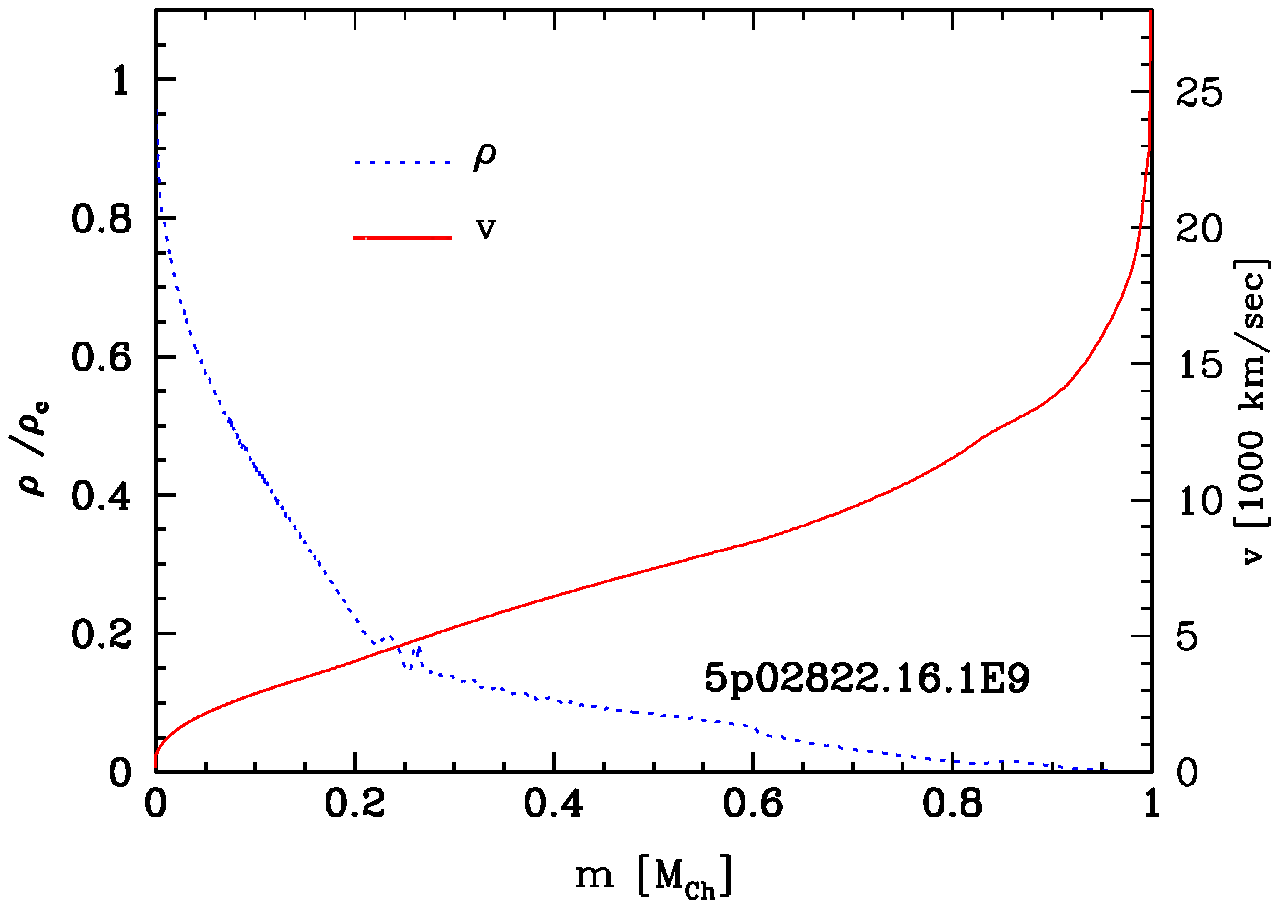} &
\includegraphics[width=0.29\textwidth]{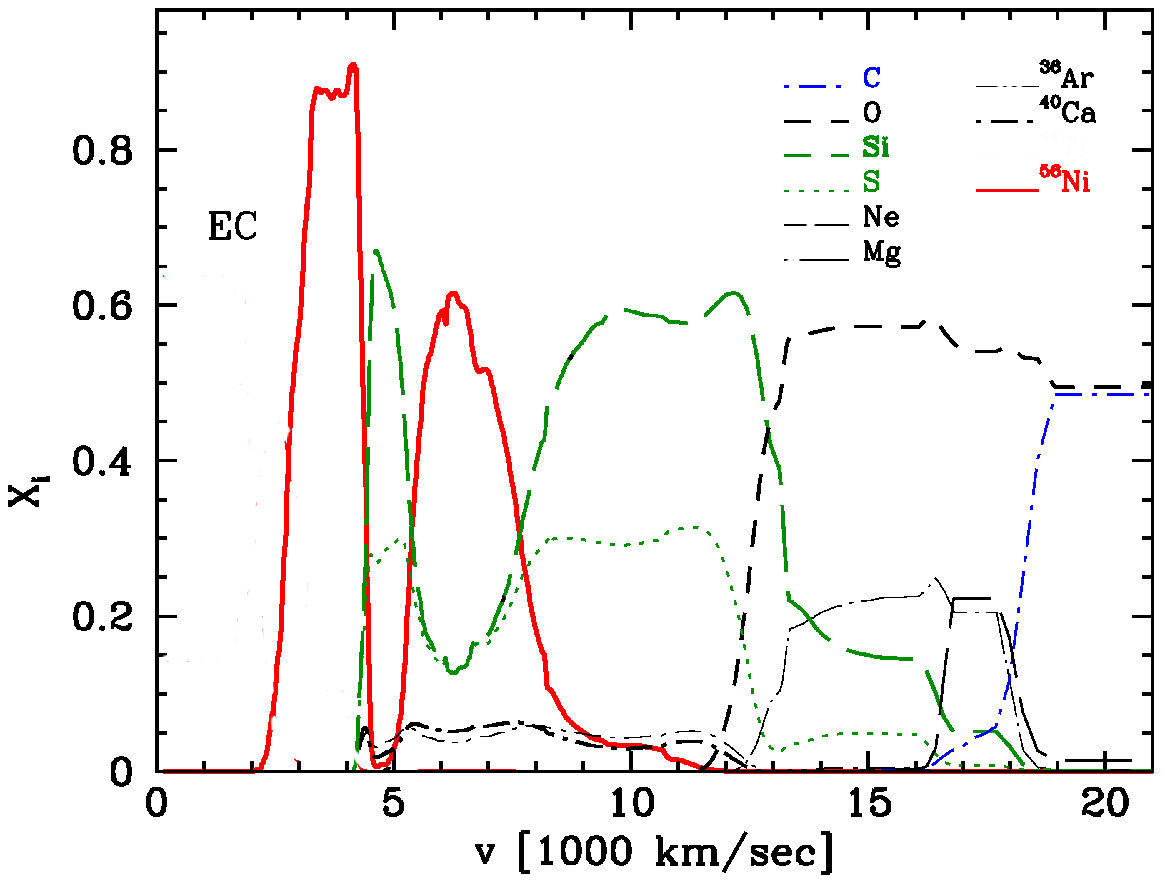} &
\includegraphics[width=0.30\textwidth]{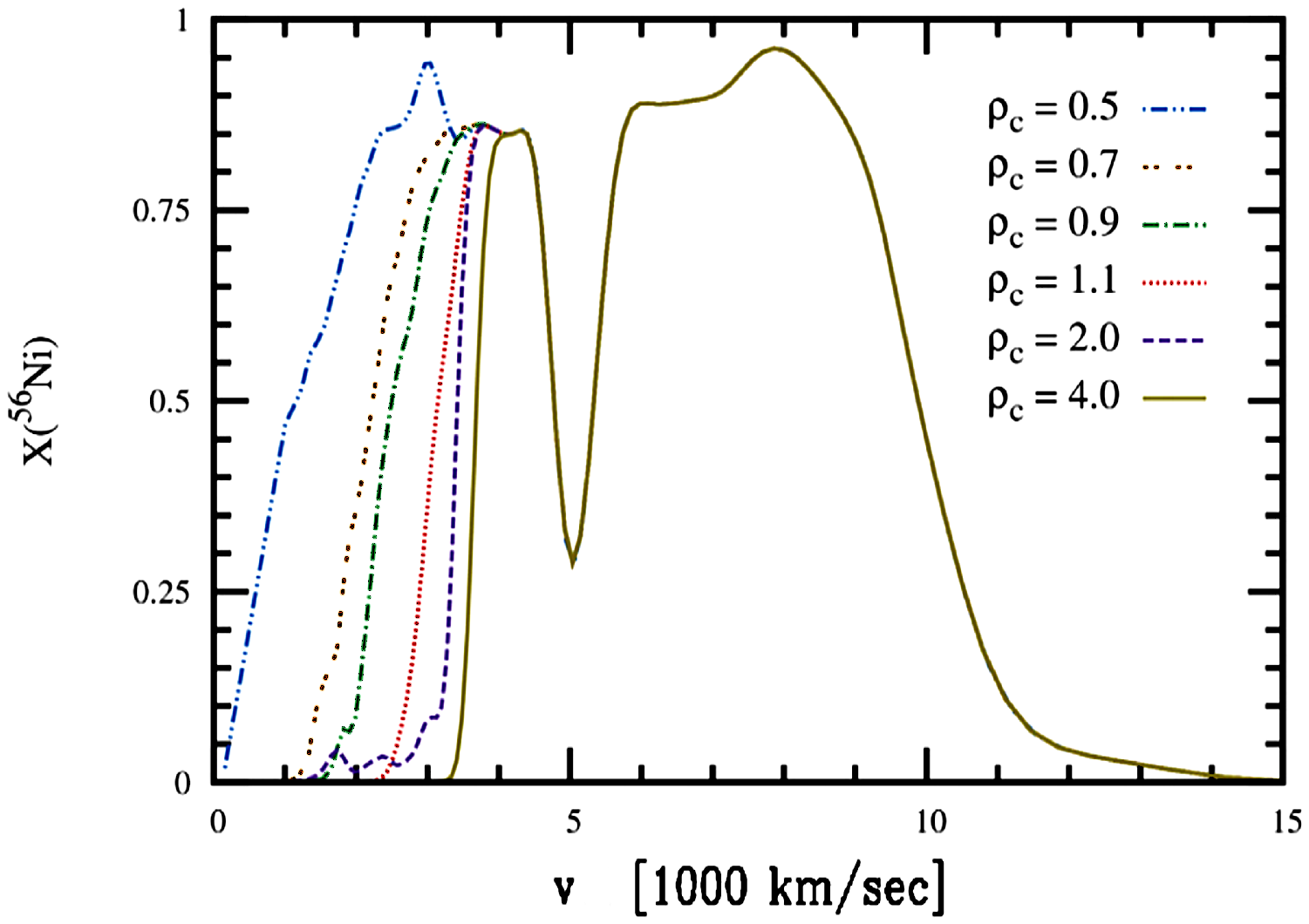}
\end{array}$
\caption{Simulation results corresponding to model  5p0z22.16.1E9 
which best-fits the observations of SN~2007on. This DD model simulates the 
disruption of a WD that explodes  with $\rho_{tr} = 1.6\times10^{7}$~g~cm$^{-3}$ 
and $\rho_{c} = 1.0\times10^9$ g~cm$^{-3}$.  \textit{(left panel:)} Both the 
density (blue dotted line) and velocity structures (red solid line) are plotted vs. 
mass.  \textit{(middle panel:)} Element abundances of the most important 
elements out of 218 isotopes  plotted in mass fraction (X$_{i}$) vs. velocity.
The composition of the central region is dominated by numerous iron-group 
electron capture (EC) isotopes \citep{1998ApJ...495..617H,2000ApJ...536..934B}. 
\textit{(right panel:)} Variation in the $^{56}$Ni distribution vs. velocity with $\rho_c$ 
varied from (0.5-4.0)$\times$10$^{9}$ g~cm$^{-3}$ for a normal-bright SN~Ia 
model \citep[see][]{2015ApJ...806..107D}.}
\end{center}
\label{modelresults}
 \end{figure*}
\clearpage
%
\section{Lack of evidence for CSM}
\label{SSC:LECSM}
\begin{figure}
\includegraphics[width=1.6\hsize]{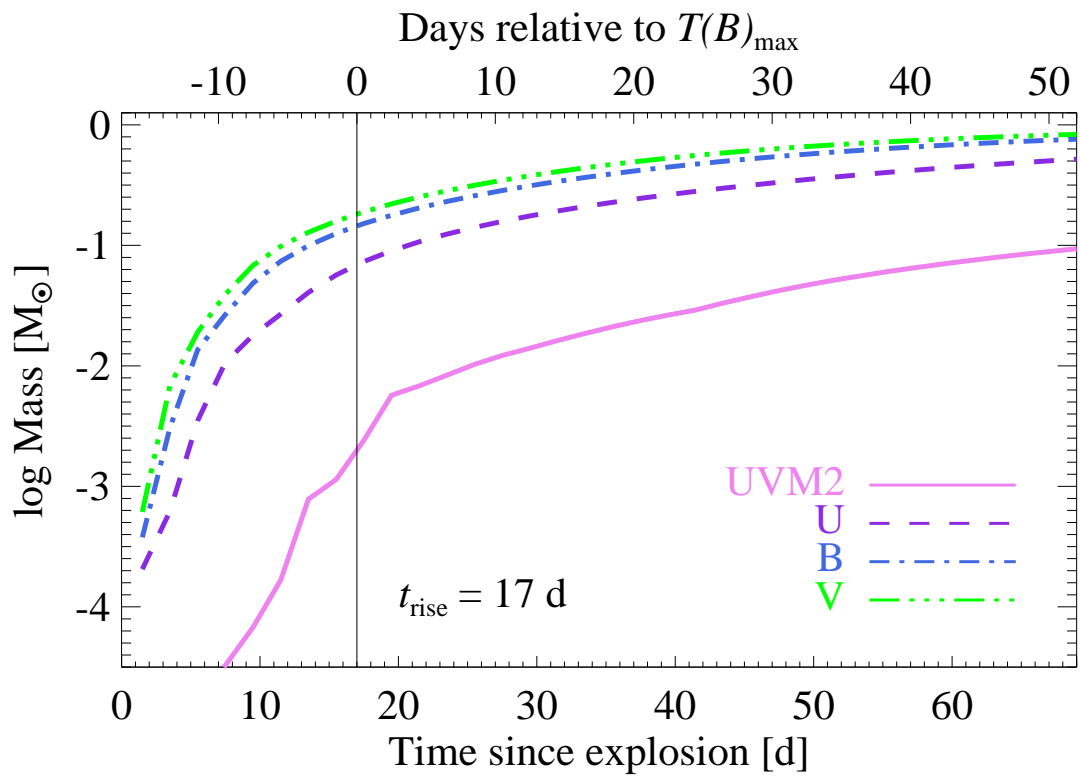}
\caption{The amount of SN~Ia ejecta mass probed by the 
emission contained within the $uvm2$, $U$, $B$, and $V$ bands 
as a function of days past explosion. These calculations are based 
on the DD model 5p0z22.18b.  
The masses are calculated by first averaging the
opacity across the filter range, and  then  the total mass 
at this opacity  is integrated from infinity
down to an optical depth of unity. 
}
\label{FIG:UVMASS}
 \end{figure}

%
As shown in Figure~\ref{FIG:UVMASS}, in the case of the DD 
model 5p0z22.18b the UV spectrum is formed in the outermost 
layers of the ejecta. By $+$10~d past explosion the model indicates 
the {\em Swift} $uvm2$ band samples a mass range $\sim 10^{-5}$--$10^{-2}$ 
M$_\odot$, while the $U$, $B$, and $V$ bands sample the 
mass range $\sim 10^{-2}$--0.7 M$_\odot$. It follows that the early 
colours of SN~2007on and SN~2011iv (see Fig.~\ref{FIG:BVCOMP} and 
Fig.~\ref{FIG:NUVC}) are driven by material from the outer ejecta. 
Since the colours of the two objects are very similar at peak, this 
suggests that the structure and condition of their outer layers are 
also similar. 

The peak $B-V$ colours of SN~2007on (0.08 mag) and SN~2011iv 
(0.01 mag) are very similar and entirely consistent with the 
predictions between the DD models 5p0z22.16 and 5p0z22.18b, 
and they suggest no interaction between the SN ejecta and their
environment. However, UV observations offer the opportunity to 
place limits on interaction expected to be produced when the rapidly 
expanding SN ejecta shocks CSM \citep{2004ApJ...607..391G, 
2007ApJ...658..396F}.

\begin{figure*}
\begin{center}
\includegraphics[width=0.45\hsize]{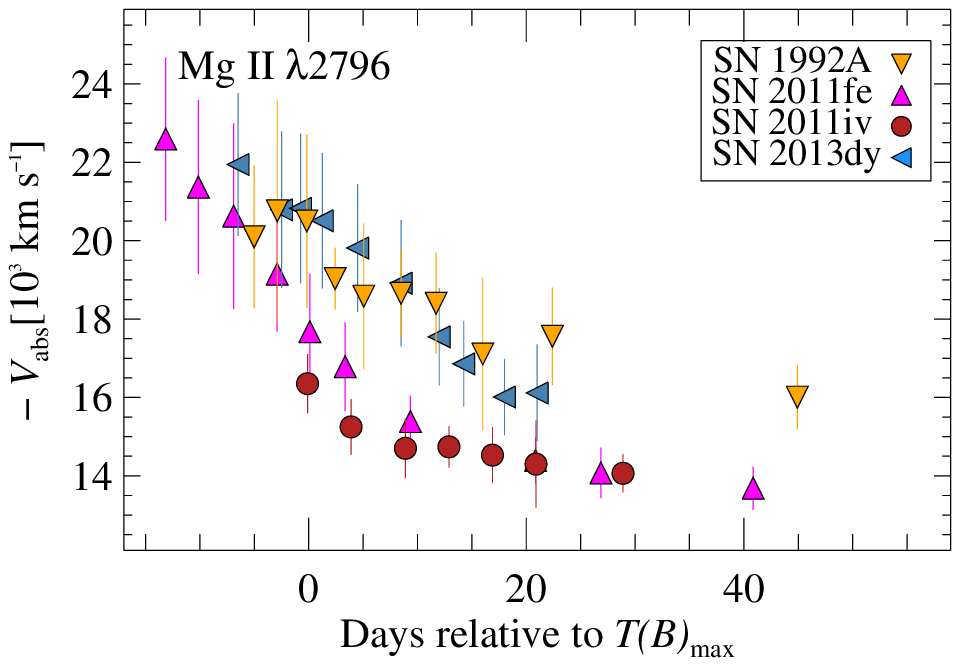}
\includegraphics[width=0.45\hsize]{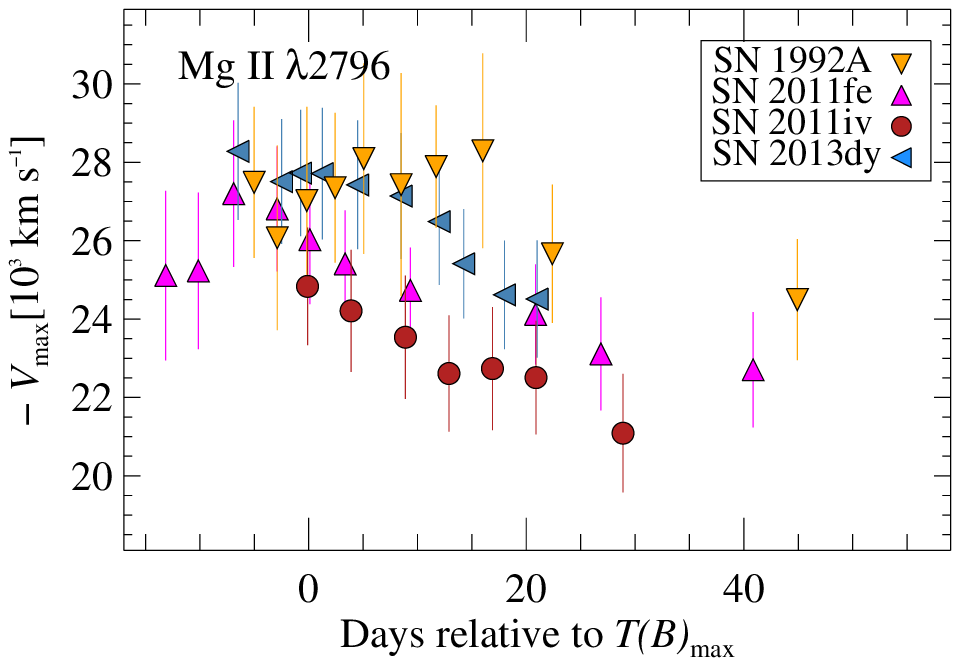}
\end{center}
\caption{
{\em (left)} Evolution of the Doppler velocity at maximum 
absorption evaluated for the pW01 feature (see Fig.~\ref{FIG:UVPWS}) 
attributed to the $\ion{Mg}{ii}$ $\lambda\lambda$2796, 2803 doublet for 
SN~1992A, SN~2011fe, SN~2011iv, and SN~2013dy. {\em (right)} Blue-wing 
velocity measurements of the $\ion{Mg}{ii}$ feature for the same objects.
}
 \label{FIG:UVVELWAVE}
 \end{figure*}

To ascertain the velocity of the material associated with the 
outermost layers of the ejecta, we examine the UV 
pW01 feature (see Fig.~\ref{FIG:UVVELWAVE}), which is 
located at the expected wavelength interval of a prevalent 
magnesium feature associated with material  in the outer ejecta. As 
magnesium is a product of carbon burning, its presence is expected 
to be more dominant in subluminous SNe~Ia since they experience 
incomplete burning compared to normal SNe~Ia. Assuming 
the pW01 feature observed in SN~2011iv is dominated by 
absorption from the \ion{Mg}{ii} $\lambda\lambda$2796, 2803 
doublet resonance transition, we plot in the left panel of 
Figure~\ref{FIG:UVVELWAVE} its corresponding Doppler velocity at 
maximum absorption ($-V_{\rm abs}$), while the right panel shows the 
evolution of the blue-wing velocity. Also included for comparison  
are identical measurements made for those SNe~Ia with published 
UV spectra: SN~1992A \citep{1993ApJ...415..589K}, 
SN~2011fe \citep{2014MNRAS.439.1959M}, and SN~2013dy 
\citep{2015MNRAS.452.4307P}.

At maximum brightness, the $-V_{\rm abs}$ of \ion{Mg}{ii} in SN~2011iv is in 
excess of $\sim 16,000$ km~s$^{-1}$ and decreases down to 
$\sim 14,000$ km~s$^{-1}$ within a month, while the highest 
velocity reached as inferred from the blue wing is $\sim 25,000$ 
km~s$^{-1}$. Interestingly,  SN~2011fe exhibits $-V_{\rm abs}$ and 
blue-wing velocity values that are consistent with SN~2011iv 
extending from maximum light to a month later, while those of 
SN~1992A and SN~2013dy are significantly higher, exhibiting 
$-V_{\rm abs}$ and blue-wing velocity values at maximum in excess of 
20,000 km~s$^{-1}$ and $\sim~27,000$ km~s$^{-1}$, respectively. 
This could be due to SN~2011fe having experienced enhanced 
mixing of carbon-burning products into the photospheric region 
\citep{2011Natur.480..344N}, thus leading  to a reduction of 
carbon-burning products at high velocity and enhanced carbon-burning 
products at lower velocities. 

In the 5p0z22.18b model considered here, magnesium forms in a 
shell of 0.2~M$_\odot$ with a velocity extent of 
$\sim 13,250$--19,350~km~s$^{-1}$. This is consistent with the  
velocity measurements of SN~2011iv. For comparison, in the 
normal SN~Ia model 5p0z22.25, the magnesium forms in a 
shell of 0.1~M$_\odot$, extending from between $\sim 15,500$ 
to $> 25,000$ km~s$^{-1}$.
In the case of the 5p0z22.18b model, the outer unburned C+O 
mass amounts to 0.04 M$_\odot$. This implies no significant 
pulsation or interaction occurred  with material in excess of 0.04 
M$_\odot$ during the explosion. For example, if another 0.04 
M$_\odot$ of material would interact with the peak blue-wing 
velocity, we would expect (owing to momentum conservation)  
to reduce the peak blue-wing velocity down to 16,000 km~s$^{-1}$. 

Finally, we note that the strength of the pW02 feature, which is 
probably produced by a multitude of \ion{Fe}{ii} lines, is determined 
by the primordial abundances of Fe-group elements at high 
velocity. We therefore expect this feature to be similar among the 
SNe, as demonstrated in the bottom panel of Figure~\ref{FIG:UVPWS}.
%
\clearpage
 
 \clearpage
 %

\end{appendix}
%
\end{document}
%

%% file: UVOT.tex
\begin{table*}
\caption{UVOT UV photometry of SN~2007on and SN~2011iv.}
\label{TAB:UVOT}
\centering          
\begin{tabular}{ccccc}    
\hline\hline    
MJD &
Phase\tablefootmark{a} &
$uvw2$   &
$uvm2$   &
$uvw1$\\
%
\hline 
\noalign{\smallskip}
\multicolumn{5}{c}{{\bf SN~2007on}}\\
\noalign{\smallskip}
\hline  
54410.7 &  $-$9.1	&	16.524(080)  & 16.783(067)  &  15.170(065)\\ 
54412.1 &  $-$7.9	&	16.068(075)  & 16.196(067)  &  14.631(050)\\ 
54414.4 &  $-$5.4	&	15.598(065)  & 15.804(066)  &  14.237(045)\\ 
54416.5 &  $-$3.3	&	15.574(067)  & 15.785(066)  &  14.205(045)\\ 
54418.9 &  $-$1.0	&	15.786(075)  & 16.005(067)  &  14.404(046)\\ 
54420.5 &  $+$0.7	&	15.975(075)  & 16.094(068)  &  14.594(050)\\ 
54422.5 &  $+$2.7	&	16.220(076)  & 16.472(070)  &  14.912(055)\\ 
54423.9 &  $+$4.1	&	16.321(094)  & 16.578(114)  &  15.150(077) \\
54424.7 &  $+$4.9	&	16.526(077)  & 16.725(072)  &  15.242(063)\\ 
54426.3 &  $+$6.5 	&	16.781(077)  & 16.991(073)  &  15.532(064)\\ 
54428.7 &  $+$8.9	&	17.052(079)  & 17.176(078)  &  15.841(065)\\ 
54430.4 &  $+$10.6	&	17.331(080)  & 17.288(080)  &  16.074(066)\\ 
54432.5 &  $+$12.7	&	17.502(081)  & 17.547(082)  &  16.314(067)\\ 
54435.5 &  $+$15.7	&	17.654(081)  & 17.629(080)  &  16.481(067)\\ 
54436.8 &  $+$17.0	&	17.704(082)  & 17.703(082)  &  16.528(067)\\ 
54438.8 &  $+$19.0	&	17.794(085)  & 17.769(088)  &  16.610(070)\\ 
54442.8 &  $+$23.0	&	17.837(083)  & 17.809(083)  &  16.801(069)\\ 
54446.1 &  $+$26.3	&	18.023(088)  & 17.957(092)  &  16.973(074)\\ 
54450.1 &  $+$30.3	&	18.101(088)  & 17.969(089)  &  17.072(073)\\ 
54454.2 &  $+$34.4	&	18.165(090)  & 18.084(097)  &  17.227(076)\\ 
54458.8 &  $+$39.0	&	18.389(093)  & 18.236(098)  &  17.350(077)\\ 
54461.3 &  $+$41.5	&	18.447(089)  & 18.353(094)  &  17.512(075)\\ 
54468.7 &  $+$48.9	&	18.746(095)  & 18.850(110)  &  17.719(079) \\
\hline  
\noalign{\smallskip}
\multicolumn{5}{c}{{\bf SN~2011iv}}\\
\noalign{\smallskip}
\hline  
55900.2 & $-$5.9 &  $\cdots$     &   $\cdots$        &  13.922(046)      \\
55900.3 & $-$5.8 &  15.386(079)  &   $\cdots$        &  $\cdots$         \\
55901.4 & $-$4.7 &  15.182(064)  &   $\cdots$        &  13.820(047)       \\
55901.8 & $-$4.3 &  15.163(068)  &   $\cdots$        &  13.820(048)       \\
55901.9 & $-$4.2 &  15.199(080)  &   $\cdots$        &  $\cdots$              \\
55902.5 & $-$3.6 &  15.143(069)  &   15.605(068)     &  $\cdots$              \\
55902.6 & $-$3.5 &  $\cdots$     &   $\cdots$        &  13.798(046)       \\
55903.3 & $-$2.8 &  15.160(065)  &   $\cdots$        &  $\cdots$              \\
55903.4 & $-$2.7 &  $\cdots$     &   15.560(066)     &  13.811(048)       \\
55904.8 & $-$1.3 &  $\cdots$     &    $\cdots$       &  13.890(048)       \\
55905.7 & $-$0.4 &  15.357(082)  &   $\cdots$        &  $\cdots$              \\
55906.6 & $+$0.5 &  15.322(067)  &   15.720(067)     &  14.049(048)       \\
55907.5 & $+$1.4 &  15.485(079)  &   $\cdots$        &  $\cdots$             \\
55909.3 & $+$3.2 &  15.708(079)  &   16.059(070)     &  14.390(055)       \\
55911.5 & $+$5.4 &  16.092(083)  &   16.486(074)     &  14.849(067)       \\
55913.3 & $+$7.2 &  16.418(088)  &   16.845(082)     &  $\cdots$              \\
55913.5 & $+$7.4 &  $\cdots$     &    $\cdots$       &  15.147(072)       \\
55915.7 & $+$9.6 &  16.827(096)  &   17.277(090)     & 15.525(075)       \\
55917.9 & $+$11.8 &  17.197(106) &   17.631(101)     &  16.014(086)      \\
55919.8 & $+$13.7 &  17.437(116) &   17.911(114)     &  16.308(094)       \\
55922.7 & $+$16.6 &  17.788(133) &   18.063(121)     &  16.545(101)      \\
55925.5 & $+$19.4 &  17.934(140) &   18.368(139)     &  16.806(114)      \\
55928.7 & $+$22.6 &  18.060(151) &   18.368(141)     &  17.087(140)       \\
55931.6 & $+$25.5 &  18.356(176) &   18.480(149)     &  17.109(140)       \\
55934.2 & $+$28.1 &  18.452(184) &   18.449(145)     &  17.139(131)       \\
\hline  
\noalign{\smallskip}
\end{tabular}
\tablefoot{
$1\sigma$ uncertainties given in parentheses are in millimag.
\tablefoottext{a}{Days relative to $T(B)_{\mathrm{max}}$.} 
}
\end{table*}

%% file: OPTICALLOCALSEQ.tex
\clearpage
\begin{table}
\tiny
\caption{Optical photometry of the local sequence stars in the field of NGC~1404 in the {\em standard} system.}
\label{tab:opticallocalseq}      
\centering
\begin{tabular}{ccccccccc} 
\hline\hline             
Star & 
$\alpha$(2000) & 
$\delta$(2000) & 
$u'$ & 
$g'$ &
$r'$ & 
$i'$ & 
$B$ &
$V$ \\
\hline
\multicolumn{9}{c}{\bf SN~2007on}\\ 
 1 & 03:38:37.30 & $-$35:33:58.6 & 16.400(021) & 14.266(005) & 13.477(005) & 13.195(006) & 14.866(005) & 13.907(004) \\ 
 2 & 03:38:41.79 & $-$35:35:39.2 & 15.778(015) & 14.261(005) & 13.681(005) & 13.510(006) & 14.718(005) & 13.996(004) \\ 
 3 & 03:38:49.88 & $-$35:32:14.1 & 15.533(014) & 14.288(005) & 13.869(005) & 13.764(006) & 14.664(005) & 14.084(004) \\ 
 4 & 03:39:07.63 & $-$35:32:46.5 & 15.619(015) & 14.552(005) & 14.179(005) & 14.090(006) & 14.903(005) & 14.373(004) \\ 
 5 & 03:39:08.62 & $-$35:32:11.7 & 16.179(019) & 14.693(005) & 14.202(005) & 14.082(006) & 15.118(006) & 14.463(004) \\ 
 6 & 03:38:57.88 & $-$35:34:08.2 & 17.093(035) & 15.945(006) & 15.514(006) & 15.398(007) & 16.319(009) & 15.733(006) \\ 
 7 & 03:39:08.80 & $-$35:33:18.9 & 17.212(041) & 16.198(007) & 15.834(006) & 15.733(008) & 16.535(010) & 16.028(007) \\ 
 8 & 03:39:06.75 & $-$35:34:48.0 &  $\cdots$   & 17.510(014) & 16.419(008) & 15.994(009) & 18.103(033) & 17.046(013) \\ 
 9 & 03:38:37.33 & $-$35:35:12.4 &  $\cdots$   & 18.255(025) & 16.911(011) & 15.538(007) & 18.865(084) & 17.658(022) \\ 
10 & 03:39:08.98 & $-$35:33:00.0 &  $\cdots$   & 18.084(021) & 17.305(014) & 16.978(016) & 18.581(057) & 17.744(023) \\ 
11 & 03:38:37.45 & $-$35:35:52.3 &  $\cdots$   & 18.011(019) & 17.369(015) & 17.186(020) & 18.336(044) & 17.675(022) \\ 
13 & 03:38:48.09 & $-$35:32:49.5 &  $\cdots$   & 19.510(084) & 18.684(048) & 18.291(057) &  $\cdots$   &  $\cdots$   \\ 
14 & 03:38:41.83 & $-$35:36:49.1 &  $\cdots$   &  $\cdots$      & 18.699(051) & 17.751(032) &  $\cdots$   &  $\cdots$   \\ 
15 & 03:38:51.21 & $-$35:31:20.5 &  $\cdots$   & 19.436(104) & 18.267(033) & 17.172(020) & 18.841(199) & 19.215(199) \\ 
16 & 03:39:04.84 & $-$35:36:40.0 &  $\cdots$   & 19.606(115) & 18.623(045) & 18.226(055) &  $\cdots$   &  $\cdots$   \\ 
17 & 03:39:06.74 & $-$35:32:10.8 &  $\cdots$   & 19.590(165) & 18.525(043) & 17.358(023) &  $\cdots$   &  $\cdots$   \\ 
18 & 03:38:40.93 & $-$35:32:06.9 &  $\cdots$   &  $\cdots$     & 19.081(082) & 17.683(031) &  $\cdots$   &  $\cdots$   \\ 
19 & 03:38:40.66 & $-$35:36:54.0 &  $\cdots$   & 19.693(179) & 19.257(089) & 18.651(151) &  $\cdots$   &  $\cdots$   \\ 
21 & 03:38:33.98 & $-$35:35:38.5 &  $\cdots$   & 19.580(110) & 19.123(101) & 18.591(193) &  $\cdots$   &  $\cdots$   \\ 
\hline
\multicolumn{9}{c}{\bf SN~2011iv}\\ 
 1 & 03:38:37.30 & $-$35:33:58.6 & 16.425(062) & 14.258(033) & 13.457(008) & 13.142(006) & 14.864(016) & 13.882(007) \\ 
 2 & 03:38:41.79 & $-$35:35:39.2 & 15.762(022) & 14.256(032) & 13.669(008) & 13.462(007) & 14.710(015) & 13.984(011) \\ 
 3 & 03:38:49.88 & $-$35:32:14.1 & 15.491(027) & 14.281(028) & 13.855(010) & 13.707(008) & 14.655(020) & 14.068(013) \\ 
 4 & 03:39:07.63 & $-$35:32:46.5 & 15.621(025) & 14.550(031) & 14.168(011) & 14.032(012) & 14.891(017) & 14.365(013) \\ 
 5 & 03:39:08.62 & $-$35:32:11.7 & 16.165(039) & 14.687(030) & 14.195(010) & 14.036(013) & 15.116(021) & 14.453(015) \\ 
 6 & 03:38:57.88 & $-$35:34:08.2 & 17.137(048) & 15.937(032) & 15.499(018) & 15.355(024) & 16.331(028) & 15.720(023) \\ 
 7 & 03:39:08.79 & $-$35:33:18.9 & 17.289(039) & 16.184(052) & 15.817(024) & 15.683(030) & 16.520(029) & 16.007(030) \\ 
 8 & 03:39:06.75 & $-$35:34:48.0 &  $\cdots$   & 17.524(054) & 16.399(043) & 15.960(030) & 18.215(035) & 17.058(042) \\ 
 9 & 03:38:37.33 & $-$35:35:12.4 &  $\cdots$   & 18.244(045) & 16.907(026) & 15.498(029) & 19.229(181) & 17.676(042) \\ 
10 & 03:39:08.98 & $-$35:33:00.0 &  $\cdots$   & 18.070(039) & 17.272(070) & 16.989(058) & 18.576(089) & 17.714(041) \\ 
11 & 03:38:37.45 & $-$35:35:52.3 &  $\cdots$   & 17.972(047) & 17.355(073) & 17.138(061) & 18.457(038) & 17.691(062) \\ 
\hline
\end{tabular}
\end{table}

%% file: NIRLOCALSEQ.tex
\clearpage
\begin{table}
\tiny
\caption{Near-IR photometry of the local sequence stars in the field of NGC~1404 in the {\em natural} system.}
\label{tab:nirlocalseq}      
\centering
\begin{tabular}{cccccc} 
\hline\hline             
Star & 
$\alpha$(2000) & 
$\delta$(2000) & 
$Y$ & 
$J$ &
$H$ \\
\hline
\noalign{\smallskip}
\multicolumn{6}{c}{\bf SN~2007on}\\ 
\noalign{\smallskip}
\hline
101 & 03:38:41.79 & $-$35:35:39.0 & 12.779(015) & 12.485(028) & 12.041(013)  \\ 
102 & 03:39:14.28 & $-$35:31:43.0 & 12.794(011) & 12.565(020) & 12.291(011)  \\ 
103 & 03:39:19.77 & $-$35:31:55.7 & 14.276(010) & 13.986(023) & 13.505(019)  \\ 
104 & 03:39:06.76 & $-$35:34:47.9 & 15.058(053) & 14.646(033) & 13.980(032)  \\ 
105 & 03:38:51.10 & $-$35:31:21.1 & 15.865(042) & 15.399(033) & 14.847(023)  \\ 
106 & 03:39:06.73 & $-$35:32:11.0 & 15.987(034) & 15.512(046) & 14.934(083)  \\ 
107 & 03:39:17.84 & $-$35:36:52.0 & 16.112(023) & 15.626(044) & 14.905(063)  \\ 
108 & 03:39:19.31 & $-$35:36:31.3 & 16.092(030) & 15.828(046) & 15.438(108)  \\ 
109 & 03:38:49.81 & $-$35:37:41.8 & 16.767(056) & 16.320(118) & 15.697(093)  \\ 
110 & 03:39:17.19 & $-$35:31:39.2 & 17.068(089) & 16.689(161) & 15.936(125)  \\ 
111 & 03:39:04.86 & $-$35:36:39.6 & 17.296(167) & 16.858(078) & 16.192(129)  \\ 
112 & 03:38:48.09 & $-$35:32:49.6 &  $\cdots$ & 17.169(107) &  $\cdots$  \\ 
113 & 03:39:15.06 & $-$35:38:18.0 & 18.099(191) & 17.516(161) &  $\cdots$  \\ 
114 & 03:39:13.60 & $-$35:37:55.5 & 17.853(106) &  $\cdots$ & 16.914(089)  \\ 
115 & 03:39:07.93 & $-$35:37:46.8 & 17.759(167) & 17.318(149) & 16.740(081)  \\ 
116 & 03:39:17.60 & $-$35:36:08.3 & 18.130(029) &  $\cdots$ & 16.949(133)  \\ 
117 & 03:39:07.54 & $-$35:36:58.9 & 18.327(058) & 17.432(166) &  $\cdots$  \\ 
118 & 03:38:57.88 & $-$35:34:08.1 & 14.823(025) & 14.580(040) & 14.238(046)  \\ 
119 & 03:38:49.86 & $-$35:32:14.6 & 13.113(013) & 12.863(021) & 12.534(013)  \\ 
120 & 03:38:50.42 & $-$35:31:46.3 &  $\cdots$ & 17.036(146) &  $\cdots$  \\ 
121 & 03:38:52.98 & $-$35:38:08.2 & 16.114(039) & 15.810(072) & 15.294(093)  \\ 
122 & 03:39:16.09 & $-$35:37:51.4 & 11.487(013) & 11.247(027) & 10.947(025)  \\ 
123 & 03:39:13.98 & $-$35:35:58.6 & 16.604(042) & 16.130(058) & 15.561(066)  \\ 
124 & 03:39:08.78 & $-$35:33:19.2 & 15.124(025) & 14.867(018) & 14.520(034)  \\ 
125 & 03:39:08.99 & $-$35:32:59.9 & 16.233(045) & 15.863(061) & 15.322(077)  \\ 
126 & 03:39:07.63 & $-$35:32:46.6 & 13.466(018) & 13.223(026) & 12.902(019)  \\ 
127 & 03:39:08.60 & $-$35:32:11.8 & 13.425(017) & 13.161(024) & 12.821(015)  \\ 
\hline
\noalign{\smallskip}
\multicolumn{6}{c}{\bf SN~2011iv}\\ 
\noalign{\smallskip}
\hline
101 & 03:38:55.09 & $-$35:36:06.4 & 11.349(088) & 11.192(127) &  $\cdots$  \\ 
102 & 03:38:50.15 & $-$35:35:32.1 & 15.076(128) & 14.696(070) & 14.112(086)  \\ 
104 & 03:38:48.55 & $-$35:35:31.2 &  $\cdots$ & 16.984(184) &  $\cdots$  \\ 
\hline
\end{tabular}
\end{table}

%% file: FINALOPTICALPHOT.tex
\clearpage
\begin{deluxetable}{ccccccc}
\tablewidth{0pt}
\tabletypesize{\scriptsize}
\tablecaption{Optical photometry of SN~2007on and SN~2011iv in the  {\em natural} system.\label{tab:opticalphotometry}}
\tablehead{
\colhead{JD$-2,450,000$} &
\colhead{$u$ (mag)}&
\colhead{$g$ (mag)}&
\colhead{$r$ (mag)}&
\colhead{$i$ (mag)}&
\colhead{$B$ (mag)} &
\colhead{$V$ (mag)} }
\startdata
\noalign{\smallskip}
\multicolumn{7}{c}{\bf SN~2007on}\\ 
\noalign{\smallskip}
\hline
4411.79 & 14.466(009) &  $\cdots$ & 14.176(008) & 14.330(010) & 14.174(007) & 14.173(006) \\ 
4413.82 & 13.971(017) & 13.584(016) & 13.633(016) & 13.805(017) & 13.655(016) & 13.650(016) \\ 
4417.72 & 13.652(009) & 13.085(006) &  $\cdots$ & 13.368(013) &  $\cdots$ & 13.128(006) \\ 
4418.75 &  $\cdots$ & 13.009(007) &  $\cdots$ &  $\cdots$ & 13.136(006) & 13.058(008) \\ 
4421.72 & 13.935(009) &  $\cdots$ &  $\cdots$ &  $\cdots$ & 13.095(004) &  $\cdots$ \\ 
4422.73 & 14.043(014) & 13.002(006) & 12.950(006) &  $\cdots$ & 13.161(006) & 12.957(006) \\ 
4423.73 & 14.217(011) &  $\cdots$ &  $\cdots$ & 13.564(013) &  $\cdots$ & 13.020(008) \\ 
4425.73 & 14.492(010) &  $\cdots$ & 13.100(004) &  $\cdots$ &  $\cdots$ &  $\cdots$ \\ 
4426.66 &  $\cdots$ & 13.313(005) & 13.149(006) & 13.706(013) & 13.588(006) & 13.118(006) \\ 
4427.68 &  $\cdots$ &  $\cdots$ & 13.201(008) &  $\cdots$ & 13.752(009) &  $\cdots$ \\ 
4428.72 & 15.102(024) &  $\cdots$ &  $\cdots$ &  $\cdots$ & 13.948(006) & 13.164(009) \\ 
4429.69 &  $\cdots$ & 13.734(005) & 13.365(005) &  $\cdots$ & 14.091(006) &  $\cdots$ \\ 
4430.67 & 15.498(023) & 13.833(006) &  $\cdots$ & 13.737(013) & 14.278(006) &  $\cdots$ \\ 
4435.66 & 16.207(027) & 14.541(007) & 13.640(007) & 13.740(013) &  $\cdots$ & 13.956(008) \\ 
4440.69 & 16.625(023) &  $\cdots$ & 14.094(007) &  $\cdots$ &  $\cdots$ &  $\cdots$ \\ 
4443.73 &  $\cdots$ & 15.269(005) & 14.411(005) &  $\cdots$ & 15.675(007) & 14.690(005) \\ 
4445.73 & 16.837(017) &  $\cdots$ & 14.528(006) &  $\cdots$ & 15.765(007) & 14.813(006) \\ 
4447.66 & 16.915(027) & 15.443(008) &  $\cdots$ & 14.673(012) &  $\cdots$ &  $\cdots$ \\ 
4449.70 & 17.052(030) & 15.536(008) & 14.768(009) &  $\cdots$ & 15.931(009) & 14.986(009) \\ 
4453.64 & 17.119(033) &  $\cdots$ & 14.963(007) & 15.005(011) & 16.067(009) & 15.154(008) \\ 
4455.62 & 17.209(036) &  $\cdots$ & 15.052(008) & 15.109(011) & 16.107(009) & 15.230(007) \\ 
4456.67 & 17.225(041) &  $\cdots$ & 15.097(006) & 15.147(011) &  $\cdots$ &  $\cdots$ \\ 
4457.63 & 17.236(028) & 15.766(008) & 15.132(007) & 15.203(011) & 16.180(010) &  $\cdots$ \\ 
4458.69 & 17.292(032) &  $\cdots$ & 15.202(008) & 15.243(011) & 16.196(010) & 15.335(009) \\ 
4462.73 & 17.409(025) &  $\cdots$ &  $\cdots$ & 15.411(011) & 16.311(009) & 15.465(007) \\ 
4467.63 & 17.577(038) &  $\cdots$ & 15.552(007) & 15.626(011) & 16.424(009) &  $\cdots$ \\ 
4468.64 & 17.653(039) &  $\cdots$ & 15.587(008) & 15.665(013) & 16.450(010) & 15.651(011) \\ 
4470.59 & 17.628(039) & 16.090(009) & 15.683(012) & 15.758(014) & 16.469(011) &  $\cdots$ \\ 
4472.61 & 17.711(040) &  $\cdots$ &  $\cdots$ & 15.822(016) & 16.537(012) &  $\cdots$ \\ 
4475.61 & 17.791(049) & 16.199(007) &  $\cdots$ & 15.948(012) & 16.569(011) &  $\cdots$ \\ 
4476.65 & 17.790(026) &  $\cdots$ & 15.917(010) &  $\cdots$ & 16.608(011) &  $\cdots$ \\ 
4484.59 & 18.048(078) &  $\cdots$ &  $\cdots$ & 16.272(014) & 16.720(013) & 16.103(010) \\ 
4488.58 & 18.219(086) &  $\cdots$ & 16.399(010) &  $\cdots$ & 16.844(014) & 16.231(010) \\ 
4490.58 & 18.232(030) & 16.485(007) & 16.477(009) &  $\cdots$ & 16.869(010) &  $\cdots$ \\ 
4494.62 & 18.344(023) &  $\cdots$ & 16.639(009) &  $\cdots$ & 16.932(009) & 16.396(009) \\ 
4497.62 & 18.408(083) &  $\cdots$ &  $\cdots$ & 16.699(014) & 16.987(013) & 16.500(009) \\ 
4499.60 & 18.468(103) & 16.682(009) & 16.841(009) & 16.750(014) & 17.026(016) & 16.557(019) \\ 
4508.61 & 18.841(065) & 16.896(009) & 17.212(012) & 17.021(014) & 17.203(013) & 16.817(011) \\ 
\hline
\noalign{\smallskip}
\multicolumn{7}{c}{\bf SN~2011iv}\\ 
\noalign{\smallskip}
\hline
5902.68 & 12.994(016) & 12.564(014) & 12.666(009) & 12.966(010) & 12.611(013) & 12.653(009) \\ 
5903.67 & 12.946(014) & 12.491(013) & 12.609(009) & 12.952(008) & 12.536(009) & 12.586(008) \\ 
5904.66 & 12.962(021) & 12.434(017) & 12.551(009) & 12.950(012) & 12.523(014) & 12.531(009) \\ 
5905.71 & 13.001(016) & 12.419(014) & 12.534(011) & 13.015(014) & 12.495(012) & 12.501(011) \\ 
5906.73 & 13.054(018) & 12.459(014) & 12.511(010) & 13.010(017) & 12.514(013) & 12.473(009) \\ 
5907.67 & 13.146(017) & 12.424(014) & 12.480(009) & 13.002(017) & 12.525(011) & 12.441(008) \\ 
5908.73 &  $\cdots$ &  $\cdots$ &  $\cdots$ & 13.053(017) &  $\cdots$ &  $\cdots$ \\ 
5909.68 &  $\cdots$ &  $\cdots$ &  $\cdots$ & 13.040(017) &  $\cdots$ &  $\cdots$ \\ 
5910.68 &  $\cdots$ &  $\cdots$ &  $\cdots$ & 13.047(017) &  $\cdots$ &  $\cdots$ \\ 
5911.63 &  $\cdots$ &  $\cdots$ &  $\cdots$ & 13.142(017) &  $\cdots$ &  $\cdots$ \\ 
5912.67 &  $\cdots$ &  $\cdots$ &  $\cdots$ & 13.116(017) &  $\cdots$ &  $\cdots$ \\ 
5915.59 & 14.406(018) &  $\cdots$ &  $\cdots$ & 13.205(017) & 13.336(012) & 12.731(011) \\ 
5916.63 & 14.647(013) &  $\cdots$ & 12.874(010) & 13.206(008) & 13.528(009) & 12.856(008) \\ 
5917.63 & 14.890(015) & 13.343(014) & 12.907(009) & 13.188(009) & 13.720(011) & 12.951(008) \\ 
5918.61 & 15.098(013) & 13.453(012) & 12.930(006) & 13.174(008) & 13.867(009) & 13.043(007) \\ 
5922.62 & 15.742(016) &  $\cdots$ &  $\cdots$ &  $\cdots$ &  $\cdots$ &  $\cdots$ \\ 
5923.58 & 15.841(016) &  $\cdots$ &  $\cdots$ &  $\cdots$ &  $\cdots$ &  $\cdots$ \\ 
5924.60 & 15.939(015) &  $\cdots$ & 13.197(012) & 13.268(016) & 14.669(013) &  $\cdots$ \\ 
5925.57 & 16.009(013) & 14.295(013) & 13.284(010) & 13.280(012) & 14.762(011) & 13.621(009) \\ 
5926.54 & 16.104(015) & 14.384(013) & 13.349(012) & 13.317(013) & 14.867(011) & 13.702(007) \\ 
5928.58 & 16.227(015) & 14.603(013) & 13.586(012) & 13.473(009) & 15.032(009) & 13.891(007) \\ 
5929.57 &  $\cdots$ & 14.538(013) &  $\cdots$ & 13.526(013) & 14.999(010) &  $\cdots$ \\ 
5930.60 &  $\cdots$ & 14.742(013) & 13.739(010) & 13.669(008) & 15.185(009) & 14.076(007) \\ 
5931.58 & 16.418(016) & 14.764(012) & 13.798(009) & 13.702(007) & 15.251(010) & 14.142(008) \\ 
5932.58 & 16.481(016) & 14.896(017) & 13.903(013) & 13.817(012) & 15.305(012) & 14.217(012) \\ 
5933.58 & 16.508(016) & 14.924(014) & 13.968(014) & 13.870(010) & 15.361(012) & 14.250(009) \\ 
5934.58 & 16.561(016) & 14.968(016) & 14.032(015) & 13.936(014) & 15.397(012) & 14.313(010) \\ 
5935.57 & 16.598(016) & 14.993(014) & 14.088(009) & 14.020(008) & 15.432(010) & 14.381(009) \\ 
5937.56 & 16.669(014) & 15.087(013) & 14.206(012) & 14.216(016) & 15.511(011) & 14.448(007) \\ 
5938.58 & 16.691(016) &  $\cdots$ &  $\cdots$ & 14.167(017) & 15.544(015) &  $\cdots$ \\ 
5939.57 & 16.737(016) & 15.126(015) & 14.283(016) & 14.180(017) & 15.571(013) &  $\cdots$ \\ 
5940.57 & 16.762(017) & 15.155(015) & 14.338(016) & 14.287(017) & 15.587(012) & 14.549(011) \\ 
5941.56 & 16.802(016) & 15.179(016) & 14.398(017) & 14.282(017) & 15.595(010) & 14.571(015) \\ 
5942.55 & 16.805(018) & 15.197(016) & 14.420(017) & 14.291(017) & 15.628(012) &  $\cdots$ \\ 
5943.57 & 16.836(016) & 15.219(013) &  $\cdots$ & 14.451(010) & 15.646(010) & 14.636(010) \\ 
5949.58 & 16.993(016) & 15.355(016) & 14.700(020) & 14.679(014) & 15.771(013) & 14.795(010) \\ 
5953.63 & 17.088(017) & 15.392(015) & 14.800(016) & 14.805(017) & 15.838(012) & 14.917(010) \\ 
5955.62 & 17.126(018) & 15.438(020) &  $\cdots$ & 14.808(017) & 15.864(013) & 14.960(016) \\ 
5957.63 & 17.184(019) & 15.488(016) & 14.940(016) & 14.988(009) & 15.920(010) & 15.020(008) \\ 
5962.60 & 17.302(018) & 15.591(016) & 15.102(014) & 15.140(015) & 15.988(010) & 15.134(010) \\ 
5963.66 & 17.301(021) & 15.610(016) & 15.155(015) & 15.137(018) & 15.996(010) & 15.163(008) \\ 
5965.66 & 17.338(023) & 15.626(019) & 15.208(015) & 15.269(016) & 16.029(011) &  $\cdots$ \\ 
5967.65 & 17.382(023) & 15.662(012) & 15.299(016) & 15.318(015) & 16.068(010) & 15.276(008) \\ 
5969.61 & 17.502(019) & 15.676(019) & 15.396(022) & 15.368(018) & 16.088(010) & 15.320(011) \\ 
5972.64 & 17.576(020) & 15.714(020) & 15.547(030) & 15.486(016) & 16.117(011) & 15.368(011) \\ 
5981.61 & 17.811(021) & 15.850(020) & 15.797(022) & 15.775(016) & 16.255(011) & 15.596(009) \\ 
5985.57 & 17.910(020) & 15.921(020) & 15.972(029) & 15.909(018) & 16.309(011) & 15.699(011) \\ 
5991.61 & 17.998(040) & 16.075(014) & 16.140(013) & 16.078(011) &  $\cdots$ &  $\cdots$ \\ 
5994.57 & 18.113(042) & 16.146(023) & 16.268(018) & 16.198(016) & 16.487(014) & 15.954(012) \\ 
5997.55 & 18.195(027) & 16.138(019) & 16.376(022) & 16.265(019) & 16.514(012) & 16.003(012) \\ 
6001.57 & 18.400(027) & 16.222(018) & 16.490(021) & 16.359(021) & 16.578(012) & 16.091(012) \\ 
6004.58 & 18.491(033) & 16.276(017) &  $\cdots$ & 16.465(018) & 16.619(012) & 16.151(009) \\ 
6006.57 & 18.525(033) & 16.305(021) &  $\cdots$ & 16.542(017) & 16.666(013) & 16.222(013) \\ 
6008.50 & 18.556(028) & 16.368(025) &  $\cdots$ & 16.574(020) & 16.696(015) & 16.235(019) \\ 
6013.53 &  $\cdots$ & 16.452(018) &  $\cdots$ & 16.625(023) & 16.783(013) & 16.388(012) \\ 
6016.53 &  $\cdots$ & 16.504(015) &  $\cdots$ & 16.710(025) & 16.860(013) & 16.438(014) \\ 
6019.51 & 18.845(049) & 16.567(017) &  $\cdots$ & 16.750(017) & 16.937(013) & 16.463(010) \\ 
6024.53 &  $\cdots$ &  $\cdots$ &  $\cdots$ & 16.909(020) &  $\cdots$ &  $\cdots$ \\ 
6026.50 &  $\cdots$ & 16.690(015) &  $\cdots$ & 16.928(015) &  $\cdots$ &  $\cdots$ \\ 
6031.51 &  $\cdots$ &  $\cdots$ &  $\cdots$ &  $\cdots$ & 17.048(016) &  $\cdots$ \\ 
6032.50 &  $\cdots$ & 16.889(016) &  $\cdots$ &  $\cdots$ &  $\cdots$ &  $\cdots$ \\ 
\enddata
\tablecomments{Values in parentheses are 1$\sigma$ measurement uncertainties in millimag.}
\end{deluxetable}

%% file: FINALNIRPHOT.tex
\begin{deluxetable}{ccccc}
\tablewidth{0pt}
\tablecaption{NIR photometry of SN~2007on and SN~2011iv in the {\it natural} system.
\label{tab:nirpot}}
\tablehead{
\colhead{JD$-2,450,000$} &
\colhead{$Y$ (mag)}&
\colhead{$J$ (mag)}&
\colhead{$H$ (mag)} &
\colhead{Telescope\tablenotemark{a}}}
\startdata
\noalign{\smallskip}
\multicolumn{5}{c}{\bf SN~2007on}\\ 
\noalign{\smallskip}
\hline
4410.73 & 14.419(018) & 14.353(016) & 14.273(017) & SWO \\ 
4410.74 & 14.450(017) & 14.312(021) & 14.259(016) & SWO\\ 
4412.73 & 13.835(017) & 13.739(010) & 13.742(014) & SWO\\ 
4412.74 & 13.850(013) & 13.723(014) & 13.737(009) & SWO\\ 
4415.75 & 13.324(014) & 13.217(011) &  $\cdots$ & SWO\\ 
4415.76 & 13.331(013) & 13.216(014) &  $\cdots$ & SWO\\ 
4421.75 & 13.435(026) & 13.304(041) & 13.365(046) & DUP \\ 
4421.76 & 13.438(027) & 13.312(041) & 13.345(046) & DUP \\ 
4422.84 & 13.481(025) &  $\cdots$  &  $\cdots$  & DUP \\ 
4422.85 & 13.511(026) & $\cdots$  &  $\cdots$  & DUP\\ 
4423.76 & 13.546(026) & 13.533(040) &  $\cdots$ & DUP \\ 
4423.76 & 13.559(027) &  $\cdots$  &  $\cdots$  & DUP\\ 
4424.73 & 13.572(012) & 13.627(013) & 13.473(011) & SWO\\ 
4424.77 & 13.589(014) & 13.649(013) & 13.456(013) & SWO\\ 
4425.77 & 13.672(025) & 13.850(040) &  $\cdots$ & DUP \\ 
4425.78 & 13.672(025) & 13.890(042) &  $\cdots$  & DUP\\ 
4426.79 & 13.696(026) & 13.989(041) & 13.542(046) & DUP\\ 
4426.80 & 13.720(026) & 14.016(041) & 13.528(046) & DUP\\ 
4427.73 & 13.704(026) & 14.125(041) & 13.535(046) & DUP\\ 
4427.73 & 13.676(026) & 14.125(041) & 13.527(046) & DUP\\ 
4428.73 & 13.684(025) & 14.220(040) & 13.517(046) & DUP\\ 
4428.73 & 13.710(026) & 14.207(040) & 13.539(046) & DUP\\ 
4431.73 & 13.534(014) & 14.229(014) & 13.434(012) & SWO\\ 
4431.74 & 13.524(013) & 14.243(014) & 13.451(009) & SWO\\ 
4432.68 & 13.483(012) & 14.239(016) & 13.391(014) & SWO\\ 
4432.69 & 13.475(016) & 14.232(015) & 13.411(012) & SWO\\ 
4437.72 & 13.185(012) & 14.040(011) & 13.425(012) & SWO\\ 
4437.73 & 13.252(014) & 14.035(014) & 13.431(008) & SWO\\ 
4439.70 & 13.181(017) & 14.062(012) & 13.542(015) & SWO\\ 
4439.70 & 13.210(016) & 14.065(015) & 13.559(008) & SWO\\ 
4442.72 & 13.400(017) & 14.429(015) & 13.840(013) & SWO\\ 
4442.73 & 13.417(018) & 14.442(015) & 13.851(010) & SWO\\ 
4444.79 & 13.562(007) & 14.712(012) & 14.026(013) & SWO\\ 
4444.80 & 13.563(008) & 14.725(011) & 14.021(013) & SWO\\ 
4446.69 & 13.719(013) & 14.928(015) & 14.179(015) & SWO\\ 
4446.70 & 13.762(016) & 14.922(016) & 14.186(013) & SWO\\ 
4448.67 & 13.839(011) & 15.141(013) & 14.300(014) & SWO\\ 
4448.67 & 13.841(008) & 15.149(014) & 14.331(010) & SWO\\ 
4454.66 & 14.237(013) & 15.733(015) & 14.678(016) & SWO\\ 
4454.67 & 14.255(013) & 15.690(021) & 14.682(014) & SWO\\ 
4456.66 & 14.388(026) &  $\cdots$ & 14.786(046) & DUP \\ 
4457.73 & 14.459(026) &  $\cdots$ & 14.845(046) & DUP\\ 
4457.73 & 14.482(026) &  $\cdots$ & 14.883(047) & DUP\\ 
4461.69 & 14.689(010) & 16.315(020) & 15.014(013) & SWO\\ 
4461.70 & 14.700(012) & 16.299(021) & 15.053(011) & SWO\\ 
4463.66 & 14.826(013) & 16.449(017) & 15.134(014) & SWO\\ 
4463.67 & 14.844(014) & 16.491(019) & 15.178(013) & SWO\\ 
4468.67 & 15.142(027) &  $\cdots$ & 15.431(046) & DUP\\ 
4468.67 & 15.138(026) &  $\cdots$ & 15.430(046) & DUP\\ 
4471.66 & 15.270(012) & 16.952(025) & 15.540(018) & SWO\\ 
4471.67 & 15.303(015) & 17.063(033) & 15.542(022) & SWO\\ 
4477.62 & 15.574(010) & 17.497(051) & 15.872(021) & SWO\\ 
4477.63 & 15.581(010) & 17.545(051) & 15.874(024) & SWO\\ 
4485.64 & 15.927(013) & 17.972(081) & 16.338(031) & SWO\\ 
4485.65 & 15.919(013) & 18.049(078) & 16.323(035) & SWO\\ 
4491.60 & 16.140(011) & 18.146(093) & 16.656(035) & SWO\\ 
4491.61 & 16.166(013) & 18.313(103) & 16.623(044) & SWO\\ 
\hline
\noalign{\smallskip}
\multicolumn{5}{c}{\bf SN~2011iv}\\ 
\noalign{\smallskip}
\hline
5902.70 & 12.943(073) & 12.967(042) & 13.043(081)  & DUP \\ 
5903.59 & 12.970(073) & 12.866(041) & 12.978(078) & DUP \\ 
5906.66 & 13.157(073) & 12.925(045) & 13.004(079) & DUP \\ 
5907.76 & 13.213(073) & 13.073(042) & 13.076(078) & DUP \\ 
5908.74 & 13.260(073) & 13.182(042) & 13.118(078) & DUP \\ 
5908.75 & 13.248(073) &  $\cdots$ &  $\cdots$ & DUP \\ 
5909.65 & 13.285(073) & 13.178(058) & 13.148(080) & DUP \\ 
5909.63 & 13.318(073) &  $\cdots$ &  $\cdots$ & DUP \\ 
5910.65 & 13.292(073) & 13.318(042) & 13.149(078) & DUP \\ 
5910.67 &  $\cdots$ &  $\cdots$ & 13.226(078) & DUP \\ 
5911.65 &  $\cdots$ & 13.505(042) & 13.305(080) & DUP \\ 
5912.71 & 13.339(073) & 13.760(043) & 13.297(081) & DUP \\ 
5913.63 & 13.351(074) & 13.989(045) & 13.451(083) & DUP \\ 
5914.66 & 13.319(074) & 14.106(045) & 13.413(082) & DUP \\ 
5915.63 & 13.262(073) & 14.120(042) & 13.439(080) & DUP \\ 
5927.54 & 12.610(073) & 13.585(042) & 12.847(078) & DUP \\ 
5927.55 & 12.592(073) &  $\cdots$ &  $\cdots$ & DUP \\ 
5928.56 &  $\cdots$ & 13.546(042) & 12.906(078) & DUP \\ 
5929.55 & 12.746(073) & 13.770(043) & 13.183(078) & DUP \\ 
5930.53 & 12.698(073) &  $\cdots$ &  $\cdots$ & DUP \\ 
5931.55 & 12.715(138) & 13.791(042) & 13.223(078) & DUP \\ 
5992.57 & 16.212(073) & 17.861(053) &  $\cdots$ & DUP \\ 
6001.55 & 16.540(073) &  $\cdots$ & 16.599(083) & DUP \\ 
6021.48 & 16.974(076) &  $\cdots$ &  $\cdots$ & DUP \\ 
6022.50 & 17.218(074) & 18.806(077) &  $\cdots$ & DUP \\
 \enddata
\tablecomments{Values in parentheses are 1$\sigma$ measurement uncertainties in millimag.}
\tablenotetext{a}{SWO and DUP correspond to the Swope and du Pont telescopes, respectively.  
NIR response functions are reported in Krisciunas et al., in prep. 
}
\end{deluxetable}

%% file: TAB_SPJ07on.tex
\begin{table*}
\caption{Journal of Spectroscopic Observations for SN~2007on.}             
\label{T:SPJ07ON}      
\centering          
\begin{tabular}{ccccc}     
\hline\hline       
Date &
JD$-$2,450,000 &
Phase\tablefootmark{a} &
Telescope &
Instrument \\
\hline 
\noalign{\smallskip}
\multicolumn{5}{c}{\bf Visual-wavelength spectroscopy} \\
\noalign{\smallskip}
\hline
2007 Nov 11.3 	& 4415.8 	& $-$4.0 		& du Pont		 & WFCCD\tablefootmark{b}	\\   
2007 Nov 14.3 	& 4418.8 	& $-$1.0   		& du Pont		 & WFCCD	\tablefootmark{b}	\\   
2007 Nov 17.2 	& 4421.7 	& $+$1.9   	& Baade		 & IMACS\tablefootmark{b}		\\   
2007 Nov 19.2 	& 4423.7 	& $+$3.9   	& NTT		 & EMMI\tablefootmark{b}		\\   
2007 Nov 25.2 	& 4429.7 	& $+$9.9   	& Clay		 & MagE\tablefootmark{b}		\\   
2007 Nov 27.2 	& 4431.7 	& $+$11.9   	& NTT		 & EMMI\tablefootmark{c}			\\   
2007 Nov 30.2 	& 4434.7 	& $+$14.9  	& ESO-3P6	 & EFOSC\tablefootmark{b}	\\   
2007 Dec 03.2	& 4437.7 	& $+$17.9  	& du Pont		 & WFCCD	\tablefootmark{b}	\\   
2007 Dec 09.2	& 4443.7 	& $+$23.9   	& du Pont		 & WFCCD	\tablefootmark{b}	\\   
2007 Dec 10.2	& 4444.7 	& $+$24.9   	& du Pont		 & WFCCD	\tablefootmark{b}	\\   
2007 Dec 18.1 	& 4452.6 	& $+$32.8   	& NTT		 & EMMI\tablefootmark{b}		\\   
2007 Dec 27.3 	& 4461.8 	& $+$42.0   	& Baade	        	 & IMACS\tablefootmark{b}		\\   
2008 Jan 03.1 	& 4468.6 	& $+$48.8  	& ESO-3P6	 & EFOSC\tablefootmark{b}	\\   
2008 Jan 05.1 	& 4470.6 	& $+$50.8   	& du Pont		 & B\&C\tablefootmark{b}		\\  
2008 Jan 10.1 	& 4475.6 	& $+$55.8  	& du Pont		 & B\&C\tablefootmark{b}		\\   
2008 Jan 27.1 	& 4492.6 	& $+$72.8   	& ESO-3P6	 & EFOSC\tablefootmark{b}	\\   
2008 Feb 28.2 	& 4493.7 	& $+$73.9  	& NTT		 & EMMI\tablefootmark{c}					\\   
2008 Feb 01.1 	& 4497.6 	& $+$77.8   	& du Pont		 & B\&C\tablefootmark{b}		\\   
2008 Feb 13.1 	& 4509.6 	& $+$89.8  	& NTT		 & EMMI\tablefootmark{b}		\\   
2008 Feb 25.1 	& 4521.6 	& $+$101.8   	& Clay		 & LDSS3\tablefootmark{b}		\\   
2008 Aug 27.3 	& 4705.8 	& $+$286.0   	& Clay		 & LDSS3\tablefootmark{b}		\\   
2008 Nov 03.0 	& 4773.5 	& $+$353.7   	& Gemini-South & GMOS-S\tablefootmark{d}	\\   
2008 Nov 30.0 	& 4800.5 	& $+$380.7   	& Gemini-South & GMOS-S\tablefootmark{d}	\\   
\hline  
\noalign{\smallskip}
\multicolumn{5}{c}{\bf NIR spectroscopy} \\
\noalign{\smallskip}
\hline
2007 Dec 19.2	& 4453.7 	& $+$33.9  	& NTT   		& SOFI\tablefootmark{c}		\\  
2008 Feb 14.1	& 4510.5 	& $+$90.7  	& NTT   		& SOFI\tablefootmark{c}		\\   
\hline                  
\end{tabular}
\tablefoot{
\tablefoottext{a}{Days relative to $T(B)_{\mathrm{max}}$, i.e., JD$-$2,454,419.8.}
\tablefoottext{b}{Published by \citet{2013ApJ...773...53F}.}
\tablefootmark{c}{This work, see Sect.~\ref{SEC:OBS}.}
\tablefoottext{d}{Previously included in analysis by \citet{2010Natur.466...82M}. Note that the $+$380d spectrum now includes a vastly improved wavelength calibration at the blue end of the spectrum.}
}
\end{table*}

%% file: TAB_SPJ11iv.tex
\begin{table*}
\caption{Journal of Spectroscopic Observations for SN 2011iv.}             
\label{T:SPJ11IV}      
\centering          
\begin{tabular}{ccccc}     
\hline\hline       
Date &
JD$-$2,450,000 &
Phase\tablefootmark{a} &
Telescope &
Instrument \\
\hline 
\noalign{\smallskip}
\multicolumn{5}{c}{\bf UV - Visual-wavelength spectroscopy}\\
\noalign{\smallskip}
\hline                
2011 Dec 05.4 	& 5901.0 	& $-$5.1    	& SWIFT   & UVOT 	\\    
2011 Dec 06.6 	& 5902.1 	& $-$4.0    	& SWIFT   & UVOT 	\\    
2011 Dec 07.6 	& 5903.1 	& $-$3.0    	& SWIFT   & UVOT 	\\    
2011 Dec 09.6 	& 5905.1 	& $-$1.0    	& SWIFT   & UVOT 	\\    
2011 Dec 11.7 	& 5907.2 	& $+$1.1    	& SWIFT   & UVOT 	\\    
2011 Dec 11.0 	& 5906.5 	& $+$0.4    	& HST     & STIS 	\\    
2011 Dec 15.0 	& 5910.5 	& $+$4.4 		& HST     & STIS 	\\    
2011 Dec 20.0 	& 5915.5 	& $+$9.4   	& HST     & STIS 	\\     
2011 Dec 24.0 	& 5919.5 	& $+$13.4  	& HST     & STIS 	\\    
2011 Dec 28.0 	& 5923.5 	& $+$17.4   	& HST     & STIS 	\\    
2012 Jan 01.0 	& 5927.5 	& $+$21.4   	& HST     & STIS 	\\    
2012 Jan 09.0 	& 5935.5 	& $+$29.4   	& HST     & STIS 	\\ 
\hline
\noalign{\smallskip}
\multicolumn{5}{c}{\bf Visual-wavelength spectroscopy} \\
\noalign{\smallskip}
\hline
2011 Dec 04.0 	& 5899.5 	& $-$6.6   		& Gemini-South	                & GMOS-S	\\   
2011 Dec 04.0 	& 5899.5 	& $-$6.6   		& du Pont			& B\&C		\\   
2011 Dec 05.0 	& 5900.6 	& $-$5.5   		& du Pont			& B\&C		\\   
2011 Dec 06.0 	& 5901.6 	& $-$4.5   		& du Pont			& B\&C		\\    
2011 Dec 18.0 	& 5913.5 	& $+$7.4  		& WHT   			& ISIS 		\\  
2011 Dec 18.1	& 5913.8	& $+$7.7		& SOAR	         		& Goodman 	\\
2011 Dec 19.0 	& 5914.5 	& $+$8.4  		& WHT  	         		& ISIS 		\\  
2011 Dec 19.2	& 5914.7 	& $+$8.6  		& NTT   			& EFOSC 		\\  
2011 Dec 20.0 	& 5915.5 	& $+$9.4  		& WHT   			& ISIS 		\\  
2011 Dec 21.0 	& 5916.5 	& $+$10.4  	        & WHT   			& ISIS 		\\  
2011 Dec 21.1 	& 5916.6 	& $+$10.5		& du Pont   		        & WFCCD 	\\  
2011 Dec 27.0   & 5922.0	& $+$15.9		& NOT			        & ALFOSC	\\
2011 Dec 31.1 	& 5926.6 	& $+$20.5		& du Pont   		        & WFCCD 	\\  
2011 Dec 31.2	& 5926.8	& $+$20.7		& SOAR			        & Goodman 	\\
2011 Jan 10.0	& 5936.0	& $+$29.9		& NOT     			& ALFOSC 	\\  	
2012 Jan 19.0 	& 5945.6 	& $+$39.5		& du Pont   		        & WFCCD 	\\  
2012 Jan 22.2	& 5948.7 	& $+$42.6		& NTT   			& EFOSC 		\\  
2012 Feb 04.1   & 5961.5	&  $+$55.4	        & Clay          		& MIKE		\\
2012 Mar 02.1   & 5988.5	&  $+$82.4	        & Clay          		& LDSS3		\\
2012 Apr 30.9 	& 6048.5 	& $+$142.4  	        & du Pont   		        & WFCCD 	\\  
2012 Aug 10.4   & 6150.0	& $+$243.9	        & du Pont       		& WFCCD       	\\   
2012 Aug 27.3	& 6166.8 	& $+$260.7	        & NTT  			        & EFOSC 		\\  
2012 Sept 11.3	& 6181.8	& $+$275.7	        & du Pont       		& WFCCD		\\
\hline  
\noalign{\smallskip}
\multicolumn{5}{c}{\bf NIR spectrscopy} \\
\noalign{\smallskip}
\hline
2011 Dec 09.0 	& 5904.5 	& $-$1.6		& Baade 	& FIRE		\\
2011 Dec 11.0 	& 5906.5 	& $+$0.4		& Baade 	& FIRE		\\
2011 Dec 11.0 	& 5906.7 	& $+$0.6 	 	& Baade   	& FIRE		\\    
2011 Dec 14.0	& 5909.5	& $+$3.4		& VLT	        & ISAAC		\\
2011 Dec 16.2 	& 5911.7 	& $+$5.6   	        & Baade   	& FIRE		\\    
2011 Dec 18.2 	& 5913.7 	& $+$7.6   	        & Baade   	& FIRE		\\    
2011 Dec 20.2	& 5915.7 	& $+$9.6  		& NTT   	& SOFI 		\\  
2011 Dec 21.2 	& 5916.7 	& $+$10.6   	        & Baade   	& FIRE		\\    
2011 Dec 29.1	& 5924.6	& $+$18.5		& VLT	        & ISAAC		\\
2012 Jan 07.1 	& 5933.6	& $+$27.5   	        & Baade   	& FIRE		\\    
2012 Jan 15.1 	& 5941.6 	& $+$35.5  	        & Baade   	& FIRE		\\    
2012 Jan 19.2 	& 5945.7 	& $+$39.6   	        & Baade   	& FIRE		\\    
2012 Feb 03.1 	& 5960.6 	& $+$54.5   	        & Baade   	& FIRE		\\    
2012 Feb 16.1	& 5973.6 	& $+$67.5  	        & NTT   	& SOFI 		\\ 
2012 Mar 03.1 	& 5989.6 	& $+$83.5   	        & Baade   	& FIRE		\\    
2012 Apr 30.0	& 6047.5        & $+$141.4	        & Baade         & FIRE		\\
\hline                  
\end{tabular}
\tablefoot{
\tablefoottext{a}{Days relative to $T(B)_{\mathrm{max}}$, i.e., JD$-$2,455,906.1.}
}
\end{table*}

%% file: TAB_LCPARAM.tex
\begin{table}
\caption{Light curve parameters of SN~2007on and SN~2011iv.}             
\label{T:LCPARAM}      
\centering                          
\begin{tabular}{l r r c }        
\hline\hline                
Parameter & 
SN 2007on &
SN 2011iv &
Unit  \\    
\hline                        
$T(B)_{\mathrm{max}}$\tablefootmark{a}	& 2,454,419.8$\pm$0.04       	& 2,455,906.1 $\pm$ 0.05     &  JD   \\
  $s_{{BV}}$\tablefootmark{a} 		& 0.57   $\pm$ 0.03    		& 0.64 $\pm$ 0.03     	&       \\
  $m_{{u}}$\tablefootmark{a,b}		& 13.62  $\pm$ 0.01          	& 12.91 $\pm$ 0.01           &  mag   \\
  $m_{{B}}$                          		& 13.03  $\pm$ 0.01          	& 12.44 $\pm$ 0.01           &  mag  \\
  $m_{{g}}$                          		& 12.92  $\pm$ 0.01          	& 12.37 $\pm$ 0.01           &  mag  \\
  $m_{{V}}$                          		& 12.92  $\pm$ 0.01          	& 12.38 $\pm$ 0.01           &  mag  \\
  $m_{{r}}$                          		& 12.98  $\pm$ 0.01          	& 12.43 $\pm$ 0.01           &  mag  \\
  $m_{{i}}$                          		& 13.35  $\pm$ 0.02          	& 12.82 $\pm$ 0.01           &  mag  \\
  $m_{{Y}}$                          		& 13.26  $\pm$ 0.01         		& 12.83 $\pm$ 0.07           &  mag  \\
  $m_{{J}}$                          		& 13.10  $\pm$ 0.02         		& 12.76 $\pm$ 0.03           &  mag  \\
  $m_{{H}}$                          		& 13.25  $\pm$ 0.02          	& 12.87 $\pm$ 0.05           &  mag  \\
\hline
   $E(B-V)_{\mathrm{host}}$\tablefootmark{c}	& $-$0.052 $\pm$ 0.07	&$-$0.07    $\pm$ 0.07	&  mag	\\
   $E(B-V)_{\mathrm{lira}}$\tablefootmark{d}	& $-$0.025 $\pm$ 0.07	& 0.10        $\pm$ 0.07 	&  mag   \\ 
   $\rm{\mu_{SNooPy}}$\tablefootmark{c}	        & 31.57 $\pm$ 0.09		& 31.17      $\pm$ 0.09	&  mag	\\
   $\Delta m_{15}(B)$\tablefootmark{e}		& 1.96  $\pm$ 0.01 		& 1.77        $\pm$ 0.01	&  mag	\\
\hline     

\end{tabular}
\tablefoot{
\tablefoottext{a}{Based on {\tt SNooPy} fits using the  ``max model'' $s_{BV}$ parameterization.}
\tablefoottext{b}{Peak magnitudes estimated from photometry $K$-corrected using the \citet{2007ApJ...663.1187H} spectral template, and corrected for Galactic extinction.} 
\tablefoottext{c}{Based on {\tt SNooPy} fits using the ``EBV\_method2'' model. The reported uncertainty was determined by adding in quadrature both statistical and systematic errors.}
\tablefoottext{d}{Based on the Lira-relation fits \citep[cf.][]{2010AJ....139..120F}.}
\tablefoottext{e}{Based on direct ``Gaussian process'' spline fits computed within the {\tt SNooPy} envirnoment.}
}
\end{table}

%% file: UVpEW.tex
\begin{table}
\caption{UV feature limits for the pseudo-equivalent width.}             
\label{T:PSWFL}      
\centering                          
\begin{tabular}{l c c c}        
\hline\hline                
Feature & 
Position /  &
Blueward &
Redward  \\    
 & 
Wavelength  &
limit range &
limit range  \\    
 & 
 [$\AA$] &
 [$\AA$] &
 [$\AA$]  \\    
\hline                        
$pW01$	& \ion{Mg}{ii} 2800	& 2580--2620	& 2700--2750 \\
$pW02$	& $\sim$2500 		& 2380--2420	& 2580--2620 \\
\hline                    
\end{tabular}
\end{table}

%% file: BOLOUVOIR_07.tex
\begin{table*}
\caption{{\tt SNooPy} UVOIR light curve of SN~2007on.}             
\label{TAB:UVOIR07on}      
\centering          
\begin{tabular}{ccc}     
\hline\hline       
Phase (days)\tablefootmark{a} &
$L_{bol}$ (erg s$^{-1}$)\tablefootmark{b}&
$Uncertainty$ (erg s$^{-1}$)\tablefootmark{c}\\
\hline 
\noalign{\smallskip}
$-$7.87  &    3.472$\times10^{47}$ &    3.145$\times10^{45} $\\
$-$6.80  &    2.308$\times10^{42}$ &    3.834$\times10^{39}$ \\
$-$5.88  &    2.988$\times10^{42}$ &    2.516$\times10^{40} $\\
$-$4.78  &    3.735$\times10^{42}$ &    1.676$\times10^{40} $\\
$-$3.21  &    4.756$\times10^{42}$ &    8.906$\times10^{40} $\\
$-$1.12  &    5.600$\times10^{42}$ &    3.607$\times10^{40} $\\
$+$0.12  &    5.726$\times10^{42}$ &    7.056$\times10^{40} $\\
$+$1.36  &    5.703$\times10^{42}$ &    1.198$\times10^{40} $\\
$+$2.85  &    5.422$\times10^{42}$ &    1.051$\times10^{40} $\\
$+$4.07  &    5.147$\times10^{42}$ &    4.063$\times10^{40} $\\
$+$4.84  &    4.779$\times10^{42}$ &    2.826$\times10^{40} $\\
$+$6.04  &    4.454$\times10^{42}$ &    5.467$\times10^{40} $\\
$+$7.03  &    4.117$\times10^{42} $&    3.410$\times10^{40} $\\
$+$7.98  &    3.794$\times10^{42} $&    1.229$\times10^{40} $\\
$+$8.71  &    3.523$\times10^{42} $&    2.079$\times10^{40} $\\
$+$10.03  &    3.253$\times10^{42}$ &    8.012$\times10^{39}$ \\
$+$10.99  &    2.974$\times10^{42}$ &    1.409$\times10^{40}$ \\
$+$11.96  &    2.775$\times10^{42} $&    1.575$\times10^{40}$ \\
$+$12.69  &    2.665$\times10^{42} $&    9.730$\times10^{39}$ \\
$+$13.94  &    2.490$\times10^{42} $&    1.162$\times10^{40}$ \\
$+$14.87  &    2.389$\times10^{42} $&    5.700$\times10^{39} $\\
$+$16.92  &    2.231$\times10^{42} $&    7.396$\times10^{39} $\\
$+$17.76  &    2.170$\times10^{42} $&    6.581$\times10^{39} $\\
$+$18.95  &    2.056$\times10^{42} $&    1.178$\times10^{40} $\\
$+$20.91  &    1.851$\times10^{42} $&    2.629$\times10^{40} $\\
$+$21.92  &    1.710$\times10^{42} $&    2.391$\times10^{40} $\\
$+$23.92  &    1.446$\times10^{42} $&    1.966$\times10^{40} $\\
$+$24.93  &    1.332$\times10^{42} $&    1.339$\times10^{40} $\\
$+$25.98  &    1.229$\times10^{42} $&    6.977$\times10^{39} $\\
$+$26.92  &    1.145$\times10^{42} $&    5.039$\times10^{39} $\\
$+$27.86  &    1.058$\times10^{42} $&    3.160$\times10^{39} $\\
$+$29.83  &    9.329$\times10^{41} $&    6.400$\times10^{39} $\\
$+$30.87  &    8.778$\times10^{41} $&    4.627$\times10^{39} $\\
$+$32.26  &    8.194$\times10^{41} $&    6.796$\times10^{39} $\\
$+$34.78  &    7.270$\times10^{41} $&    1.390$\times10^{39} $\\
$+$35.78  &    6.929$\times10^{41} $&    2.545$\times10^{39} $\\
$+$36.33  &    6.682$\times10^{41} $&    1.823$\times10^{39} $\\
$+$37.78  &    6.379$\times10^{41} $&    1.755$\times10^{39} $\\
$+$38.74  &    6.131$\times10^{41} $&    2.644$\times10^{39} $\\
$+$39.80  &    5.876$\times10^{41} $&    3.588$\times10^{39} $\\
$+$40.91  &    5.640$\times10^{41} $&    3.475$\times10^{39} $\\
$+$42.77  &    5.244$\times10^{41} $&    3.100$\times10^{39} $\\
$+$43.39  &    5.041$\times10^{41} $&    3.261$\times10^{39} $\\
$+$44.72  &    4.822$\times10^{41} $&    2.960$\times10^{39} $\\
$+$48.68  &    4.111$\times10^{41} $&    1.128$\times10^{39} $\\
$+$49.68  &    3.958$\times10^{41} $&    4.050$\times10^{39}$ \\
$+$50.84  &    3.828$\times10^{41} $&    1.457$\times10^{39} $\\
$+$51.63  &    3.688$\times10^{41} $&    2.102$\times10^{39} $\\
$+$52.68  &    3.558$\times10^{41} $&    3.110$\times10^{38} $\\
$+$53.63  &    3.435$\times10^{41} $&    1.297$\times10^{39} $\\
$+$56.62  &    3.127$\times10^{41} $&    1.760$\times10^{39} $\\
$+$57.64  &    3.037$\times10^{41} $&    3.112$\times10^{39} $\\
$+$58.59  &    2.946$\times10^{41} $&    3.324$\times10^{39} $\\
$+$65.54  &    2.412$\times10^{41} $&    1.422$\times10^{39} $\\
$+$66.56  &    2.294$\times10^{41} $&    1.818$\times10^{39} $\\
$+$69.49  &    2.090$\times10^{41} $&    2.981$\times10^{38} $\\
\hline
\end{tabular}
\tablefoot{
\tablefoottext{a}{Days relative to $L_{\mathrm{max}}$, i.e., JD$-$2,454,421.6.}
\tablefoottext{b}{Assuming $\mu = 31.27$ mag to set the absolute flux scale.}
\tablefoottext{c}{Uncertainties represent {\tt SNooPy} fitting errors obtained via Monte Carlos  simulations, and do not include the error in the adopted distance.}
}
\end{table*}

%% file: BOLOUVOIR_11iv.tex
\begin{table*}
\caption{ {\tt SNooPy} UVOIR light curve of SN~2011iv.}             
\label{TAB:UVOIR11iv}     
\centering          
\begin{tabular}{ccc}    
\hline\hline       
Phase (days)\tablefootmark{a} &
$L_{bol}$ (erg s$^{-1}$)\tablefootmark{b}&
$Uncertainty$ (erg s$^{-1}$)\tablefootmark{c}\\
\hline 
\noalign{\smallskip}
$-$2.60  &    9.327$\times10^{42}$ &    3.920$\times10^{40}$\\
$-$1.78  &    9.224$\times10^{42}$ &    2.312$\times10^{40}$\\
$-$0.99  &    9.309$\times10^{42}$ &    1.307$\times10^{40}$\\
$+$0.40  &    9.186$\times10^{42}$ &    2.010$\times10^{39}$\\
$+$1.35  &    8.996$\times10^{42}$ &    1.005$\times10^{39}$\\
$+$2.29  &    8.891$\times10^{42}$ &    3.116$\times10^{40}$\\
$+$3.38  &    8.430$\times10^{42}$ &    4.221$\times10^{40}$\\
$+$4.30  &    8.045$\times10^{42}$ &    6.834$\times10^{40}$\\
$+$4.98  &    7.731$\times10^{42}$ &    7.739$\times10^{40}$\\
$+$6.30  &    6.972$\times10^{42}$ &    7.638$\times10^{40}$\\
$+$7.16  &    6.574$\times10^{42}$ &    6.734$\times10^{40}$\\
$+$8.28  &    5.978$\times10^{42}$ &    5.829$\times10^{40}$\\
$+$8.95  &    5.662$\times10^{42}$ &    7.638$\times10^{40}$\\
$+$10.22  &    5.177$\times10^{42}$ &    3.518$\times10^{40}$\\
$+$11.26  &    4.735$\times10^{42}$ &    3.518$\times10^{40}$\\
$+$12.25  &    4.445$\times10^{42}$ &    1.910$\times10^{40}$\\
$+$13.22  &    4.242$\times10^{42}$ &    1.809$\times10^{40}$\\
$+$15.41  &    3.875$\times10^{42}$ &    2.010$\times10^{40}$\\
$+$17.22  &    3.691$\times10^{42}$ &    2.111$\times10^{40}$\\
$+$18.17  &    3.593$\times10^{42}$ &    2.915$\times10^{40}$\\
$+$19.17  &    3.437$\times10^{42}$ &    4.824$\times10^{40}$\\
$+$20.14  &    3.318$\times10^{42}$ &    4.020$\times10^{40}$\\
$+$21.07  &    3.170$\times10^{42}$ &    3.618$\times10^{40}$\\
$+$22.10  &    2.990$\times10^{42}$ &    2.714$\times10^{40}$\\
$+$23.11  &    2.759$\times10^{42}$ &    9.347$\times10^{40}$\\
$+$24.09  &    2.633$\times10^{42}$ &    9.045$\times10^{40}$\\
$+$25.08  &    2.368$\times10^{42}$ &    1.266$\times10^{41}$\\
$+$26.09  &    2.240$\times10^{42}$ &    6.935$\times10^{40}$\\
$+$27.11  &    2.072$\times10^{42}$ &    4.322$\times10^{40}$\\
$+$28.10  &    1.957$\times10^{42}$ &    4.020$\times10^{40}$\\
$+$29.09  &    1.832$\times10^{42}$ &    3.015$\times10^{40}$\\
$+$29.72  &    1.726$\times10^{42}$ &    3.015$\times10^{40}$\\
$+$32.05  &    1.479$\times10^{42}$ &    1.709$\times10^{40}$\\
$+$33.07  &    1.432$\times10^{42}$ &    2.211$\times10^{40}$\\
$+$34.04  &    1.366$\times10^{42}$ &    1.608$\times10^{40}$\\
$+$35.03  &    1.273$\times10^{42}$ &    1.206$\times10^{40}$\\
$+$36.03  &    1.227$\times10^{42}$ &    1.809$\times10^{40}$\\
$+$37.01  &    1.186$\times10^{42}$ &    1.809$\times10^{40}$\\
$+$38.01  &    1.100$\times10^{42}$ &    1.407$\times10^{40}$\\
$+$43.99  &    8.850$\times10^{41}$ &    1.508$\times10^{40}$\\
$+$48.01  &    7.733$\times10^{41}$ &    8.442$\times10^{39}$\\
$+$50.00  &    7.357$\times10^{41}$ &    1.809$\times10^{39}$\\
$+$51.99  &    6.680$\times10^{41}$ &    7.035$\times10^{39}$\\
$+$56.93  &    5.841$\times10^{41}$ &    6.030$\times10^{38}$\\
$+$57.98  &    5.739$\times10^{41}$ &    2.513$\times10^{39}$\\
$+$59.97  &    5.387$\times10^{41}$ &    2.010$\times10^{38}$\\
$+$61.94  &    5.108$\times10^{41}$ &    1.106$\times10^{39}$\\
$+$63.89  &    4.831$\times10^{41}$ &    8.040$\times10^{38}$\\
$+$66.90  &    4.451$\times10^{41}$ &    6.030$\times10^{38}$\\
$+$75.82  &    3.552$\times10^{41}$ &    2.010$\times10^{39}$\\
$+$79.75  &    3.200$\times10^{41}$ &    2.613$\times10^{39}$\\
$+$85.76  &    2.768$\times10^{41}$ &    1.608$\times10^{39}$\\
$+$86.70  &    2.690$\times10^{41}$ &    6.030$\times10^{38}$\\
$+$88.69  &    2.538$\times10^{41}$ &    1.005$\times10^{38}$\\
$+$91.66  &    2.394$\times10^{41}$ &    1.005$\times10^{39}$\\
\hline                  
\end{tabular}
\tablefoot{
\tablefoottext{a}{Days relative to $L_{\mathrm{max}}$, i.e., JD$-$2,455,907.6.}
\tablefoottext{b}{Assuming $\mu = 31.27$ mag to set the absolute flux scale.}
\tablefoottext{c}{Uncertainties represent {\tt SNooPy} fitting errors obtained via Monte Carlos  simulations, and do not include the error in the adopted distance.}
}
\end{table*}

%% file: BOLOUVOIR_SPECS_11iv.tex
\begin{table*}
\caption{UVOIR light curve of SN~2011iv derived from spectra.}             
\label{TAB:UVOIR11iv2}      
\centering          
\begin{tabular}{cccc}     
\hline\hline       
Phase (days)\tablefootmark{a} &
$L_{bol}$ (erg s$^{-1}$)\tablefootmark{b}&
Wavelength range (\AA\ )&
Instrument\\
\hline 
\noalign{\smallskip}
$-$0.12  &    $8.813\times10^{42}$  &      1615.50--24776.80 &       STIS, FIRE\\
$+$3.88  &   $7.068\times10^{42}$  &      1626.50--24776.80 &       STIS, FIRE\\
$+$6.88  &    $5.806\times10^{42}$  &      3347.32--24776.80 &       ISIS, FIRE\\
$+$7.88  &    $5.481\times10^{42}$  &      3331.15--24776.80 &       ISIS, FIRE\\
$+$8.88  &    $5.168\times10^{42}$  &      1620.00--24776.80 &       STIS, FIRE\\
$+$8.88  &    $5.055\times10^{42}$  &      3293.07--24776.80 &       ISIS, FIRE\\
$+$9.88  &    $4.632\times10^{42}$  &      3296.12--24776.80 &       ISIS, FIRE\\
$+$10.13  &    $4.447\times10^{42}$  &      3698.07--24776.80 &      WFCCD, FIRE\\
$+$12.88  &    $3.981\times10^{42}$  &      1624.50--24776.80 &       STIS, FIRE\\
$+$16.88  &    $3.380\times10^{42}$  &      1624.50--24776.80 &       STIS, FIRE\\
$+$20.13  &    $3.011\times10^{42}$  &      3632.00--24776.80 &      WFCCD, FIRE\\
$+$20.88  &    $3.033\times10^{42}$  &      1627.50--24776.80 &       STIS, FIRE\\
$+$28.88  &    $1.856\times10^{42}$  &      1635.00--24776.80 &       STIS, FIRE\\
$+$38.88  &    $1.019\times10^{42}$  &      3628.00--24776.80 &      WFCCD, FIRE\\
\hline                  
\end{tabular}
\tablefoot{
\tablefoottext{a}{Days relative to $L_{\mathrm{max}}$, i.e., JD$-$2,455,907.6.}
\tablefoottext{b}{Assuming $\mu = 31.27$ mag to set the absolute flux scale.}
}
\end{table*}